\newif\ifdraft
\newif\ifpreprint
\preprinttrue     %version for arxiv
\def\snntitle{$\snn$}
\ifpreprint
\def\snntitle{$\snnbf$}
\fi

\def\dtitle{Measurement of the radius dependence of charged-particle jet suppression in Pb--Pb collisions at \snntitle = 5.02 TeV}
\def\stitle{$R$-dependence of jet suppression} 
\documentclass[ALICE,manyauthors]{cernphprep}
\usepackage[comma,square,numbers,sort&compress]{natbib}

\usepackage{hyperref}
\usepackage{graphicx}  % needed for figures
\usepackage{dcolumn}   % needed for some tables
\usepackage{bm}        % for math
\usepackage{amssymb}   % for math
\usepackage{amsfonts}
\usepackage{graphics}
\usepackage{grffile}   % to handle dots in graphics file names
\usepackage{epsfig}
\usepackage{units}
\usepackage{comment}
\usepackage[usenames]{color}
\usepackage[normalem]{ulem} % for strikethroughs (\sout{})
\usepackage{color}
\usepackage[utf8]{inputenc}
\usepackage[T1]{fontenc}
\usepackage{subfigure}
\usepackage{orcidlink}

\newcommand{\com}[1]       {\relax}

\newcommand{\pp}           {pp}

\newcommand{\AaAa}         {\mbox{AA}}
\newcommand{\PbPb}         {\mbox{Pb--Pb}}

\newcommand{\pt}           {\ensuremath{p_{\mathrm{T}}}{ }}

\newcommand{\snn}          {\ensuremath{\sqrt{s_{\mathrm{NN}}}}}
\newcommand{\snnbf}        {\ensuremath{\mathbf{{\sqrt{\textit{s}_{NN}}}}}}

\newcommand{\Fig}[1]       {Fig.~\ref{#1}}
\newcommand{\Figure}[1]    {Figure~\ref{#1}}
\newcommand{\Sec}[1]       {Sec.~\ref{#1}}

\newcommand{\Refe}[1]      {Ref.~\cite{#1}}

\newcommand{\red}[1]       {\textcolor{red}{#1}}

\newcommand{\warn}[1]      {{\small\textbf{\red{(!}\footnote{\textbf{\red{(!)}}~#1}\red{)}}}\marginpar{\textbf{\red{---}}}}

\renewcommand{\xout}[1]    {}

\newcommand{\pT}           {\ensuremath{p_{\mathrm{T}}}}

\newcommand{\FastJet}      {\ensuremath{\mathrm{FastJet}}}
\newcommand{\kT}           {\ensuremath{k_{\mathrm{T}}}}
\newcommand{\jetraw}       {\ensuremath{{p}_{\mathrm{T,\,jet}}^{\mathrm{raw}}}}

\newcommand{\GeVc}         {GeV/$c$}

\graphicspath{{./img/}}
\ifdraft
\usepackage{lineno}
\linenumbers
\else
\renewcommand{\warn}[1]{}

\fi

\begin{document}
%==========================================================%
\newlength{\figlen}
\setlength{\figlen}{0.95\linewidth}
\begin{titlepage}
\PHyear{2023}
\PHnumber{027}                  % required, obtained from PH
\PHdate{28 February}            % required
\title{\dtitle}
\ShortTitle{\stitle}
\Collaboration{ALICE Collaboration\thanks{See Appendix~\ref{app:collab} for the list of collaboration members}}
\ShortAuthor{ALICE Collaboration} % appears on left page headers, do not change
%==========================================================%
\begin{abstract}
The ALICE Collaboration reports a differential measurement of inclusive jet suppression using \pp\ and \PbPb\ collision data at a center-of-mass energy per nucleon--nucleon collision $\snn = 5.02$ TeV. 
Charged-particle jets are reconstructed using the anti-$k_{\rm T}$ algorithm with resolution parameters $R =$ 0.2, 0.3, 0.4, 0.5, and 0.6 in \pp\ collisions and $R =$ 0.2, 0.4, 0.6 in central (0--10\%), semi-central (30--50\%), and peripheral (60--80\%) \PbPb\ collisions. A novel approach based on machine learning is employed to mitigate the influence of jet background. This enables measurements of inclusive jet suppression in new regions of phase space, including down to the lowest jet $\pT \geq 40$ GeV/$c$ at $R = 0.6$ in central \PbPb\ collisions. This is an important step for discriminating different models of jet quenching in the quark--gluon plasma.
The transverse momentum spectra, nuclear modification factors, derived cross section, and nuclear modification factor ratios for different jet resolution parameters of charged-particle jets are presented and compared to model predictions.
A mild dependence of the nuclear modification factor ratios on collision centrality and resolution parameter is observed. The results are compared to a variety of jet-quenching models with varying levels of agreement.
\end{abstract}
\end{titlepage}

\setcounter{page}{2}

%==========================================================%
%==========================MAIN============================%
%==========================================================%
%%%%%%%%%%%%%%%%%%%%%%%%%%%%%%%%%%%%%%%%%%%%%%%%%%%%%%%%%%%%%%%%%%%%%%%%%%%%%%%%%%%%%%%%%%%%%%%%%%%%
\section{Introduction}
\label{sec:intro}
%%%%%%%%%%%%%%%%%%%%%%%%%%%%%%%%%%%%%%%%%%%%%%%%%%%%%%%%%%%%%%%%%%%%%%%%%%%%%%%%%%%%%%%%%%%%%%%%%%%%
Lattice quantum chromodynamics (QCD) calculations predict that strongly-interacting matter at very high
temperature exists in a phase called the quark--gluon plasma (QGP), where the partonic constituents, quarks and gluons, are not confined to hadrons. There is compelling evidence from observations reported by experiments at the Relativistic Heavy Ion Collider~(RHIC)~\cite{Arsene20051,Back200528,Adcox2005184,Adams2005102} and at the Large Hadron Collider~(LHC)~\cite{Aamodt:2010pb,Aamodt:2010cz,Chatrchyan:2011pb,Aamodt:2011mr,Aamodt:2010pa,ATLAS:2011ah,Chatrchyan:2012wg,ALICE:2011ab,Aad:2013xma,Chatrchyan:2013kba,Aamodt:2010jd,Chatrchyan:2011sx, ALICE:2022wpn} that the QGP is created in high-energy nuclear collisions.

High momentum transfer (hard) QCD scatterings of partons occur early in the heavy-ion collision evolution, producing high transverse momentum~(\pT) partons that propagate through the medium and eventually fragment into collimated sprays of hadrons known as jets. 
Since jet production in proton--proton (\pp) collisions is well described by perturbative QCD~\cite{ATLAS:2017ble, CMS:2016jip, Marzani:2019hun, CMS:2020caw}, measuring modifications to jet production and jet properties in heavy-ion collisions offers a powerful way to characterize the properties of the QGP. The high-$\pT$ partons within the jet experience in-medium interactions through elastic scatterings and induced gluon radiation, a phenomenon called jet quenching (see Ref.~\cite{Cunqueiro:2021wls} for a recent review). % ~\cite{Gyulassy:1990ye,Baier:1994bd}
Jet quenching leads to several observable consequences: parton energy loss, modification of the jet substructure, and medium-induced acoplanarity.
Jet quenching has been measured via inclusive yield and correlation measurements of high-$\pT$ hadrons and reconstructed jets, semi-inclusive jet measurements, jet shapes, and recently via jet substructure measurements at RHIC~\cite{Adcox:2001jp,Adler:2002tq,Adler:2002xw,Adcox:2002pe,Adler:2003qi,Adams:2003kv,Adams:2003im,Back:2003qr,Arsene:2003yk,Adams:2006yt,Adare:2006nr,Adare:2008cg,Adamczyk:2013jei,Adamczyk:2017yhe} and at the LHC~\cite{Aamodt:2010jd,Aad:2010bu,Chatrchyan:2011sx,Chatrchyan:2011pb,Aamodt:2011vg,CMS:2012aa,Chatrchyan:2012nia,Chatrchyan:2012gw,Chatrchyan:2012gt,Aad:2012vca,Chatrchyan:2013exa,Chatrchyan:2013kwa,Chatrchyan:2014ava,Aad:2014wha,Aad:2014bxa,Adam:2015doa,Acharya:2017goa,Acharya:2018uvf,Sirunyan:2018gct,ATLAS:2018bvp,Acharya:2019ssy, CMS:2017qlm, ALICE:2022vsz}.

Jet energy loss results in a suppression of the jet yield at a fixed value of the jet $\pT$. 
Jet suppression is quantified using the nuclear modification factor, 
\begin{equation}
\label{eq:RAA}
R_\mathrm{AA} = \frac{1}{\langle T_\mathrm{AA}\rangle} \frac{\mathrm{d}^{2}N/\mathrm{d}{p}_\mathrm{T} \mathrm{d}{\eta}}{{\mathrm{d}^{2}\sigma_\mathrm{pp}/\mathrm{d}{p}_\mathrm{T} \mathrm{d}{\eta}}},
\end{equation}
which is the ratio of the measured per-event inclusive jet yield in heavy-ion (\AaAa{}) collisions and the inclusive cross sections in \pp\ collisions scaled by the nuclear overlap in a given centrality class $T_\mathrm{AA}$~\cite{dEnterria:2020dwq}. 
The value of $R_{\rm AA}$ is expected to be one in the absence of nuclear effects. 

It is important to measure jet suppression over a wide range of parameters, including the jet $\pT$ and the resolution parameter (so-called radius), $R$, of the clustering algorithm since the influence of in-medium effects and the medium response are expected to vary with these parameters~\cite{Pablos:2019ngg, KunnawalkamElayavalli:2017hxo, Li:2018xuv, He:2018xjv}. 
Measuring jets at large $R$ is especially interesting because more of the larger-angle medium-induced modification will be recovered relative to jets with smaller $R$.
Additionally, the contribution of the medium response relative to other effects is expected to vary with $R$~\cite{Pablos:2019ngg}.
These competing effects may help to discriminate the mechanisms underlying energy loss and elucidate the energy transport properties of the QGP. 
While jet suppression has been measured over a large range in jet $\pT{}$ at both the LHC and RHIC~\cite{STAR:2020xiv, ALICE:2013dpt, ALICE:2019qyj, ATLAS:2014ipv, CMS:2016uxf}, of particular interest are the low-$\pT$ and large-$R$ regions.
A recent measurement by the CMS collaboration~\cite{CMS:2021vui} studied jet suppression up to $R=1.0$ for jets with high $\pT>200$~\GeVc. 
Measurements of jet suppression as a function of $R$ were found to have excellent discriminating power when compared to various jet quenching models~\cite{CMS:2021vui}. However, no significant radial dependence was observed, which implies that there may be a combination of competing jet-quenching effects. The ATLAS collaboration~\cite{ATLAS:2012tjt} also studied the $R$-dependence of the ratios of jet spectra measured in central and peripheral collisions ($R_{\rm CP}$) at lower \pT, $40 < \pT < 200$~\GeVc, and found a dependence on $R$ where jets with larger $R$ up to $R=0.5$ exhibit less suppression. Measuring the $R$-dependence of energy loss~\cite{Pablos:2019ngg,Li:2018xuv, He:2018xjv, KunnawalkamElayavalli:2017hxo} at low $\pT$ will probe the expectation that the $R$-dependence is larger in this region~\cite{Pablos:2019ngg}, and will connect to inclusive jet measurements at RHIC~\cite{STAR:2020xiv}.

The ALICE experiment at the LHC is well-suited to perform jet measurements at low jet $\pT$ at the LHC due to high-precision tracking in the Time Projection Chamber~(TPC)~\cite{Alme:2010ke} and Inner Tracking System~(ITS)~\cite{ALICE:2010tia}. 
However, reconstructing the jet $\pT$ in nucleus--nucleus collisions is challenging due to the large background fluctuations from the underlying event (UE), which can be a significant fraction of the jet $\pT$ itself. 
Jet measurements in heavy-ion collisions require a procedure to account for this background which involves both a correction of the jet $\pT$ and a suppression of combinatorial (or fake) jets. 

One common procedure applied for the jet $\pT$-smearing is to correct for the average background via a pedestal subtraction of the event-wise momentum density~(herein referred to as the area-based or AB method~\cite{Abelev:2012ej, Cacciari:2011ma}). 
This is accompanied by an additional correction for event-averaged residual smearing effects via an unfolding procedure. 
Contributions from combinatorial jets to the inclusive jet yield can be further suppressed via additional requirements on the jet acceptance, such as a leading hadron $\pT$ requirement. 
The drawback of such requirements is a bias of the jet population.

A generalization of this procedure is to utilize machine learning~(ML) techniques to include multi-dimensional information in calculating the corrected jet momentum, as explored in Ref. ~\cite{Haake:2018hqn}.
This approach uses specific properties of the jet and its constituents in addition to the area-based corrected jet $\pT$ to reduce the residual fluctuations and remove combinatorial jets from the inclusive jet yield.
An unfolding procedure must be applied to correct for the contributions of residual smearing effects, which can be performed down to lower jet $\pt$ due to the improved jet $p_{\rm T}$ resolution.
Unique to this procedure is that the correction for the jet energy scale and the removal of combinatorial jets from the inclusive jet yield is done in one step. 
This additional constraining power is achieved by creating a mapping between the jet properties and the corrected jet $\pT$, which may also provide the opportunity to measure jets in \PbPb\ collisions with lower jet  $\pT$ and larger $R$ than is possible with the AB method.
However, including constituent information in the training of the ML model introduces a dependence on fragmentation patterns of the training sample, which may differ from those in \PbPb\ collision data, whose effect on the results needs to be addressed.

In this manuscript, we present an analysis of inclusive charged-particle jet production at a center-of-mass energy per nucleon--nucleon collision of \snn\ = 5.02 TeV. 
Jets are measured with resolution parameters $R =$ 0.2, 0.3, 0.4, 0.5, and 0.6 in \pp\ and $R$ = 0.2, 0.4, and 0.6 in \PbPb\ collisions. The jets in \PbPb\ collisions are also measured for different centrality classes. 
The analysis of the \PbPb\ collision data in the 0--10\% and the 30--50\% centrality classes utilizes the novel approach to the correction of the underlying event contribution based on the ML techniques described above, while the analysis of the \PbPb\ collision data in the 60--80\% uses the standard area-based subtraction. The jet transverse momentum spectra, jet nuclear modification factors, as well as ratios of jet cross sections and $R_\mathrm{AA}$ are presented and compared to model calculations in central (0--10\%), semi-central (30--50\%), and peripheral (60--80\%) \PbPb\ collisions.
The dependence on the jet fragmentation model used to train the ML algorithm was studied and incorporated as a systematic uncertainty.
The new analysis extends previous measurements of inclusive charged-particle jet suppression at the LHC to both lower $\pT$ and larger $R$, measuring jets down to $\pT=30$~\GeVc\ for $R=0.4$ and to $\pT=40$~\GeVc\ for $R=0.6$. 

The article is structured as follows:
details on the detector and data reconstruction are given in \Sec{sec:setup}.
The jet reconstruction is described in \Sec{sec:reconstruction}.
The jet background correction method based on ML techniques is introduced in \Sec{sec:mlestimator}.
The dependence of the new background estimator on the fragmentation pattern used in the training and unfolding is discussed in \Sec{sec:fragdependence}.
The systematic uncertainties are discussed in \Sec{sec:systematics}.
The results and comparison with model calculations are presented in \Sec{sec:results}.
A summary concludes the paper in \Sec{sec:summary}. Appendix~\ref{sec:appendix} describes the insensitivity of the ML correction to correlated background fluctuations.

%%%%%%%%%%%%%%%%%%%%%%%%%%%%%%%%%%%%%%%%%%%%%%%%%%%%%%%%%%%%%%%%%%%%%%%%%%%%%%%%%%%%%%%%%%%%%%%%%%%%
\section{Experimental setup}
\label{sec:setup}
%%%%%%%%%%%%%%%%%%%%%%%%%%%%%%%%%%%%%%%%%%%%%%%%%%%%%%%%%%%%%%%%%%%%%%%%%%%%%%%%%%%%%%%%%%%%%%%%%%%%
A detailed description of the ALICE detector can be found in Ref.~\cite{Aamodt:2008zz}, and its performance is described in Ref.~\cite{Abelev:2014ffa}.

The analyzed dataset for \PbPb\ collisions at $\snn=5.02$ TeV was collected in 2015, with online triggers that utilize the hit multiplicity measured by forward V0 detectors. The V0 detectors are segmented scintillators covering the full azimuth over the pseudorapidity ranges $2.8<\eta<5.1$~(V0A) and $-3.7<\eta<-1.7$~(V0C).
The accepted events, reconstructed as described in \Refe{Abelev:2012hxa}, were required to have a reconstructed primary vertex within $\pm 10$~cm from the nominal interaction point along the beam axis and the obtained sample corresponds to an integrated luminosity of about 6.5 $\mu$b$^{-1}$.
Events were characterized with V0 multiplicities corresponding to the 0--10\%, 30--50\%, and 60--80\% centrality ranges using the centrality determination described in \Refe{Abelev:2013qoq}. 
The 0--10\% centrality range corresponds to the most central 10\% of the Pb--Pb inelastic cross section and 60--80\% corresponds to more peripheral collisions. 

The analyzed dataset for \pp\ collisions at $\sqrt{s}=5.02$ TeV was collected in 2017 during Run 2 of the LHC with an integrated luminosity of about 19 nb$^{-1}$~\cite{ppXsec}. 
Events were triggered using the V0 detector by having signals in both the V0A and V0C. Accepted events were required to have a reconstructed primary vertex within $\pm 10$~cm from the nominal interaction point along the beam axis, the same as the events in \PbPb\ collisions.

This analysis utilizes the ALICE tracking system in the central rapidity region, which is located inside a large solenoidal magnet with a field strength of $0.5$~T aligned with the beam axis. 
This system consists of the ITS, a high-precision six-layer cylindrical silicon detector with the innermost layer at $3.9$~cm and the outermost layer at $43$~cm radial distance from the beam axis; and the TPC with radial extent of $85$--$247$~cm, which provides up to 159 independent space points per track.
To ensure good track-momentum resolution for jet reconstruction, reconstructed tracks are required to have at least three hits in the ITS.
For tracks without any hit in the Silicon Pixel Detector (SPD), comprising the two innermost layers of the ITS, the location of the primary vertex is used to constrain the track. This approach improves the track momentum resolution and reduces the azimuthal variation in the track reconstruction efficiency arising from the non-uniform SPD acceptance.
Accepted tracks are required to have $\pT > 0.15$~\GeVc\ and $|\eta|<0.9$, 
and to have at least 70 TPC space-points, comprising no fewer than 80\% of the geometrically findable space-points in the TPC. 

For \pp\ collisions, the single-track reconstruction efficiency is estimated using \pp\ events generated with the PYTHIA 8 (Monash 2013 tune)~\cite{Sjostrand:2014zea} generator together with the GEANT 3-based detector simulation and response model of ALICE~\cite{geant3ref2}.
The efficiency is approximately 67\% for track $\pT = 0.15$~GeV/$c$, rising to approximately 84\% at track $\pT = 1$~GeV/$c$ and remaining above 75\% at higher track \pT~\cite{ALICE:2019qyj}.
The tracking efficiency in 0--10\% \PbPb\ collisions as compared to that in pp collisions was estimated by comparing central to peripheral HIJING+GEANT 3~\cite{Wang:1991hta} events, resulting in an approximately 2\% reduction in the tracking efficiency as compared to pp,  independent of the track $\pT$.
The momentum resolution in pp collisions at the primary vertex, which is determined on a track-by-track basis using a Kalman filter approach~\cite{Belikov:689414}, is about $1\%$ at a track $p_{\rm T}$ of $1$~\GeVc\ 
and about $4$\% at $50$~\GeVc. In heavy-ion collisions, the momentum resolution at high track $p_{\rm T}$ is approximately 10--15\% worse than in \pp\ collisions.
The contamination by secondary particles~\cite{ALICE-PUBLIC-2017-005} produced in particle--material interactions, conversions, and weak decays of long-lived particles, is a few percent of the yield.

For the reference spectra from \pp\ collisions, the jet measurement is carried out as described in \Refe{Acharya:2019tku} on the 2017 pp dataset at $\sqrt{s}=5.02$~TeV. This dataset is 10 times larger than the dataset used in \Refe{Acharya:2019tku}, extending the charged-particle jet spectra for all considered $R$ values up to jet $p_{\rm T} =$140~GeV/$c$.

The values for the $\langle T_\mathrm{AA} \rangle $ in the nuclear modification factor Eq. (\ref{eq:RAA}) were computed in a Glauber model~\cite{Loizides:2017ack} to be 23.26$\pm 0.168$, 3.917$\pm 0.0645$, and 0.4188$\pm 0.0106$ $\text{mb}^{-1}$ in central~(0--10\%), semi-central~(30--50\%), and peripheral~(60--80\%) collisions, respectively. 

%%%%%%%%%%%%%%%%%%%%%%%%%%%%%%%%%%%%%%%%%%%%%%%%%%%%%%%%%%%%%%%%%%%%%%%%%%%%%%%%%%%%%%%%%%%%%%%%%%%%
\section{Jet reconstruction and unfolding}
\label{sec:reconstruction}
%%%%%%%%%%%%%%%%%%%%%%%%%%%%%%%%%%%%%%%%%%%%%%%%%%%%%%%%%%%%%%%%%%%%%%%%%%%%%%%%%%%%%%%%%%%%%%%%%%%%

Charged-particle jets are reconstructed using the anti-\kT\ algorithm with $E$-scheme recombination~\cite{Cacciari:2008gp, catchment}  in the \FastJet\ package~\cite{Cacciari:2011ma} with resolution parameters $R =$ 0.2, 0.3, 0.4, 0.5, and 0.6 in pp collisions and $R=0.2$, $0.4$, and $0.6$ in Pb--Pb collisions.
Jet candidates are accepted for further analysis if their axis, defined using the standard axis~\cite{Cal:2019gxa}, is reconstructed within the pseudorapidity range $|\eta_{\rm jet}|<0.9-R$ to assure that the nominal jet cone is fully contained within the track acceptance of $|\eta|<0.9$. 
A jet area cut of $A_\mathrm{jet} > 0.56 \pi R^2$ is applied to suppress contamination by non-physical jets~\cite{ALICE:2013dpt, Adamczyk:2017yhe, Adam:2015doa}. Jets containing a track with $p_{\rm T} > 100$~GeV/$c$ are additionally removed due to reduced momentum resolution in this region. 
In this paper, jets are corrected with either an area-based or ML-based background correction. 

The transverse momentum of reconstructed jets is affected both by residual fluctuations from the UE remaining due to imperfect background subtraction as well as detector effects (mainly the tracking efficiency and the track $p_{\rm T}$ resolution). 
To account for these effects, pp collision events were simulated with the PYTHIA 8 generator using the Monash 2013 tune~\cite{Skands:2014pea} (particle-level) and the particles were then propagated through a model of the ALICE detector using GEANT 3 particle
transport framework~\cite{geant3ref2}~(detector-level). 
These events were then embedded into \PbPb\ minimum bias data to form hybrid events (hybrid-level). 
In these events, the same detector configuration is simulated as that utilized during the data taking of the above-mentioned \PbPb\ dataset. 
To account for a reduction in tracking efficiency for central and semi-central \PbPb\ collisions relative to pp and peripheral \PbPb\ collisions (where the tracking efficiency is similar), 2\% of tracks were randomly rejected, independent of the track $p_{\rm T}$.

Particle-level and hybrid-level jets are matched by the following two-step procedure. First, the hybrid-level jet is matched geometrically to a detector-level jet, where only matches with a maximum distance of $0.75\times R$ are accepted. 
The hybrid and detector-level jets are additionally required to share particles responsible for at least 50\% of the jet $p_{\rm T}$. 
Then the detector-level jet is matched geometrically with a maximum distance of $0.75\times R$ to a particle-level jet. 
These matched jets form a correspondence between the true and reconstructed-level (hybrid-level) jet $p_{\rm T}$, which is then used to fill a response matrix to reflect this mapping. The jet reconstruction efficiency, defined as the ratio of the number of accepted detector-level jets geometrically matched to a particle-level jet and the number of particle-level jets in a given $p_{\rm T}^{\rm true}$ interval,
\begin{equation}
  \varepsilon_\mathrm{rec} (p_\mathrm{T,\;ch\;jet}^\mathrm{true}) =  N_\mathrm{matched} (p_\mathrm{T,\;ch\;jet}^\mathrm{true})/N_\mathrm{generated} (p_\mathrm{T,\;ch\;jet}^\mathrm{true}),
\end{equation}
accounts also for the efficiency of matching jets. The jet reconstruction efficiency is high (above 95\%) in all regions of phase space. 
It is used to correct the unfolded spectrum.

The spectra are unfolded using the iterative Bayesian approach~\cite{DAgostini:2010hil} implemented in the RooUnfold package~\cite{Adye:2011gm}, with the response matrix described above. The prior distribution for the unfolding is the PYTHIA particle-level distribution.  
The number of iterations, which is the regularization parameter, was selected to be the value where the unfolded result becomes stable compared to further iterations, balanced against the increasing statistical errors. This selected value is referred to as the \textit{nominal} result, while a variation on this value is taken as a systematic uncertainty (see Section~\ref{sec:systematics}). 
A value of 8 iterations is used for the nominal result for the most central collisions. 
The lower limit in the measured transverse jet momentum that serves as input to the unfolding procedure corresponds to five times the width of the distribution of residual fluctuations remaining after background subtraction, to avoid contamination from combinatorial jets ~\cite{ALICE:2013dpt}. 
Some particle-level jets will migrate outside of the measured kinematic range, which is corrected with the so-called kinematic efficiency correction. 
A requirement of a minimum kinematic efficiency of $60\%$ is also imposed on the considered jet $p_\mathrm{T}$ intervals, while lower efficiency regions are rejected.

%%%%%%%%%%%%%%%%%%%%%%%%%%%%%%%%%%%%%%%%%%%%%%%%%%%%%%%%%%%%%%%%%%%%%%%%%%%%%%%%%%%%%%%%%%%%%%%%%%%%
\section{Machine learning-based background correction}
%%%%%%%%%%%%%%%%%%%%%%%%%%%%%%%%%%%%%%%%%%%%%%%%%%%%%%%%%%%%%%%%%%%%%%%%%%%%%%%%%%%%%%%%%%%%%%%%%%%%
\label{sec:mlestimator}
To expand the jet $\pT$ and $R$ reach of the measurement, a novel estimator based on machine learning is used to correct the $\pT$-smearing effects caused by the background.
This new background estimator, introduced and described in detail in \Refe{Haake:2018hqn}, is used for the first time and follows an alternative approach to the established area-based method.
The method utilizes the properties
of each individual jet candidate to assign a correction for the background contribution to
the measured $\pT$ of the jet. While the background is dominated by low-$\pT$ particles, the particles in the jet signal are distributed towards higher $\pT$ constituents~\cite{Abelev:2012ej,ALICE:2018ype}.
However, the relationship between the relevant input features of the jet candidate and the true jet $\pT$ is complex.
Machine learning techniques are powerful tools to approximate this mapping by learning from simulation instead of deriving the relation from expert knowledge alone. This problem represents a regression task, which aims to predict a reconstructed jet $\pT$ value for each jet candidate.
The main physics motivation and performance are described here, while further details on the implementation and validation of this approach are described in Appendix~\ref{sec:machineLearningMethods}.

Measurements of jet shapes have shown that certain features of quenched jets are similar to unmodified jets in vacuum, notably that the core of the jet is mildly modified due to quenching~\cite{Acharya:2018uvf,Acharya:2019ssy}.
This observation motivates the strategy adopted in this analysis to train on jets produced by the PYTHIA 8 generator for pp collisions. 
We assess the systematic uncertainty due to possible variation in the fragmentation in \PbPb\ collisions with respect to \pp\ collisions generated with PYTHIA 8 in \Sec{sec:fragdependence}.

\begin{figure}[t!]
\begin{center}
  \includegraphics[width=0.48\textwidth]{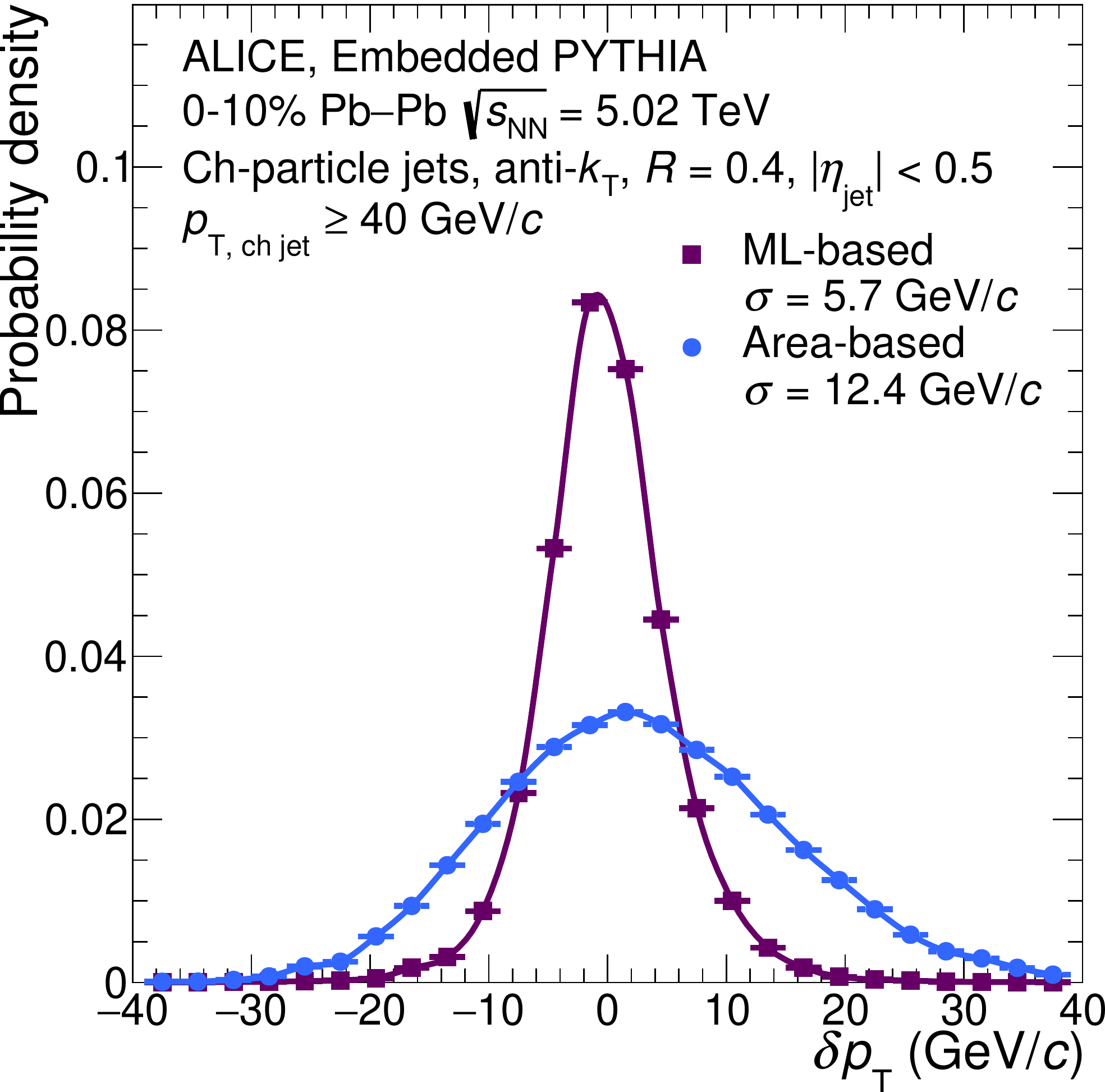}
  \includegraphics[width=0.48\textwidth]{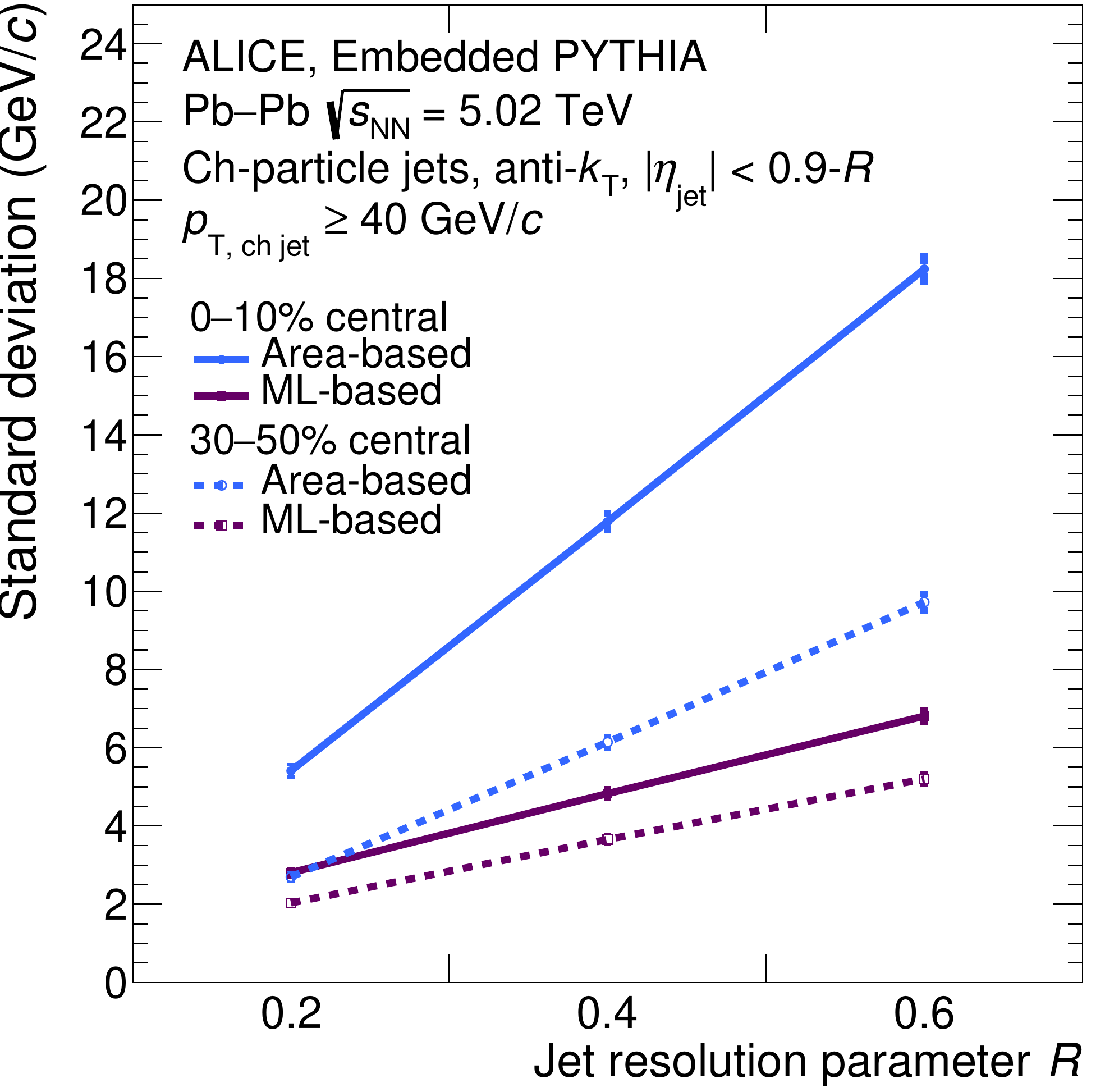}
  \caption{Residual $\pT$-distributions of embedded jet probes of known transverse momentum into \PbPb\ collision data. Left: Comparison of the distributions for the area-based and ML-based background estimators. Note the lines connecting the points do not represent a fit and are only present to guide the eye. Right: Radius dependence of the width of the distributions where the error bars come from the uncertainty in calculating the width. } 
  \label{fig:deltaPt}
\end{center}
\end{figure}

Simulated jets with
known transverse momentum are embedded into real \PbPb\ events to compare the ML background estimator to the area-based estimator.
The left panel of \Fig{fig:deltaPt} shows the distribution of the residual difference ($\delta \pT$) between background-corrected $\pT$ and target detector-level probe $\pT$, which measures how precisely the background is approximated.
The plot compares $\delta \pT$ distributions of the area-based and ML-based background estimators for $R=0.4$. The right panel of \Fig{fig:deltaPt} provides the standard deviation of the residual distributions of both background estimators versus $R$ for different centralities. 
The ML-based estimator has an approximately two times narrower $\delta \pT$ distribution than the area-based estimator, indicating a reduction in residual fluctuations. The performance of the ML-based estimator also has no dependence on the angle between the jet axis and the event plane, which is briefly discussed in Appendix~\ref{sec:appendix}, indicating that the estimator is insensitive to correlated fluctuations in the background.

%%%%%%%%%%%%%%%%%%%%%%%%%%%%%%%%%%%%%%%%%%%%%%%%%%%%%%%%%%%%%%%%%%%%%%%%%%%%%%%%%%%%%%%%%%%%%%%%%%%%
\section{Quenched jet fragmentation dependence}
\label{sec:fragdependence}
\label{sec:toystudies}
%%%%%%%%%%%%%%%%%%%%%%%%%%%%%%%%%%%%%%%%%%%%%%%%%%%%%%%%%%%%%%%%%%%%%%%%%%%%%%%%%%%%%%%%%%%%%%%%%%%%
%%Some of the text below is taken from the original systematic section on the frag bias (which was only evaluated on q/g)
The machine learning-based background estimator is trained on jets generated with PYTHIA 8 simulations for pp collisions, where the fragmentation is known to differ quantitatively from the fragmentation in \PbPb\ data to which the estimator is applied~\cite{CMS:2014jjt,ATLAS:2018bvp,CMS:2018mqn,Acharya:2019ssy}.  
The inclusion of specific fragmentation properties in the learning step of the ML-based background correction procedure introduces an explicit fragmentation dependence. 
In comparison, although the area-based correction itself is not strongly fragmentation-dependent,
this method is often combined with a requirement on the $p_{\rm T}$ of the leading charged constituent to suppress the background contribution, which biases the fragmentation of the jet sample. 
There are three points where the ML-based procedure is sensitive to jet fragmentation: the measured input spectra, the response matrix, and the training. 
In this section, we explore this dependence using model studies and quantify it as a component of the systematic uncertainty in the measured spectra. 
In these studies, the training and the response matrix were varied, allowing for the effect of a different fragmentation to be quantified through the full procedure. 
The procedure to estimate the systematic uncertainties on the inclusive jet spectra due to this fragmentation dependence is discussed in the following. The results are summarized in \Sec{sec:systematics}.

The sensitivity of the ML method to the fragmentation distribution of the Monte Carlo sample used for training generated with PYTHIA 8 is explored by modifying the fragmentation distribution in physics-motivated ways. One way to vary the fragmentation model is by utilizing quark or gluon-initiated jets.
Quark jets tend to be narrower and have fewer constituents, each of them carrying a significant fraction of the jet's momentum (harder fragmentation), while gluon jets tend to be wider and have more constituents carrying smaller fractions of the jet's momentum, $z = \frac{p_{\rm T, track}}{p_{\rm T, jet}}$.  %~\cite{Ali:2010tw}.
In practice, the inclusive jet population contains a mixture of quark and gluon jets, so using only quark or gluon-initiated jets provides significant variation to the fragmentation.
Recent analyses suggest that the properties of quenched jets, excluding the enhancement at low $z$, may result primarily from the different quenching of quarks and gluons~\cite{Spousta:2015fca}.

 Additionally, in-medium parton interactions lead to a variety of physical effects. Phenomenological modifications are performed by branching off additional hadrons from existing jet constituents, changing the final-state hadron distribution. 
The modifications are governed by tunable parameters specifying $p_{\text{loss}}$, the probability of branching off a particle; $f_{\text{loss}}$, the fraction of the jet constituent $p_{\rm T}$ to radiate; and $\Delta R$, the maximum angle of the emission relative to the jet constituent.
For each jet constituent, a particle is radiated with probability $p_\text{loss}$, carrying $\pT\ = f_{\text{loss}} \pT^{\text{const.}}$ at an angle randomly sampled from a uniform distribution between 0 and $\Delta R$. Three different shower modifications were studied using this framework, with each variation modifying the final-state hadron distribution to model a different aspect of in-medium jet modification.

 Below is a summary of all the fragmentation modifications used in this analysis: 
\begin{enumerate}
 \item \textbf{Quark Only:} jets originating from a quark in the PYTHIA 8 simulation are used.
     \item \textbf{Gluon Only:} jets originating from a gluon in the PYTHIA 8 simulation are used.
    \item \textbf{Fractional Collinear:} the radiated particle carries a specific fraction of the original constituent's energy and is emitted predominantly within the jet cone by setting $\Delta{R}$ to 0.1, 0.2, and 0.4 for $R=0.2$, $0.4$, and $0.6$, respectively. 
    The three-momentum of the radiated particle is then subtracted from the original jet constituent three-momentum, and the radiated particle is added to the list of jet constituents if it falls within the jet cone.
    \item \textbf{Fractional Large Angle:} the radiated particle carries a specific fraction of the original constituent's energy and is frequently emitted outside the jet cone by setting $\Delta{R}$ to 0.4, 0.6, and 0.8 for $R=0.2$, $0.4$, and $0.6$, respectively. 
    The three-momentum of the radiated particle is then subtracted from the original jet constituent three-momentum, and the radiated particle is added to the list of jet constituents if it falls within the jet cone.
    \item \textbf{Medium Response:} the emission occurs as described for the Fractional Collinear case, but the original jet constituent \pT\ is unmodified, emulating the addition of particles from the medium into the jet.
\end{enumerate}

The kinematic modifications vary both the momentum scale and the angular distribution of jet constituents and, thereby, the jet distributions themselves. 
Existing measurements guided the values of the tunable parameters used in the phenomenological modifications.  
Specifically, the $p_{\rm loss}$ values were determined by evaluating the excess particle yield for jets in \PbPb\ collisions compared to those in pp collisions using the jet radial profiles for $R=0.4$ inclusive jets above 100 GeV/$c$~\cite{CMS:2018zze}. 
Each modification attributes the $p_{\rm loss}$ value to individual effects, whereas in reality, the overall observed modification combines contributions from all of them. 
Therefore, this approach overestimates the contribution of each individual effect, which is a conservative choice to account for the fact that the $p_{\rm loss}$ values are not extrapolated when applied to lower energy jets.
The excess yield outside of the jet cone was used to fix the value of $p_{\rm loss}$ for the Fractional Large Angle model, and the excess inside of the jet cone was used to fix the value of $p_{\rm loss}$ for the Fractional Collinear and Medium Response models. 
The modifications were then compared to existing fragmentation measurements, as discussed below. 
The values of $f_{\rm loss}$ were set to 25\% and 10\%, which provide a charged hadron $R_{\rm AA}$ of comparable magnitude to the measured values in the 0--10\% and 30--50\% centrality ranges, respectively~\cite{ALICE:2018vuu}.

\begin{figure}[t!]
\begin{minipage}{0.5\textwidth}
  \includegraphics[width=\textwidth]{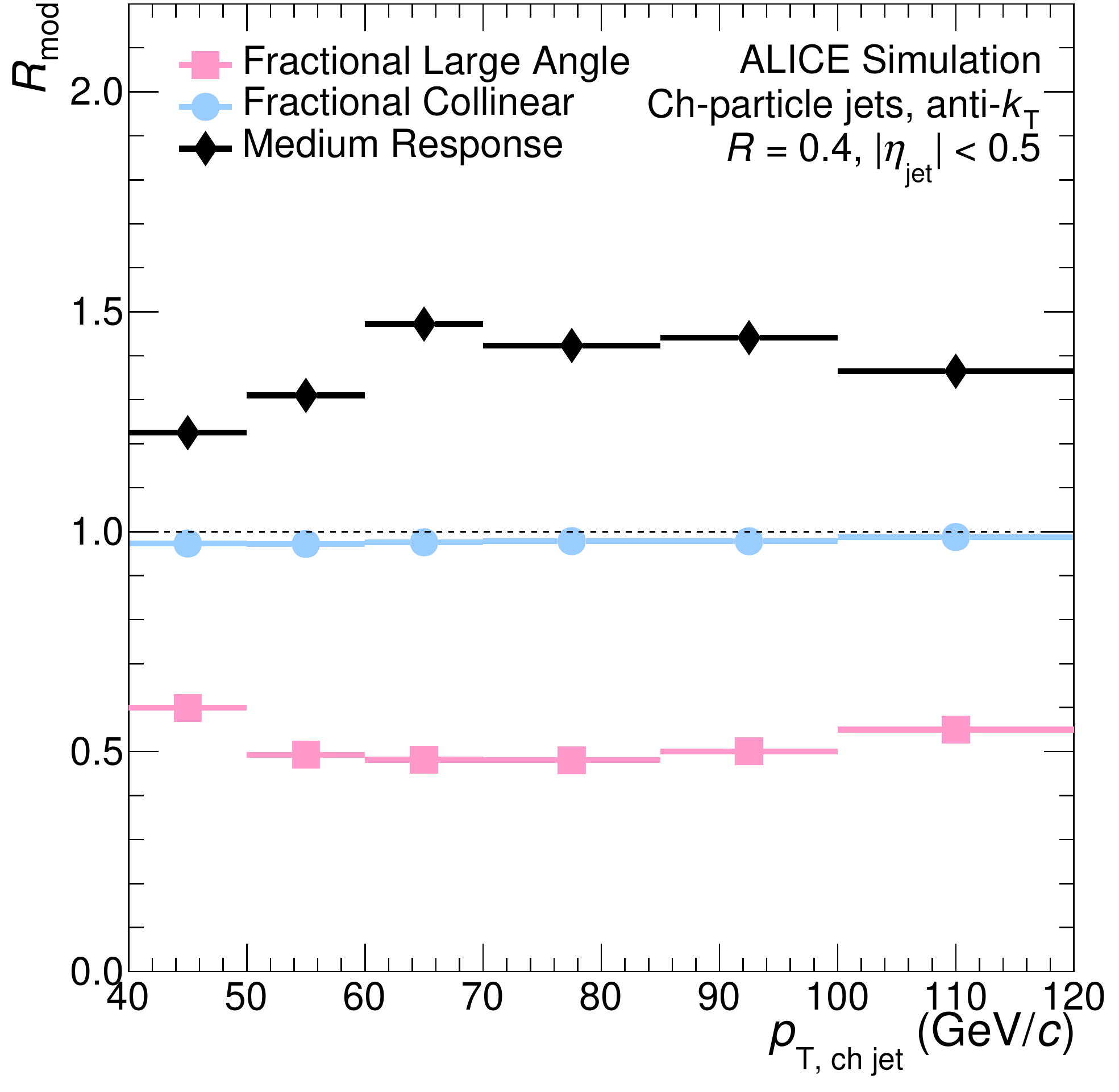}
\end{minipage}
\begin{minipage}{0.5\textwidth}
  \includegraphics[width=\textwidth]{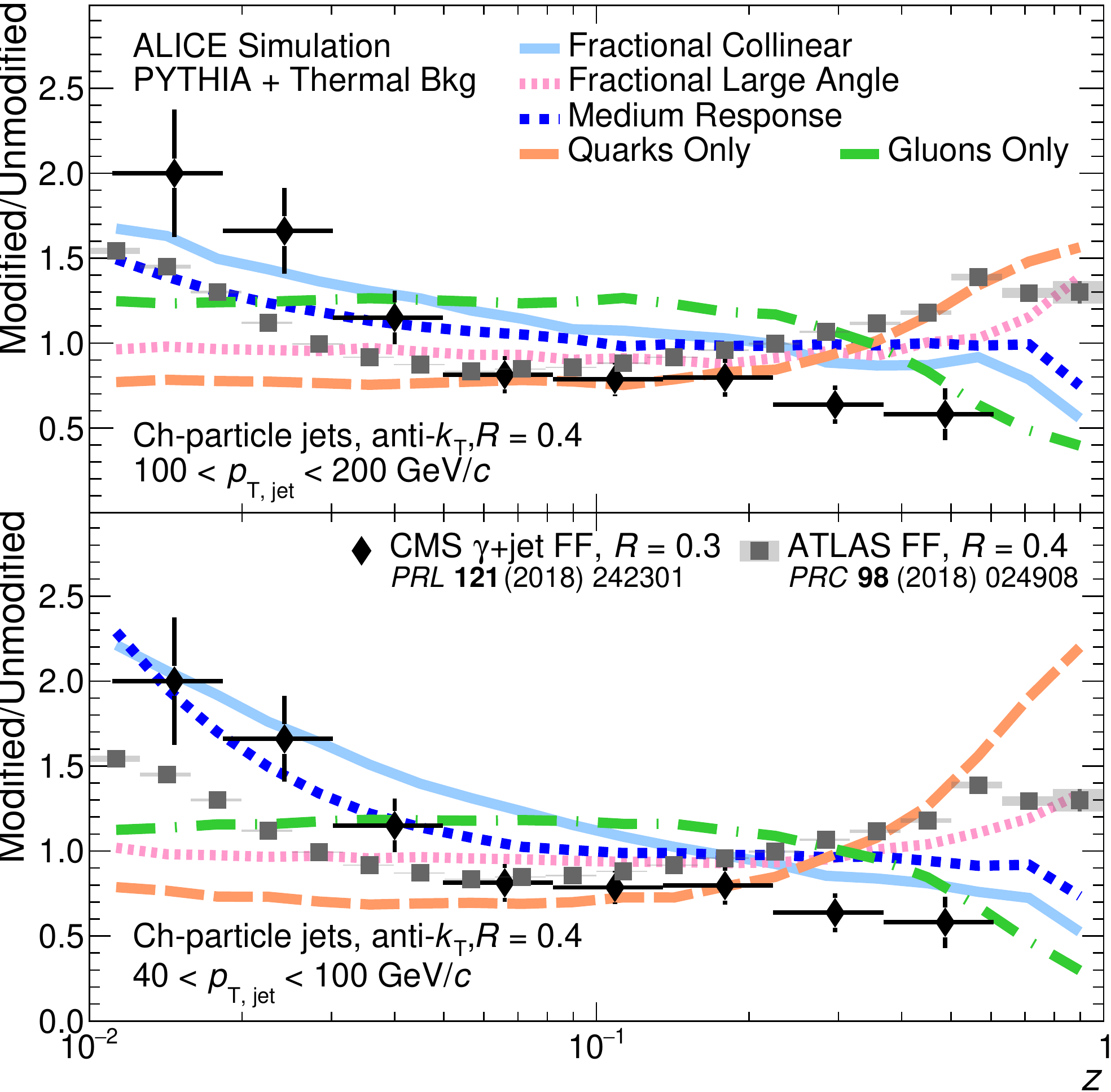}
\end{minipage}
  \caption{Left: Comparison of toy model modifications for $R = 0.4$ jets using the $R_{\rm mod}$ as defined in Eq.~\ref{raaToy} Right: The ratio of the modified to unmodified fragmentation functions at low jet $p_{\rm T}$ ($ 40 < p_{\rm T, true} < 100 \text{ GeV}/c$, lower right panel) and high jet $p_{\rm T}$ ($ 100 < p_{\rm T, true} < 200 \text{ GeV}/c$, upper right panel) for 0--10\% central \PbPb\ collisions. In the fractional collinear and fractional large angle case, $f_{\rm loss} = 25\%$  and $p_{\rm loss} = 100\%$. In the medium response case, $f_{\rm loss} = 10\%$ and $p_{\rm loss} = 50\%$.
  The ratio of the fragmentation functions measured in Pb--Pb and pp collisions are shown for jets with $R=0.4$ and $\pt>100$~GeV/$c$~\cite{ATLAS:2018bvp} (ATLAS), and for jets with $R=0.3$ and $p_{\rm T} > 30$~GeV/$c$ recoiling from a photon with $E_{\rm T}>60$~GeV/$c$~\cite{CMS:2018mqn} (CMS).
}
  \label{fig:ToyStudies}
\end{figure}

We compare the modified and the unmodified jet distributions by their ratio, $R_{\rm mod}$, 
\begin{equation}\label{raaToy}
    R_{\rm mod} = \frac{\text{Y}_{\text{modified}}}{\text{Y}_{\text{unmodified}}}\,,
\end{equation}
shown for $R = 0.4$ jets as a function of \pT\ in the left panel of \Fig{fig:ToyStudies}. 
The medium response adds energy to the jet cone, resulting in $R_{\rm mod} > 1$. 
For the fractional collinear model, the jet does not lose energy most of the time, which results in $R_{\rm mod} \approx 1$.
In the case of fractional large-angle radiation, the jet will lose energy, resulting in $R_{\rm mod} < 1$.
Note that the modifications shown here do not directly translate into the associated systematic uncertainty, which instead corresponds to the propagation of this yield modification through the ML-based correction and unfolding.

To quantify the modifications introduced by the various fragmentation scenarios, the ratio between modified and unmodified jet fragmentation functions as a function of $z$ is shown in the right panels of \Fig{fig:ToyStudies}. 
Both panels include comparisons to the measured ratio of fragmentation functions in \PbPb\ and \pp\ collisions for $R= 0.4$ inclusive jets with $\pt>100$~GeV/$c$ from ATLAS~\cite{ATLAS:2018bvp} and $R =0.3$ photon-tagged jets with $\pt>30$~GeV/$c$ from CMS~\cite{CMS:2018mqn}. The kinematic region of the ATLAS measurement is the only one where the fragmentation function of inclusive jets has been measured, so this is a possible region to check that the toy modifications cover the full phase space of modifications as observed in the data. The CMS measurement is a quark-dominated sample and does not fully describe the phase space measured in this analysis, but it is still useful for the purpose of comparing the magnitude of the induced variations in the toy models.
The top right panel shows the modifications for $R=0.4$ jets in the 0--10\% centrality class with $ 100 < p_{\rm T} < 200$ GeV/$c$. 
The Medium Response and Fractional Collinear models describe the measured low-$z$ enhancement of soft particles.
The quark-only case describes the intermediate-$z$ suppression and high-$z$ enhancement. 
Additionally, the ratio between modified and unmodified jet fragmentation functions is shown in the bottom right panel of \Fig{fig:ToyStudies} for $R=0.4$ jets with $40 < p_{\rm T} < 100$ GeV/$c$ in 0--10\% central collisions.
The fragmentation in this region has not been measured, so no direct comparison is possible, but the features are qualitatively the same as for the $R=0.4$ jets at high $p_{\rm T}$, albeit with more significant modification.
The toy model variations introduced here cover the modification of the photon-recoiling jet fragmentation measured by CMS from pp to Pb--Pb collisions over a similar kinematic region.

For each variation, both the training and the response matrix were varied to quantify the effect of a different fragmentation model through the full analysis chain. To ensure realistic variations for the systematic uncertainties, the unfolded spectrum corrected using the ML estimator was refolded with the response matrix filled with jets corrected using the AB method. The result was then compared to the spectrum obtained with the AB method at the hybrid level to ensure the result was similar to the one that would have been achieved with the AB method. For a variation to be considered, we required an agreement comparable to the size of the unfolding uncertainties.

In principle, any unfolded heavy-ion measurement could have a fragmentation bias inherent to the unfolding procedure and, by definition, assumes a fragmentation model. 
The method developed for this paper introduces a new range of variations that could be considered for such studies in the future. 
Unique to the ML method is the fragmentation dependence of the training, but the results varied minimally when only the training sample fragmentation was varied, indicating that the main effect is due to the fragmentation in the response matrix.

%%%%%%%%%%%%%%%%%%%%%%%%%%%%%%%%%%%%%%%%%%%%%%%%%%%%%%%%%%%%%%%%%%%%%%%%%%%%%%%%%%%%%%%%%%%%%%%%%%%%
\section{Systematic uncertainties}
\label{sec:systematics}
%%%%%%%%%%%%%%%%%%%%%%%%%%%%%%%%%%%%%%%%%%%%%%%%%%%%%%%%%%%%%%%%%%%%%%%%%%%%%%%%%%%%%%%%%%%%%%%%%%%%
The systematic uncertainties of the inclusive jet spectra arise from the tracking efficiency, the unfolding procedure, and the model dependence in the ML method. 
The full systematic uncertainty is given by the quadratic sum of the individual uncertainties, where the single uncertainties are taken to be symmetric about the nominal value unless otherwise specified. 
The following sources were taken into account for the measurements in Pb--Pb collisions:

{\bf Tracking efficiency uncertainty}: the loss of tracks due
to tracking efficiency less than unity results in a reduction in jet \pT, which corresponds to a significant
reduction in measured yield in a given $p_{\rm T}$ interval due to the steeply falling jet spectrum. Tracking efficiency is consequently one of the largest sources of systematic uncertainty in this measurement. 
Detailed studies of the tracking efficiency have been performed to determine an appropriate variation for the systematic uncertainty~\cite{Abelev:2014ffa, Abelev:2013kqa}.
Those studies motivate the systematic variation, which corresponds to a modified response matrix where 4\% of the tracks are randomly discarded.

{\bf Regularization parameter}: to regularize in the Bayesian unfolding procedure, a number of iterations is chosen for the nominal result where the unfolded results are stable. 
For the systematic uncertainty, variations of the parameter by $\pm 1$ are taken into account.

{\bf Prior}: for the Bayesian unfolding procedure, a prior distribution is needed.
The PYTHIA 8 jet $\pT$ spectrum was taken as the nominal prior. 
For the systematic uncertainty the sensitivity of the unfolded result to the prior was evaluated by scaling the PYTHIA 8 spectrum by the parameterized ratio of the hybrid-level MC to Pb--Pb collision data, accounting for any shape differences between the two. Then, the difference between the result unfolded by the scaled response and the nominal response was taken as a systematic uncertainty.

{\bf Measured $\pT$-range}: the minimum transverse momentum of the jets that enter the unfolding procedure is determined by the requirement that it is five times the residual fluctuations $\sigma$, which suppresses the fake jet yield (see Sec~\ref{sec:reconstruction}).
The low-$\pT$ cut-off of the data that serves as input to the unfolding procedure is varied by $\pm 5$~\GeVc. Small-$R$ jets are expected to be most sensitive to this cut due to their low $p_{\rm T, jet}$ reach. 

{\bf Fragmentation}: the background estimator is trained using the jet spectrum from simulated pp collisions utilizing PYTHIA 8 where the clustered hadrons were constructed following the Lund fragmentation model as discussed at length in Sec.~\ref{sec:fragdependence}. The systematic uncertainty was estimated from the variations in the results obtained using different fragmentation models.  In particular, the following alternatives were considered: q/g fragmentation, Medium Response, Fractional Collinear, and Fractional Large Angle. 
The variations are added in quadrature and the corresponding uncertainties are considered to be asymmetric.

\begin{table}[t!]
\centering
\caption{\label{tab:Systematics1} Relative systematic uncertainties (\%) for jet spectra for 0--10\% central Pb--Pb collisions and all resolution parameters. The maximum uncertainties for low $\pT$ ($p_{\rm T, jet} < 50$ GeV/$c$) and high $\pT$ ($p_{\rm T, jet} > 50$ GeV/$c$) are shown. The direction of asymmetric uncertainties is indicated with a $+$ or $-$ sign. The combined uncertainty is the sum in quadrature of individual uncertainties. }
\begin{tabular}[t]{l|l|l|l|l|l|l}
Resolution parameter ($R$)          & \multicolumn{2}{c}{$0.2$} & \multicolumn{2}{c}{$0.4$} & \multicolumn{2}{c}{$0.6$} \\ \hline
$p_{\rm T}$ & Low $p_{\rm T}$ & High $p_{\rm T}$ & Low $p_{\rm T}$ & High $p_{\rm T}$ & Low $p_{\rm T}$ & High $p_{\rm T}$ \\ \hline
Tracking eff. & 21 & 18 & 24 & 12 & 12 & 34\\
Regularization param. & (< 2) & (< 2) & (< 2) &  2 & 2 & (< 2) \\
Unfolding prior & 6 & 16 & 8 & (< 2) & 4 & (< 2) \\
Measured $\pT$ range & 46 & 4 & 8 &  (< 2) & 6 & 8\\
Fractional Collinear & +30 & +12 & +12 & +16 & +8 & +20 \\
Fractional Large Angle & +10 & +10 & +6 & +10 & +8 & +14 \\
Fractional Medium Response & +28 & +14 &  +20 & +14 & +22 & +14 \\
Quarks/Gluon & -8 & -6 & -12 & -12 & -14 & -12 \\ \hline
Combined & 58 & 32 & 32 & 26 & 28 & 44 \\
\end{tabular}
\end{table}

\begin{table}[t!]
\centering
\caption{\label{tab:Systematics2} Relative systematic uncertainties (\%) for jet spectra for 30--50\% central Pb--Pb collisions and all resolution parameters. The maximum uncertainties for low $\pT$ ($p_{\rm T, jet} < 50$ GeV/$c$) and high $\pT$ ($p_{\rm T, jet} > 50$ GeV/$c$) are shown. The direction of asymmetric uncertainties is indicated with a $+$ or $-$ sign. The combined uncertainty is the sum in quadrature of individual uncertainties. }
\begin{tabular}[t]{l|l|l|l|l|l|l}
Resolution parameter ($R$)           & \multicolumn{2}{c}{$0.2$} & \multicolumn{2}{c}{$0.4$} & \multicolumn{2}{c}{$0.6$} \\ \hline
$p_{\rm T}$ & Low $p_{\rm T}$ & High $p_{\rm T}$ & Low $p_{\rm T}$ & High $p_{\rm T}$ & Low $p_{\rm T}$ & High $p_{\rm T}$ \\ \hline
Tracking eff. & 10 & 12 & 12 & 16 & 12 & 14 \\  
Regularization param. & $(< 2)$ & $(< 2)$ & $(< 2)$ & $(< 2)$ & $(< 2)$ & $(< 2)$ \\
Unfolding prior & $(< 2)$ & 4 & 6 & 2 & $(< 2)$ & 6 \\
Measured $\pT$ range & 28 & $(< 2)$ & 10 & $(< 2)$ & 20 & $(< 2)$ \\
Fractional Collinear & +12 & +6 & +22 & +14 & +26 & +24 \\
Fractional Large Angle & +8  & +8 & +8 & +10 & +26 & +24 \\
Fractional Medium Response & +8 & +8 & +14 & +10 & +22 & +20 \\
Quarks/Gluon &  -8 & -6 & -8 & -6 & -12 & -6 \\ \hline
Combined & 34 & 22 & 32 & 26 & 42 & 40 \\
\end{tabular}

\end{table}

\begin{table}[t!]
\centering
\caption{\label{tab:Systematics3} Relative systematic uncertainties (\%) for jet spectra for 60--80\% central \PbPb\ collisions and all used resolution parameters. The maximum uncertainties for low $\pT$ ($p_{\rm T, jet} < 50$ GeV/$c$) and high $\pT$ ($p_{\rm T, jet} > 50$ GeV/$c$) are shown. In this centrality interval the spectra are measured with the area-based method, the uncertainties related to the fragmentation functions adopted in the machine learning algorithm are not included for this case. The combined uncertainty is the sum in quadrature of individual uncertainties.}
\begin{tabular}[t]{l|l|l|l|l|l|l}
Resolution parameter ($R$)           & \multicolumn{2}{c}{$0.2$} & \multicolumn{2}{c}{$0.4$} & \multicolumn{2}{c}{$0.6$} \\ \hline
$p_{\rm T}$ & Low $p_{\rm T}$ & High $p_{\rm T}$ & Low $p_{\rm T}$ & High $p_{\rm T}$ & Low $p_{\rm T}$ & High $p_{\rm T}$ \\ \hline
Tracking eff. & 4 & 10 & 2 & 10 & 10 & 14\\
Regularization param. & $(< 2)$ & $(< 2)$ & $(< 2)$ & $(< 2)$ & $(< 2)$ & $(< 2)$ \\
Unfolding prior & 6 & $(< 2)$ & 10 & 6 & $(< 2)$ & 8 \\
Measured $\pT$ range & 32 & 4 & 6 & 4 & 36 & 14 \\
Quarks/Gluon & -2 & -4 & -4 & -4 & $(< -2)$ & -4 \\\hline
Combined & 46 & 14 & 14 & 10 & 50 & 16 \\ 
\end{tabular}
\end{table}

For the measurements in pp collisions, several sources of uncertainty were taken into account.
For the unfolding, the SVD ~\cite{Hocker:1995kb} algorithm was used for the central value of the results, and the relative variation in the results when using the Bayesian method was taken as the uncertainty.
The regularization parameters were further varied by $\pm1$ from the nominal values, as in the \PbPb\ case. The unfolding uncertainties, including the algorithm and regularization variations, are estimated from the root-mean-square of these variations.
The tracking efficiency uncertainty was estimated in a similar manner as for the \PbPb\ spectra, but in this case, disregarding $3\%$ of the tracks due to the smaller systematic uncertainty on the tracking efficiency in pp collisions.
Additionally, there is an uncertainty from secondary track contamination due to weak decays, which was estimated by comparing the secondary track fraction in data and MC.
Note that the SVD and secondary track contamination uncertainties are small in \PbPb\ collisions, and therefore are neglected.

A summary of the systematic uncertainties discussed above for the spectra is given in Tabs.~\ref{tab:Systematics1},~\ref{tab:Systematics2}, and~\ref{tab:Systematics3} for \mbox{0--10\%}, 30--50\%, and 60--80\% Pb--Pb collisions, respectively. 
The uncertainties of the $R_\mathrm{AA}$ are calculated using the systematic uncertainties of the Pb--Pb spectra and the uncertainties of the pp reference~(evaluated as in~\cite{Acharya:2019tku}). 
Included in the $R_\mathrm{AA}$ uncertainty is an additional uncertainty associated with the calculation of the $\left<T_\mathrm{AA}\right>$, given in Ref.~\cite{Loizides:2017ack}.
The systematic uncertainties on the $R_{\rm AA}$ double ratios and the cross section ratios for different $R$ values are evaluated by separately treating correlated and uncorrelated uncertainties. 
All unfolding uncertainties for the spectra in Pb--Pb collisions are treated as uncorrelated and added in quadrature. 
The tracking uncertainty and the uncertainties due to the fragmentation dependence are treated as correlated by evaluating the double ratio for each variation and calculating the difference from the nominal double ratio. 
The deviations from the nominal value for each of these cases are then added in quadrature to obtain the final correlated uncertainty on the $R_{\rm AA}$ double ratios and the cross section ratios.

%%%%%%%%%%%%%%%%%%%%%%%%%%%%%%%%%%%%%%%%%%%%%%%%%%%%%%%%%%%%%%%%%%%%%%%%%%%%%%%%%%%%%%%%%%%%%%%%%%%%
\section{Results}
\label{sec:results}
\begin{figure}[t!]
  \includegraphics[width=0.49\textwidth]{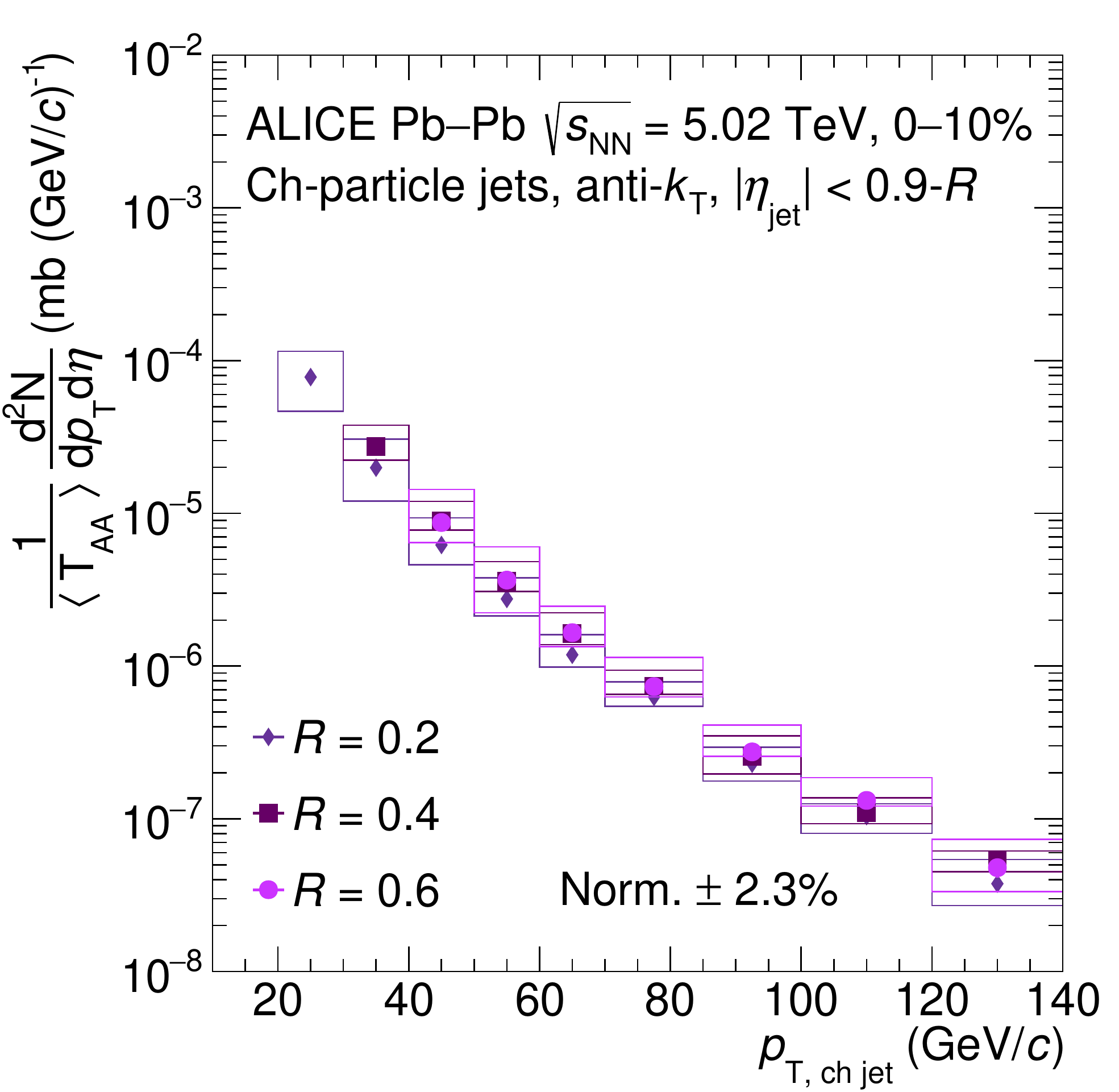}
  \includegraphics[width=0.49\textwidth]{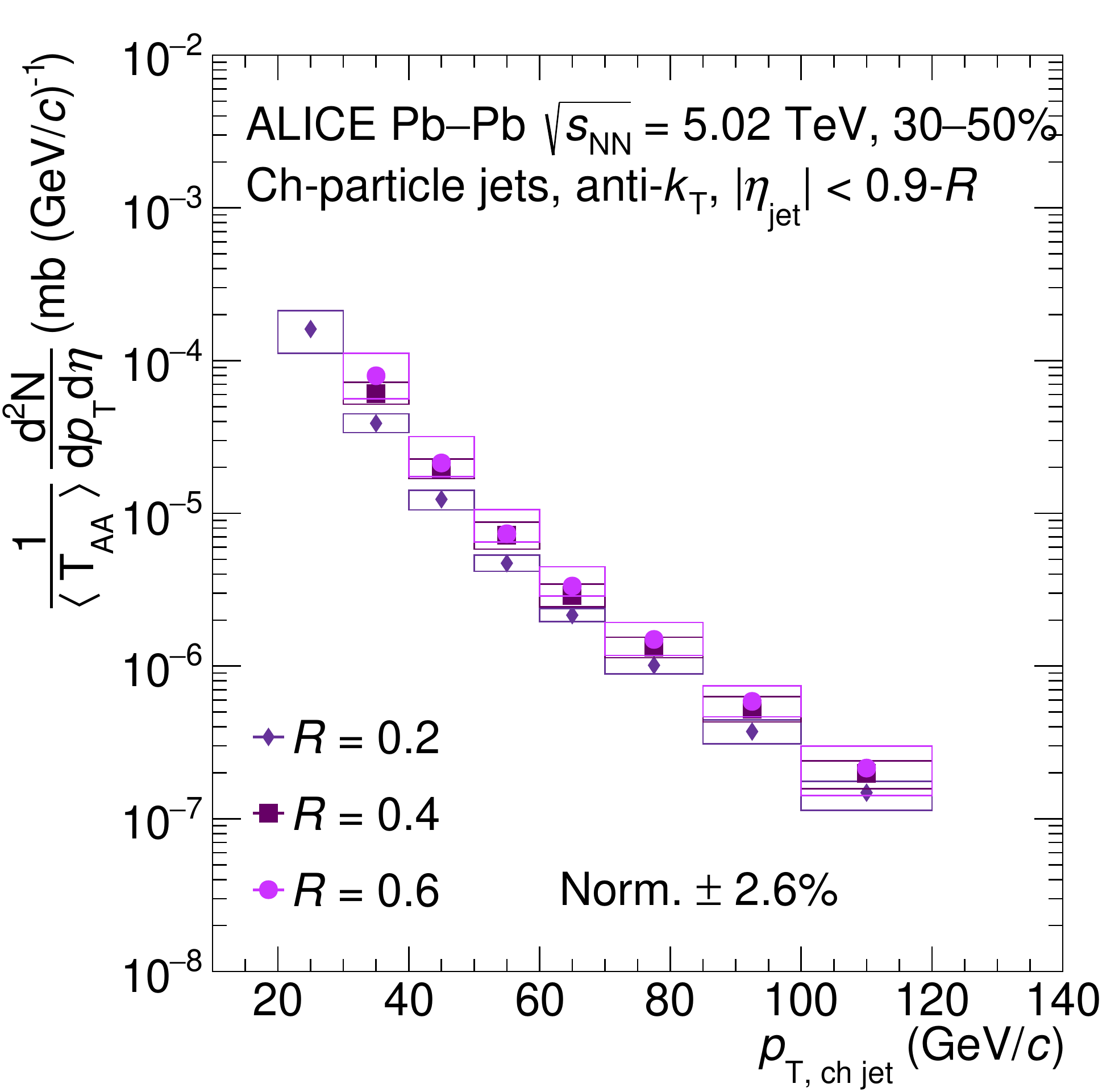}
  \includegraphics[width=0.49\textwidth]{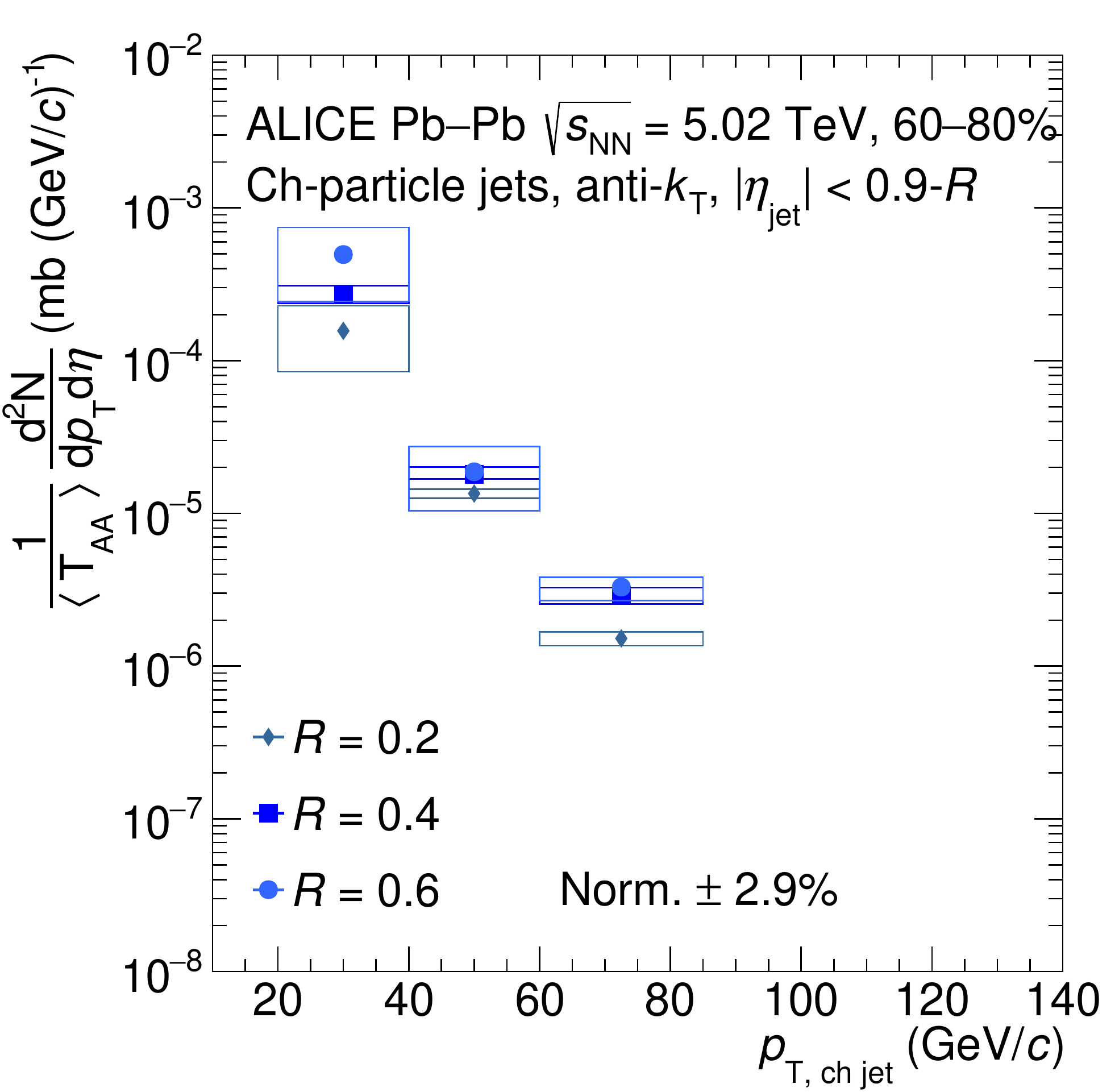}
  \includegraphics[width=0.49\textwidth]{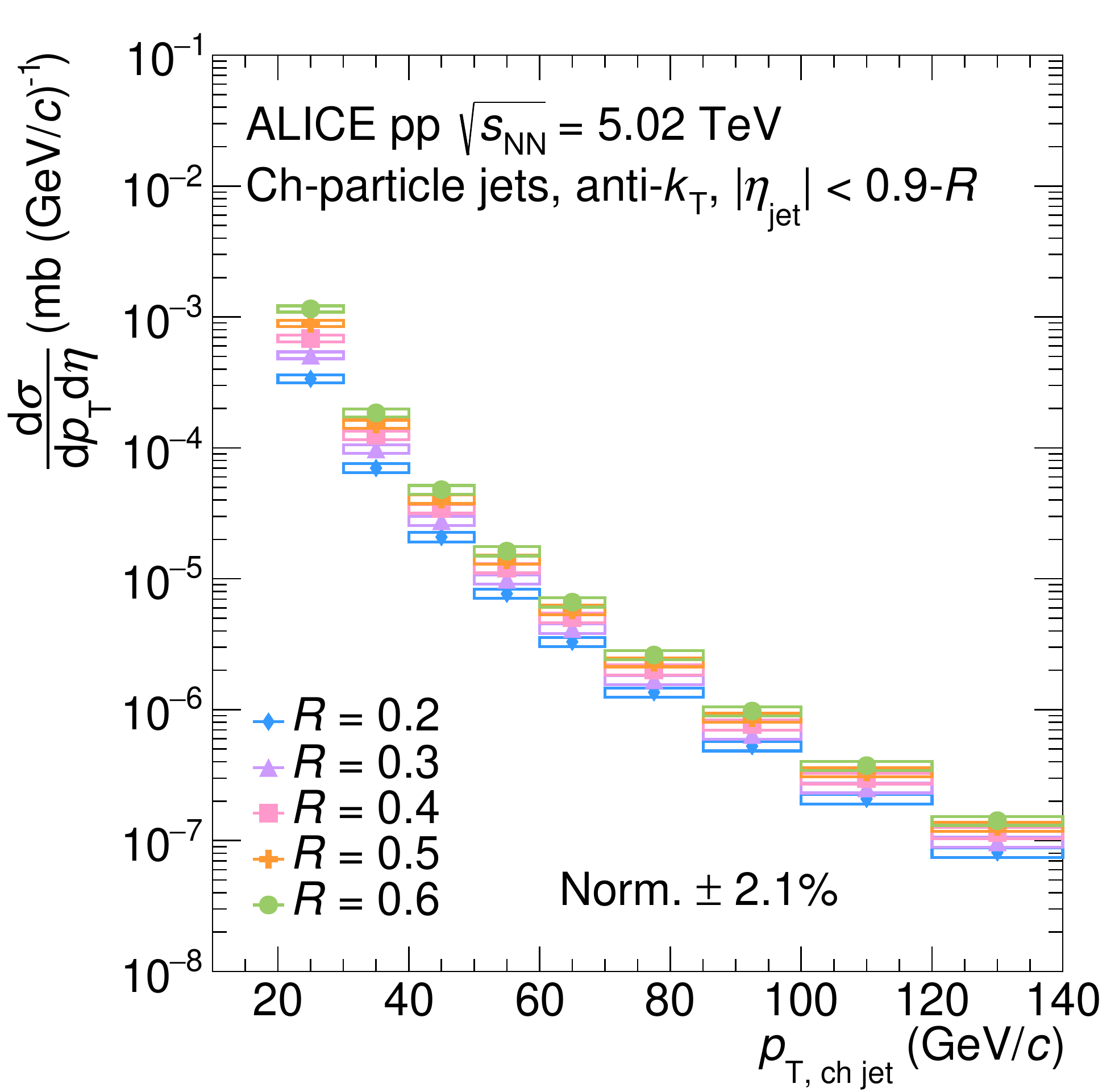}
  \caption{The \pT\/-differential inclusive charged-particle jet yield distributions as a function of \pT\ for different values of $R$  in three centrality classes: Top Left: 0--10\%, top right: 30--50\%, bottom left: 60--80\%. The peripheral spectra were measured using the area-based method for the background correction. All other reported spectra were corrected with the ML-based background estimator. In the bottom right panel, the production cross sections in pp collisions are shown. The vertical bars denote statistical uncertainties and the vertical extent of the boxes denotes systematic uncertainties. Note that the data points are plotted horizontally at the bin center.}
  \label{fig:Spectrum_1}
\end{figure}

\begin{figure}[th!]
\centering
  \includegraphics[width=0.32\textwidth]{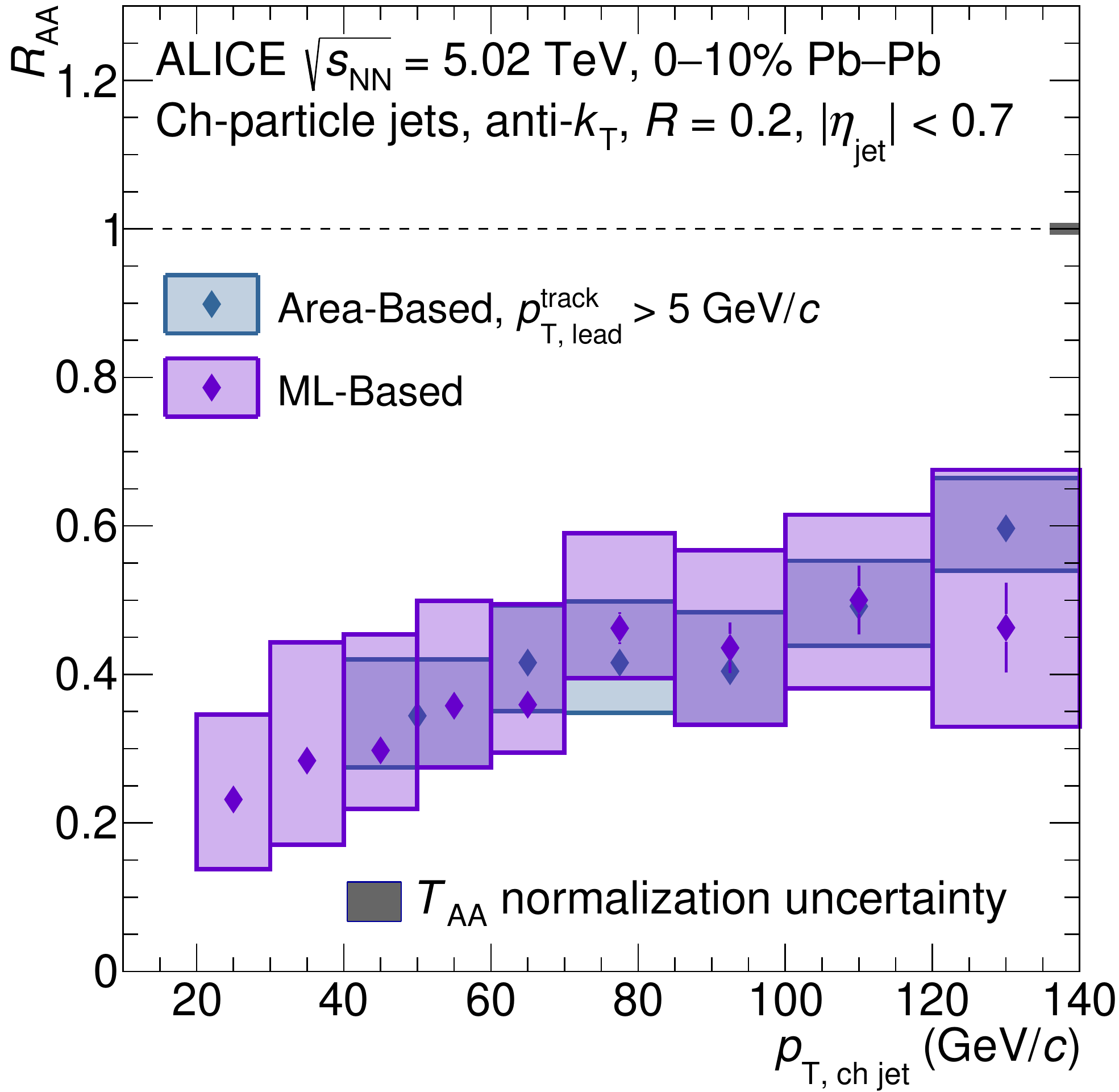}
  \includegraphics[width=0.32\textwidth]{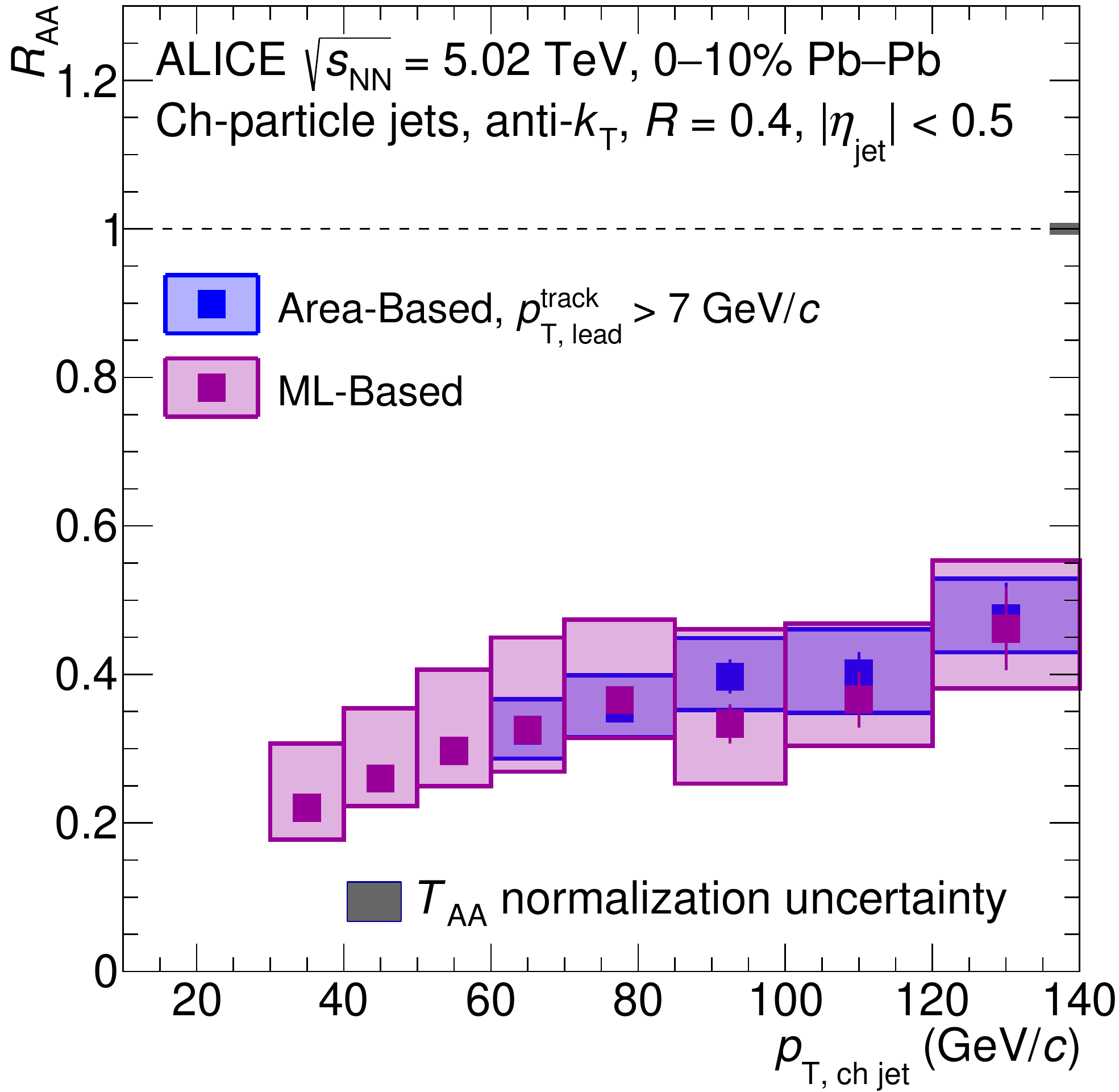}
  \includegraphics[width=0.32\textwidth]{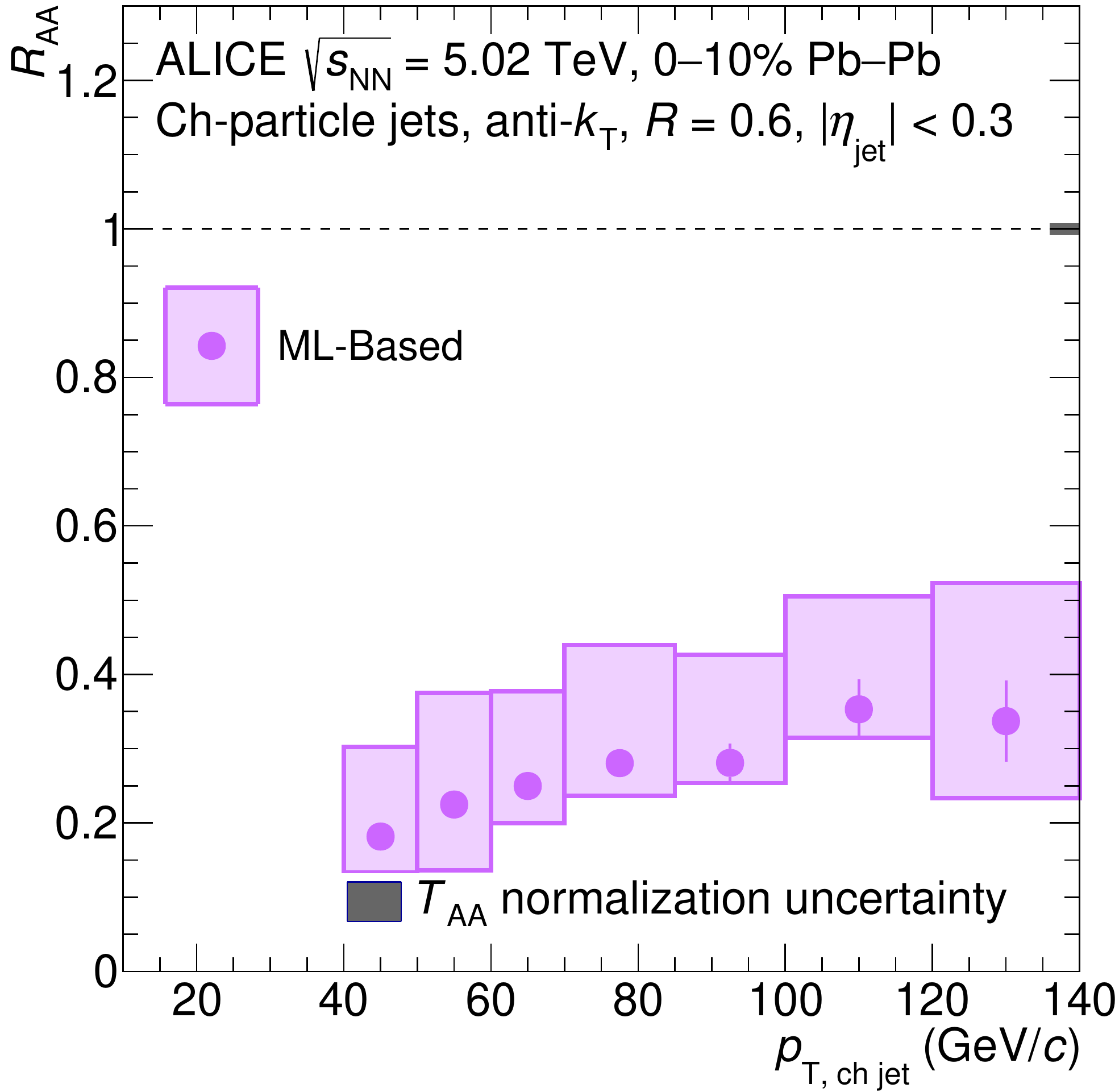}
  \includegraphics[width=0.32\textwidth]{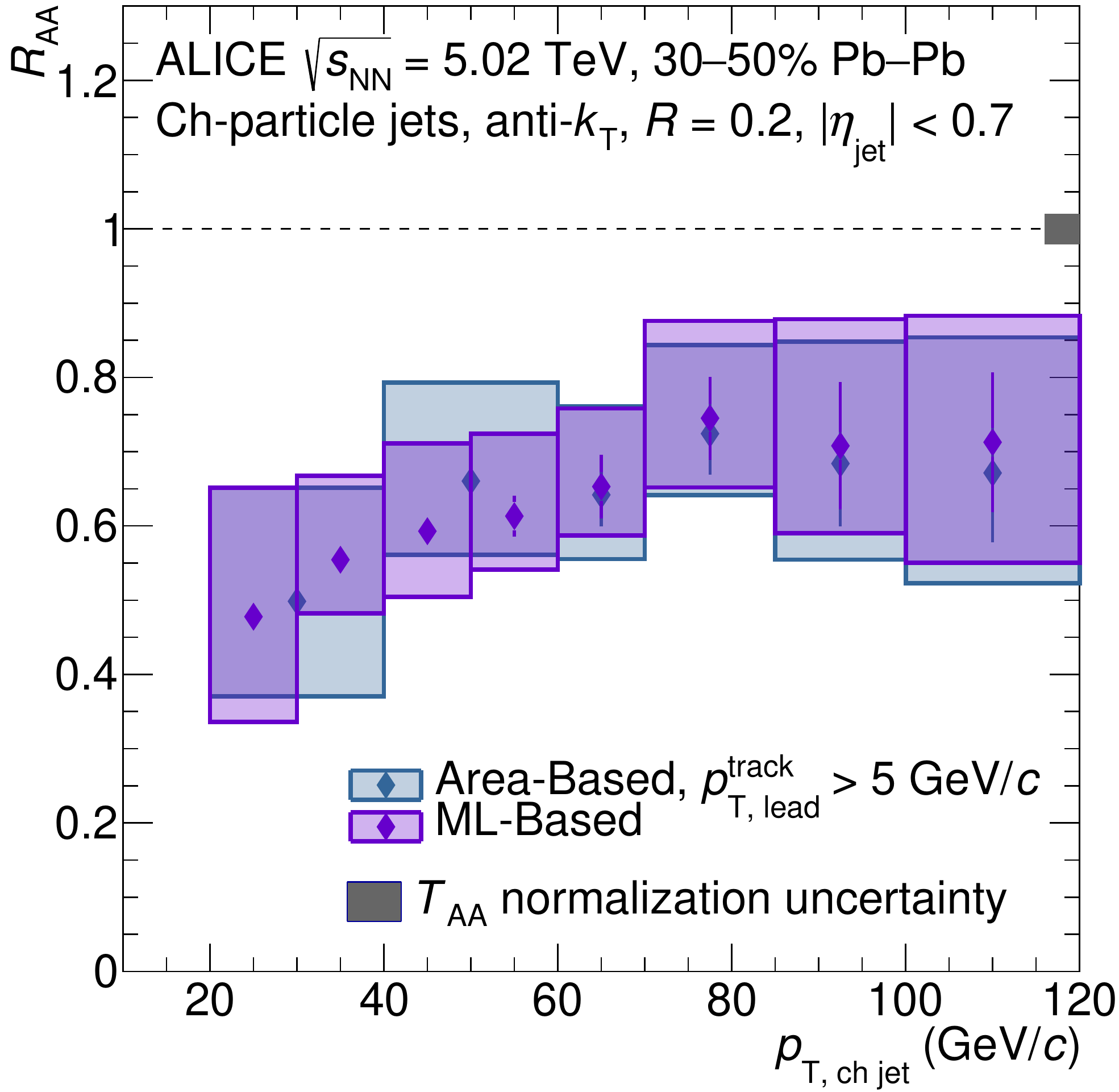}
  \includegraphics[width=0.32\textwidth]{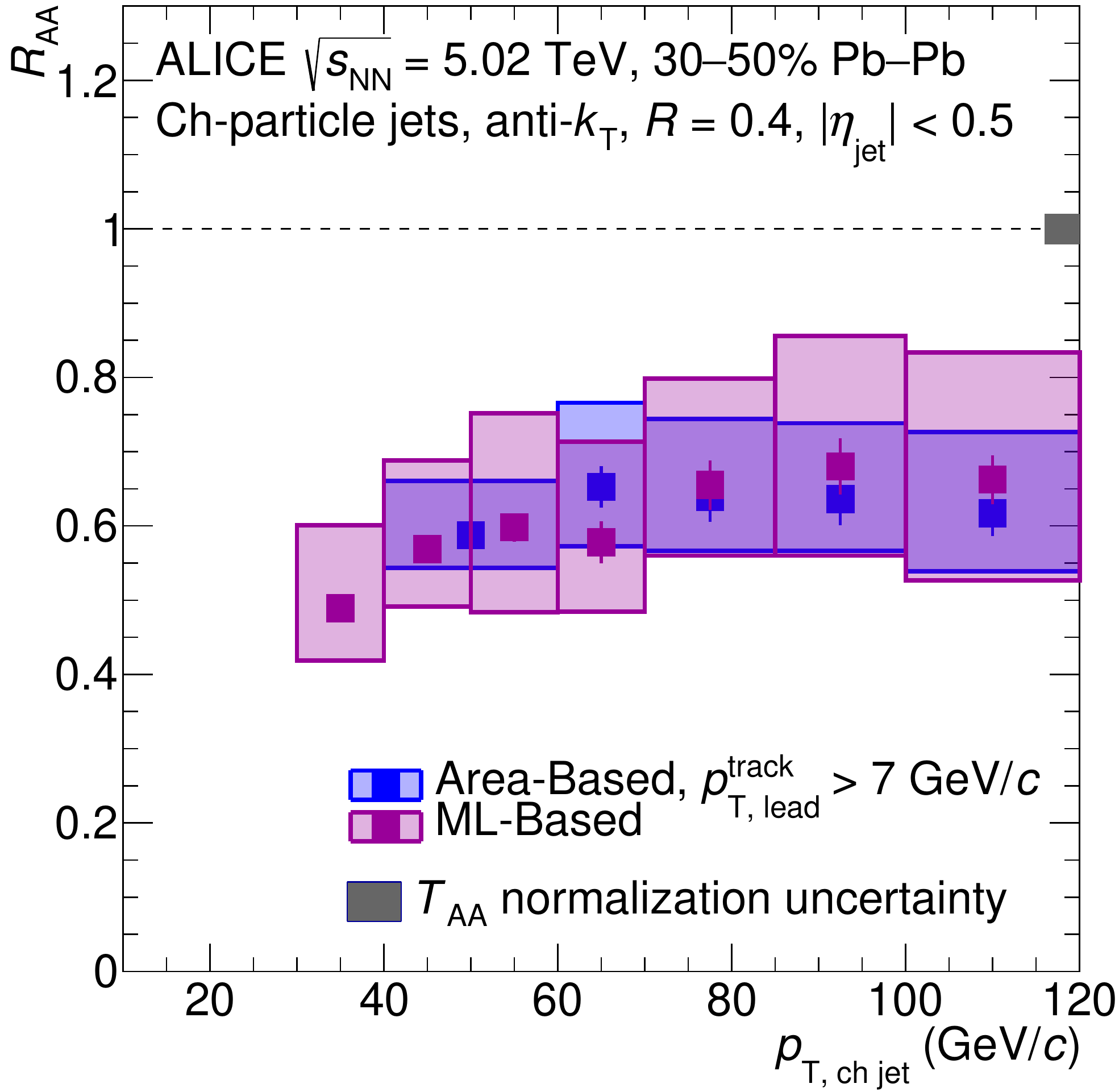}
  \includegraphics[width=0.32\textwidth]{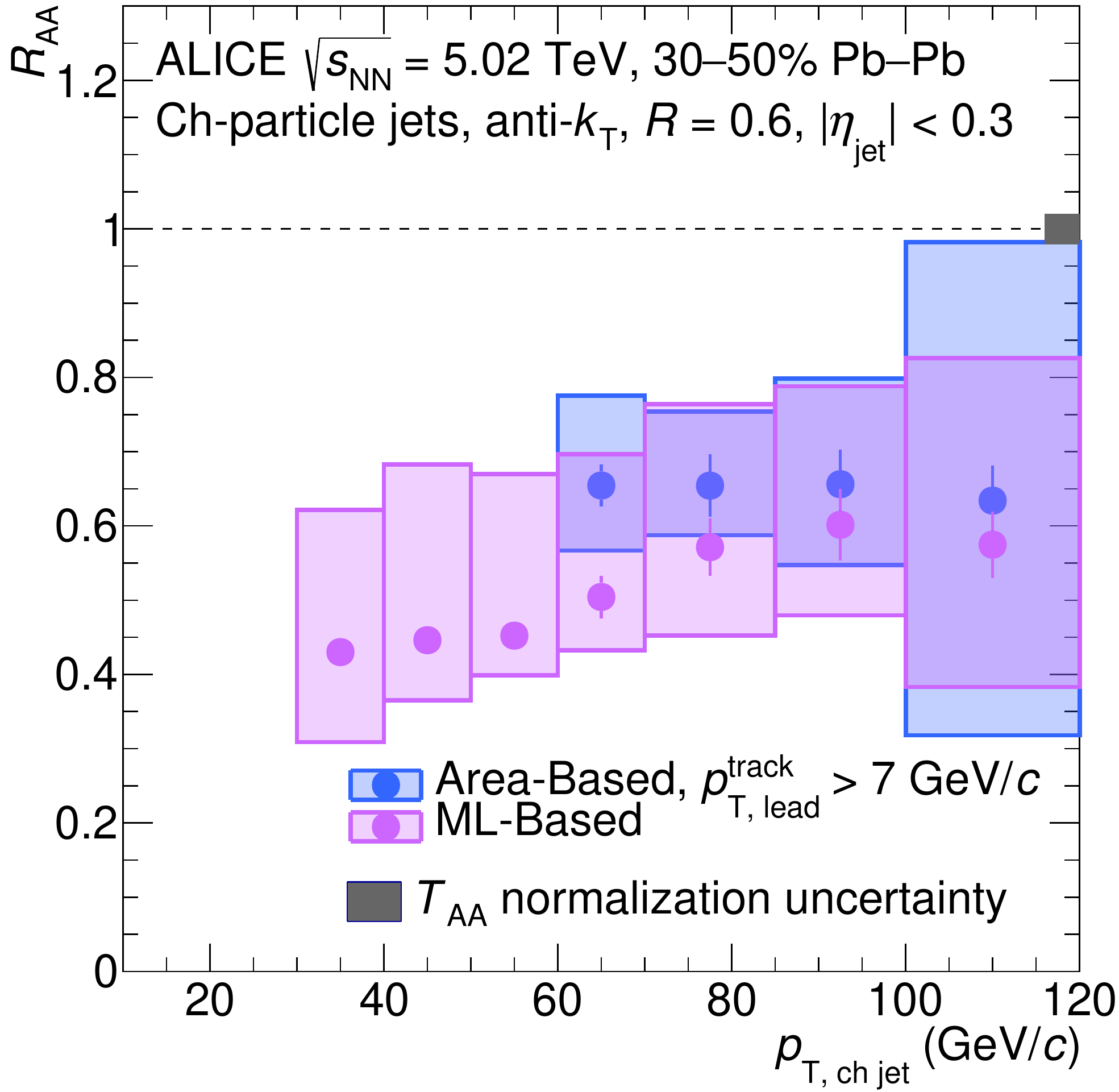}
  \includegraphics[width=0.32\textwidth]{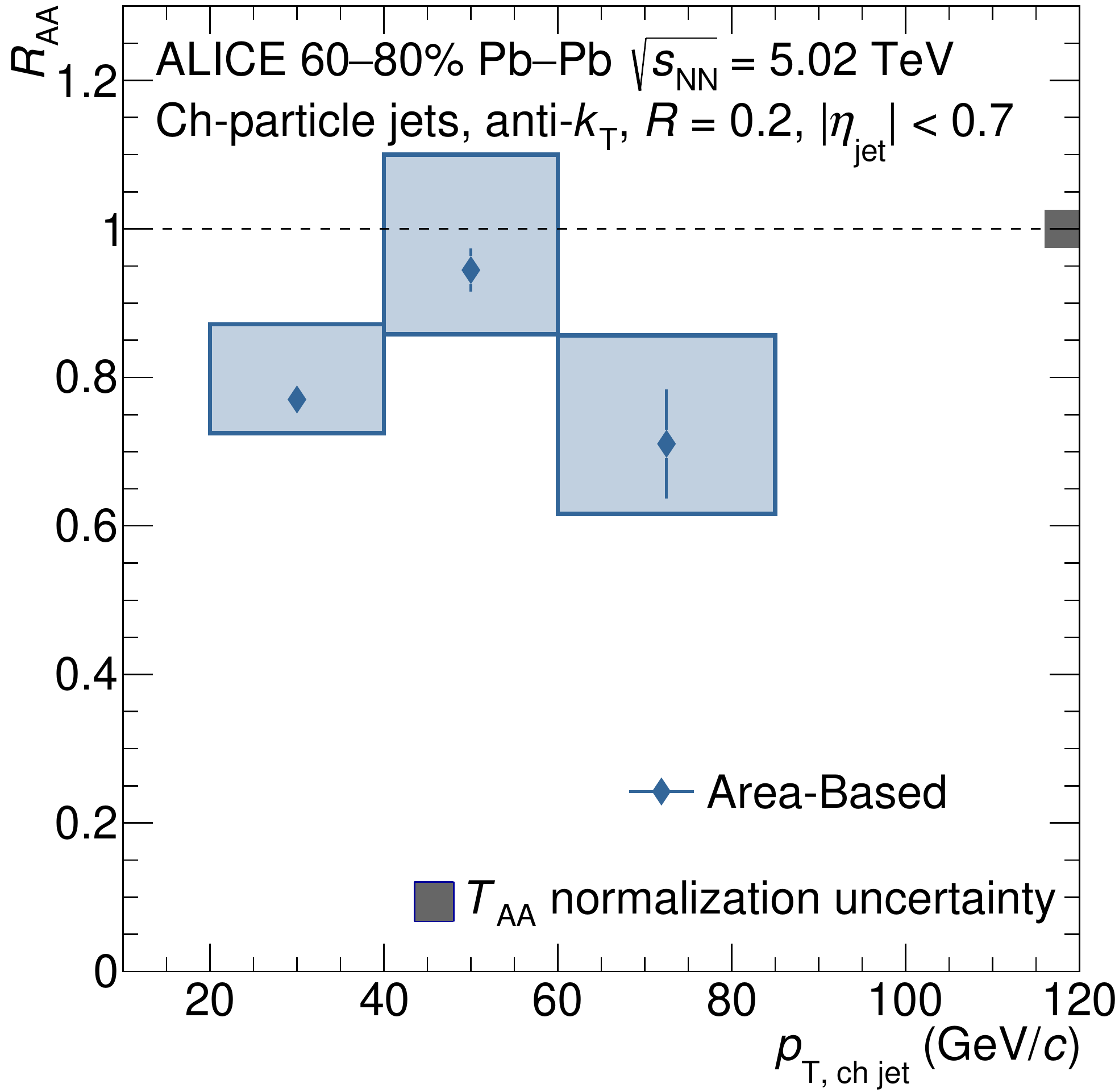}
  \includegraphics[width=0.32\textwidth]{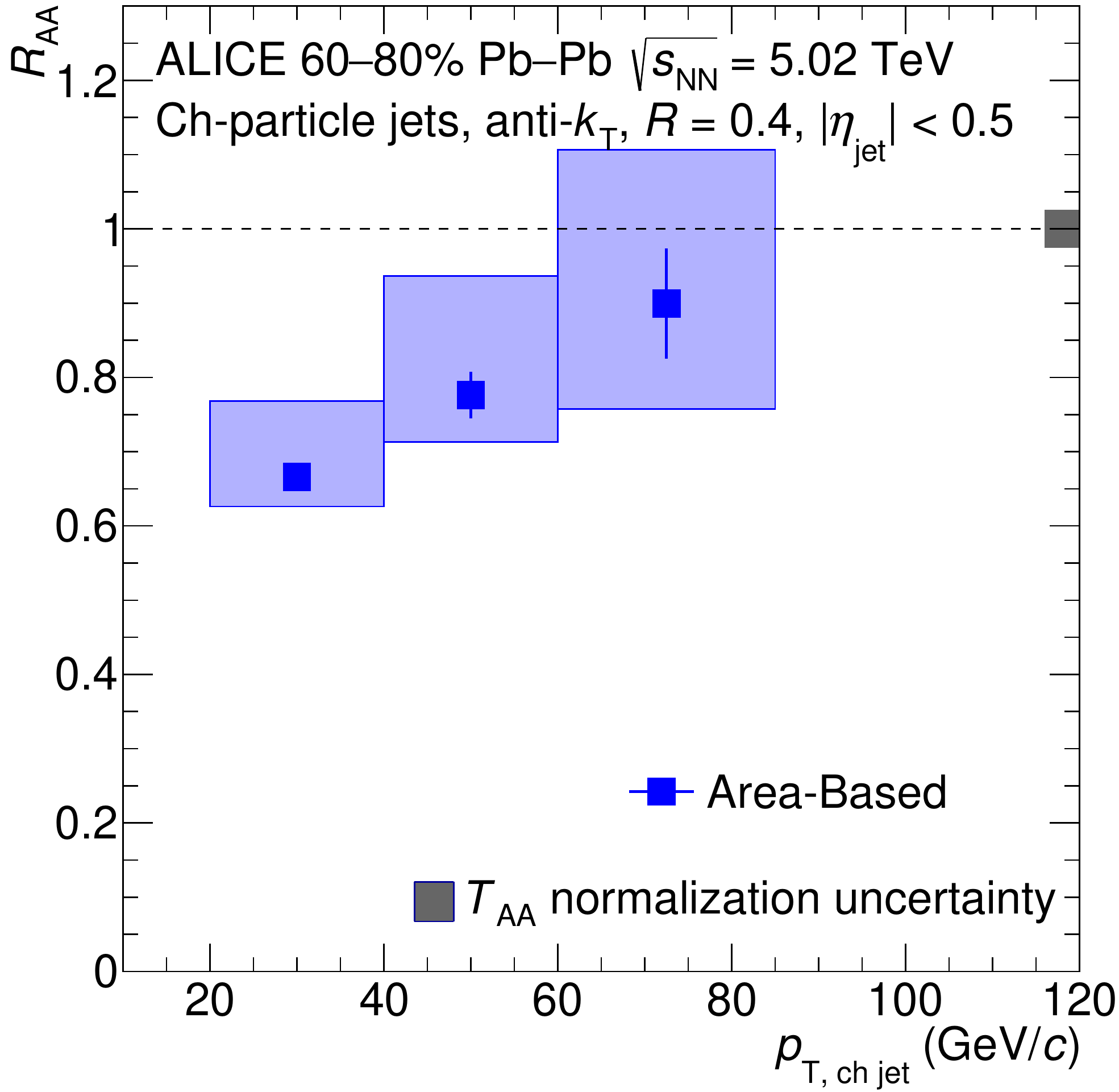}
  \includegraphics[width=0.32\textwidth]{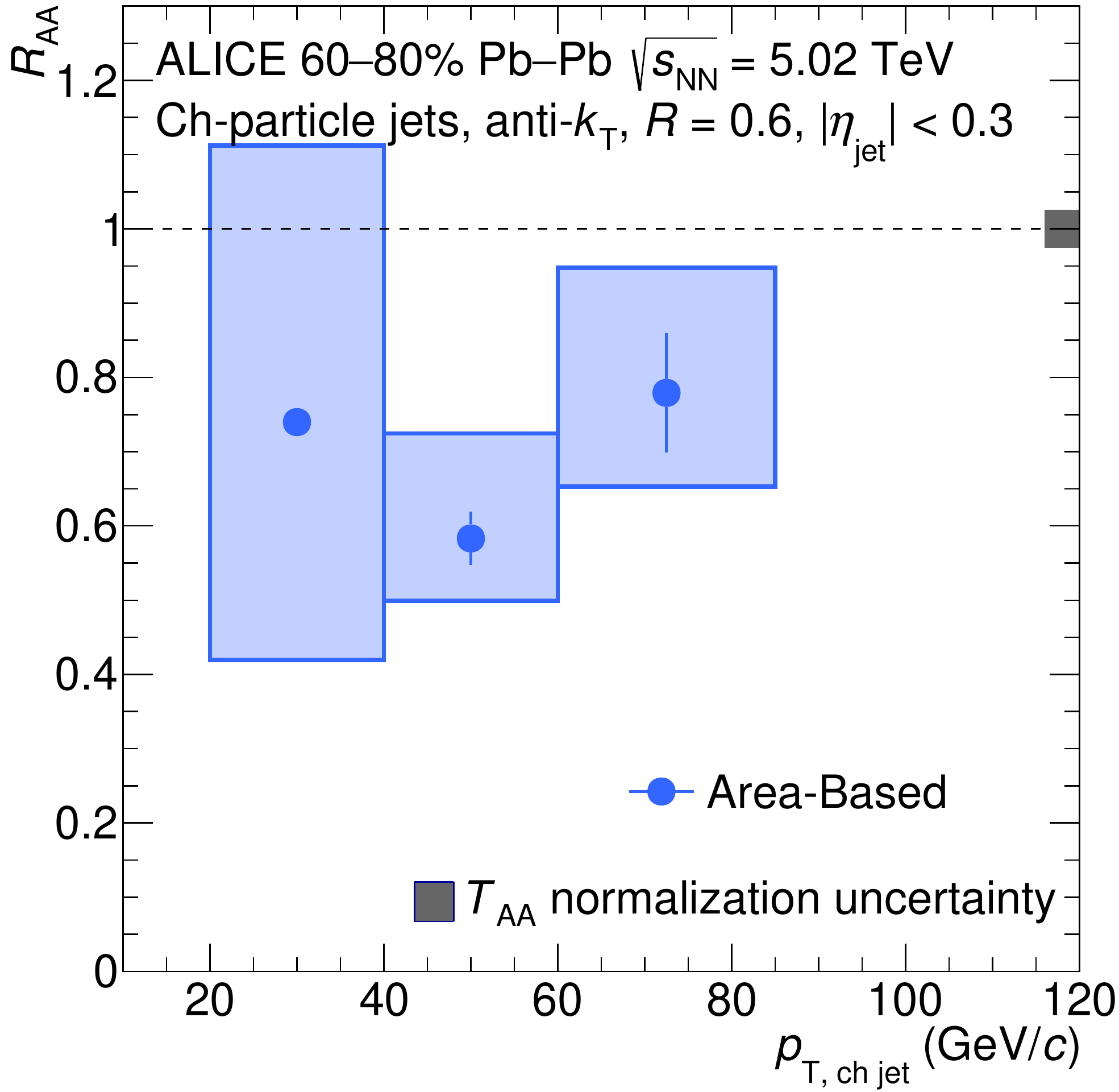}
  \caption{Nuclear modification factors of inclusive charged-particle jets as a function of \pT\ for $R=0.2$, $R=0.4$, and $R=0.6$, shown for 0--10\%, 30--50\% and 60--80\% central \PbPb\ collisions for the ML-based method compared to results obtained with the area-based method where applicable.}
  \label{fig:RAA_2}
\end{figure}

\begin{figure}
  \includegraphics[width=0.49\textwidth]{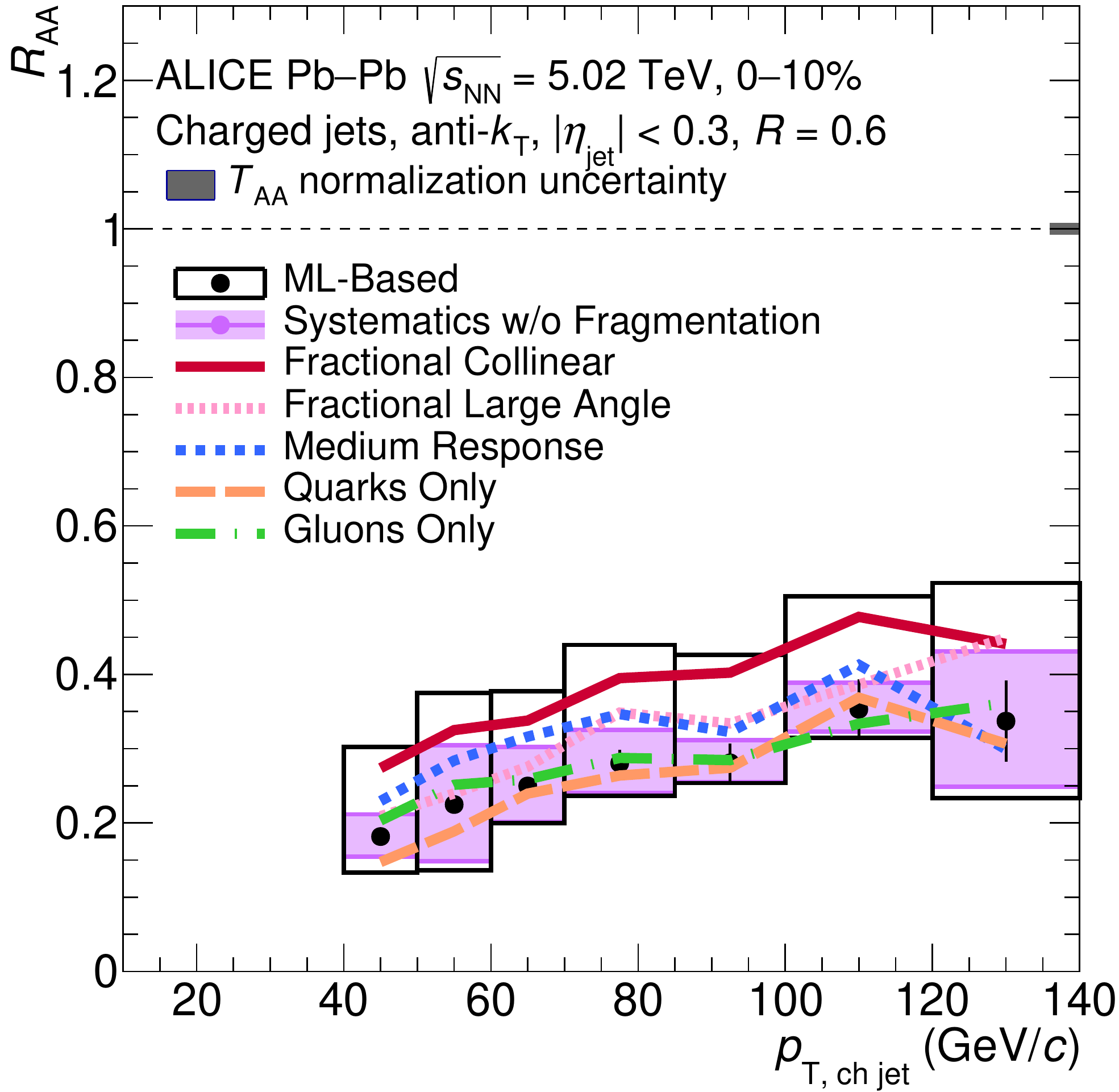}
  \includegraphics[width=0.49\textwidth]{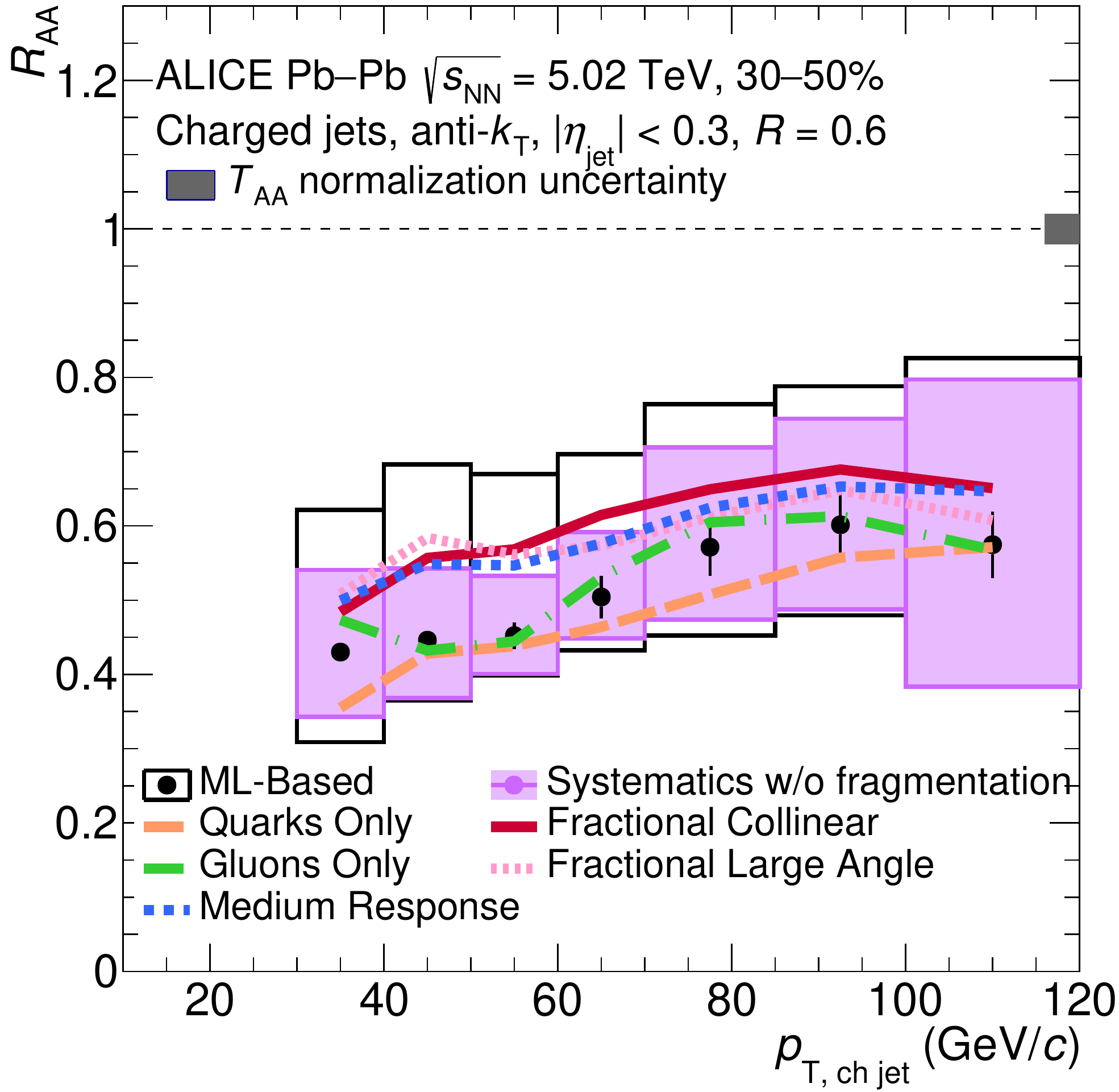}
  \centering
  \caption{Nuclear modification factors for jets with $R=0.6$ in 0--10\% (left) and 30--50\% (right) central \PbPb\ collisions outlining the impact of the various fragmentation models on the final result. Note that the systematic uncertainties, described in Sec.~\ref{sec:systematics}, are drawn both with and without the fragmentation uncertainties in the empty and filled boxes, respectively.}
  \label{fig:RAA_1}
\end{figure}

\begin{figure}[p!]
\centering
  \includegraphics[width=0.49\textwidth]{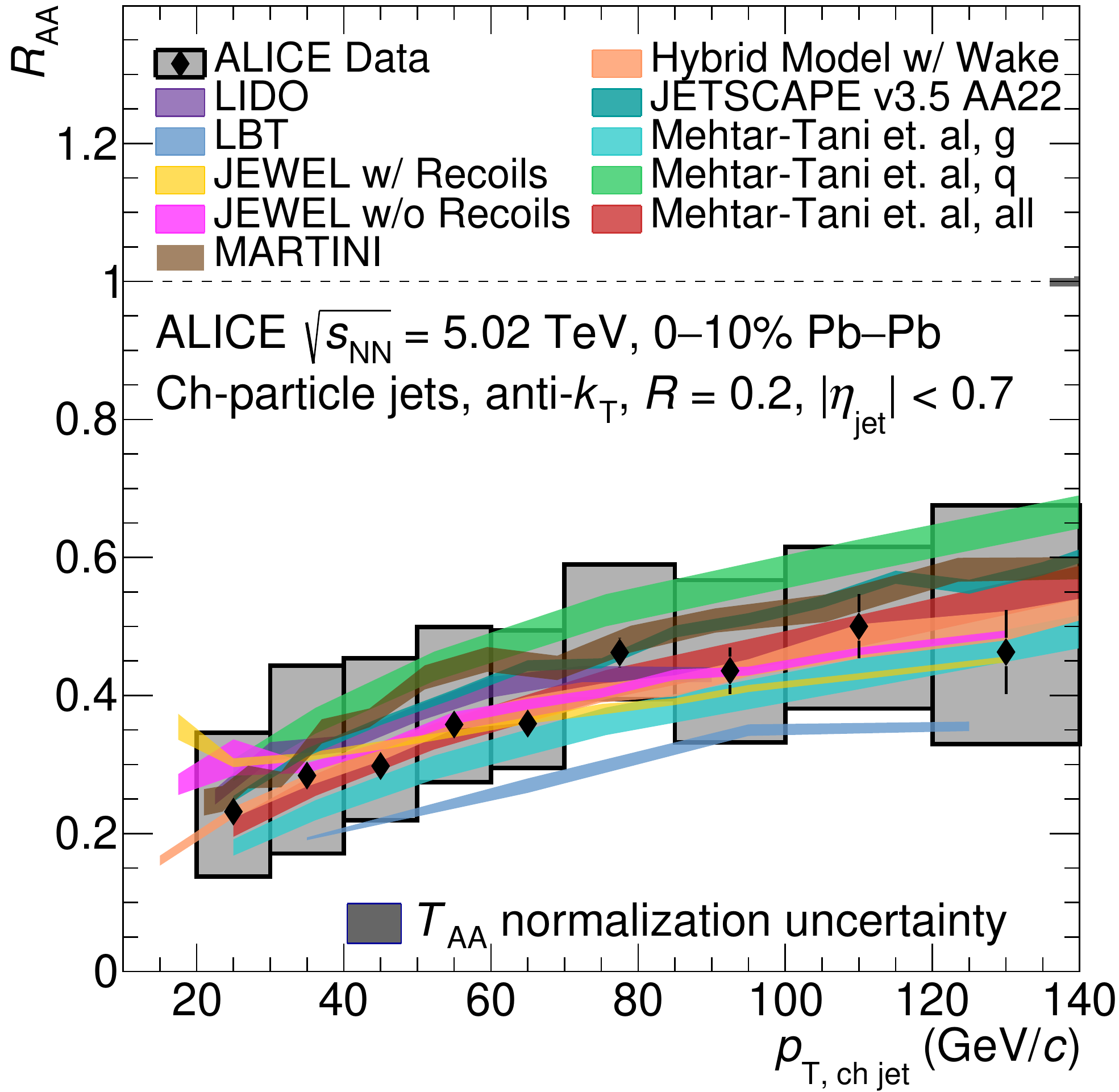}
  \includegraphics[width=0.49\textwidth] {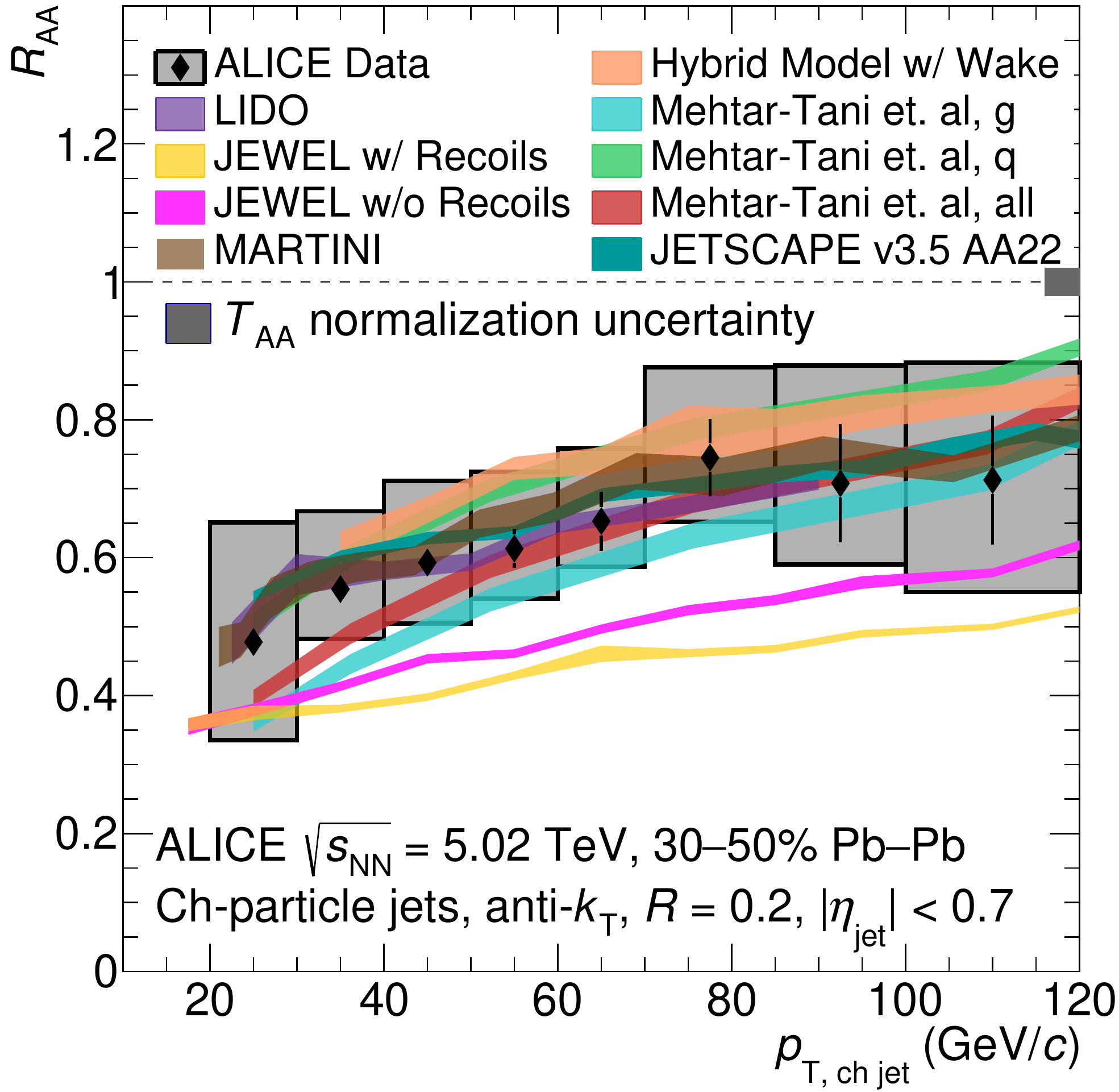}
  \includegraphics[width=0.49\textwidth]
  {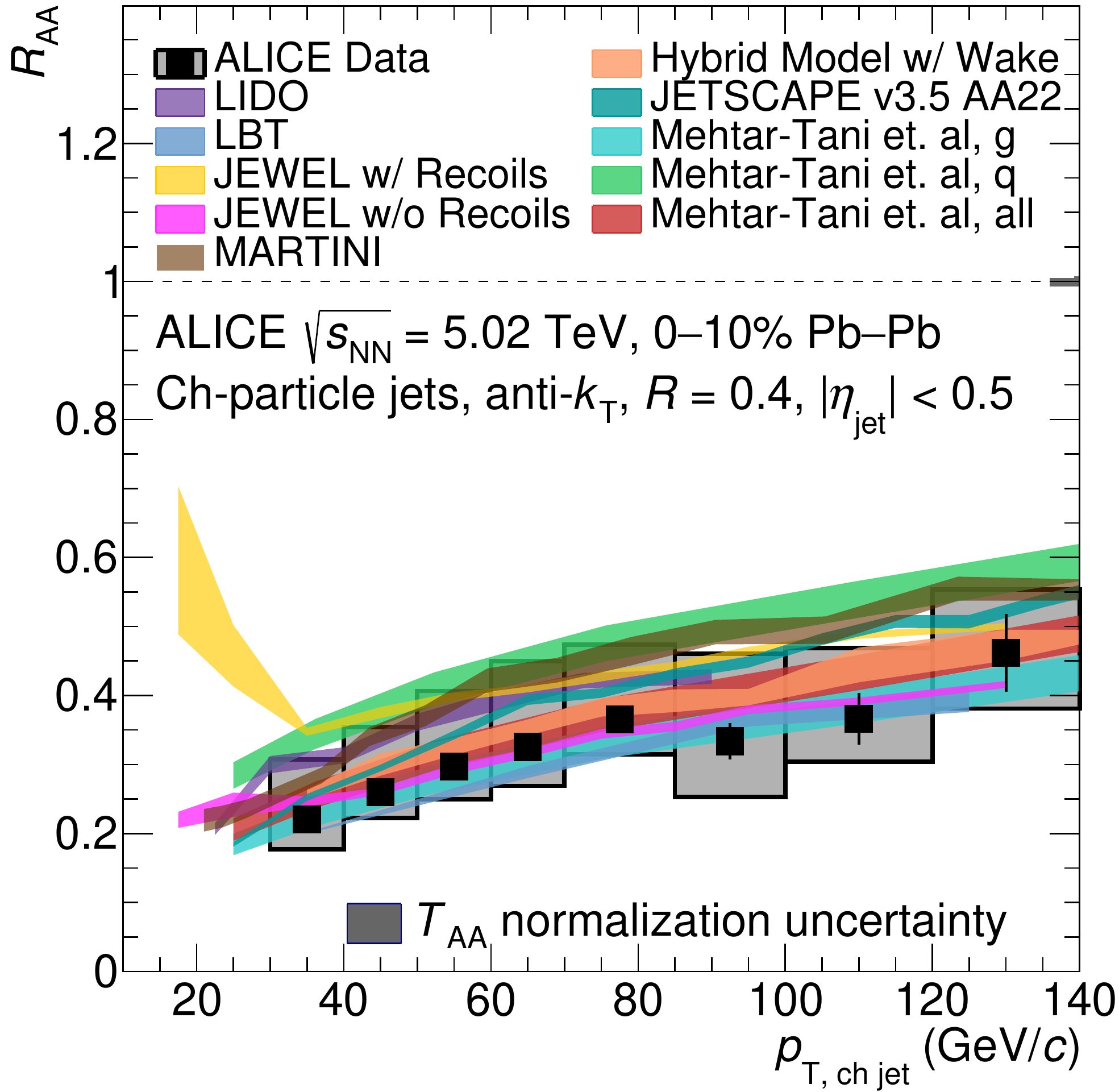}
  \includegraphics[width=0.49\textwidth]{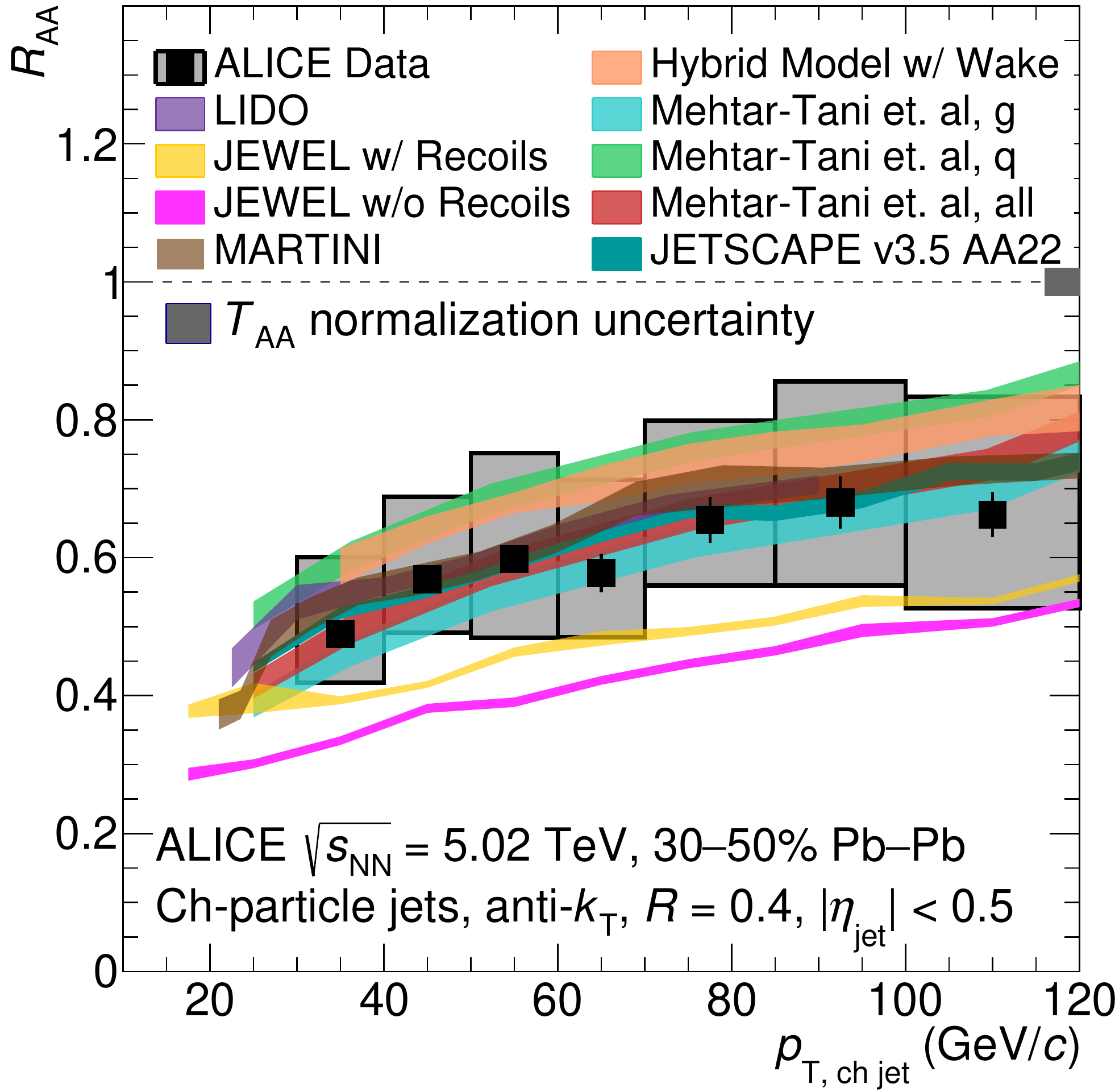}
  \includegraphics[width=0.49\textwidth]
  {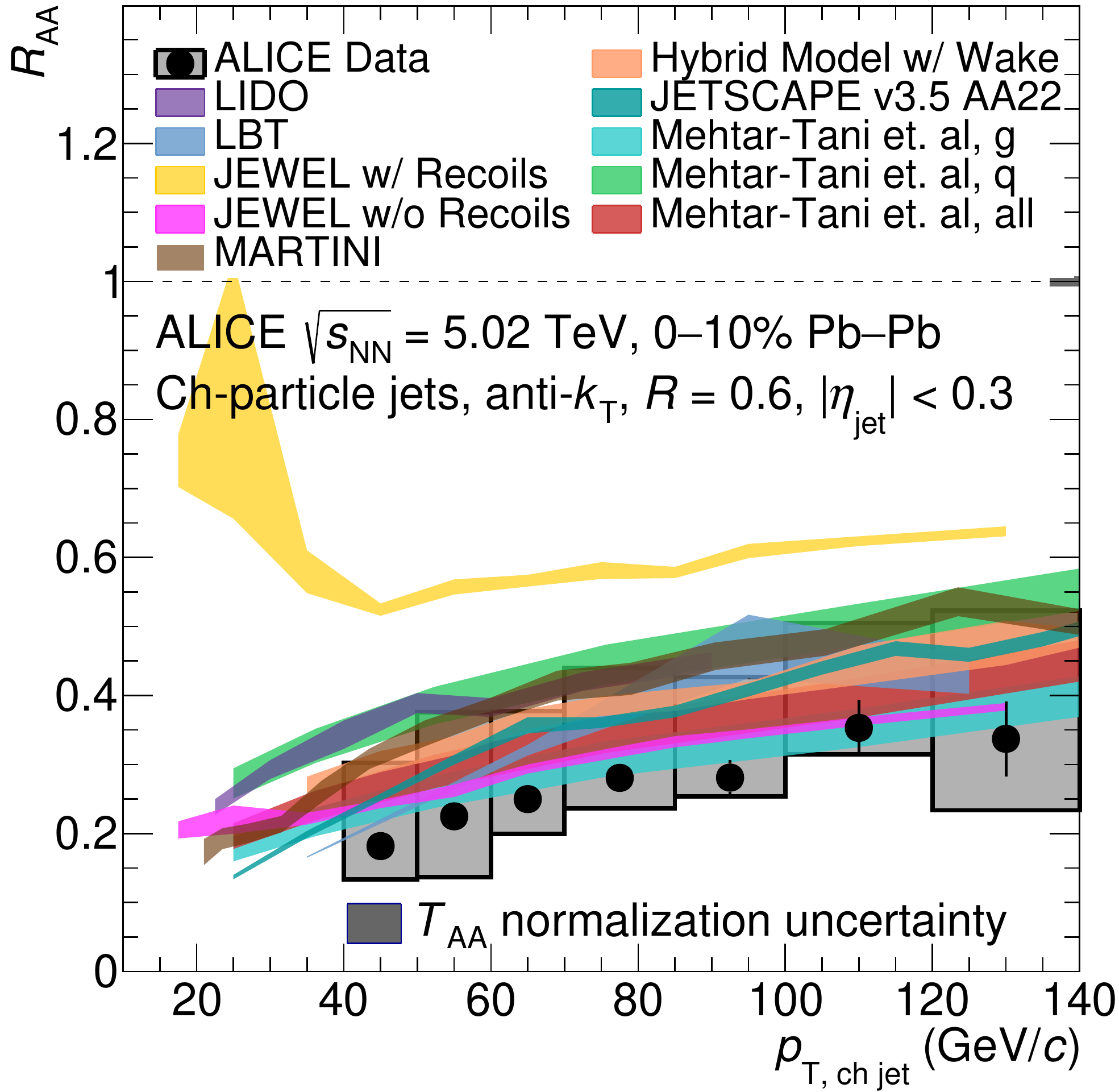}
  \includegraphics[width=0.49\textwidth]{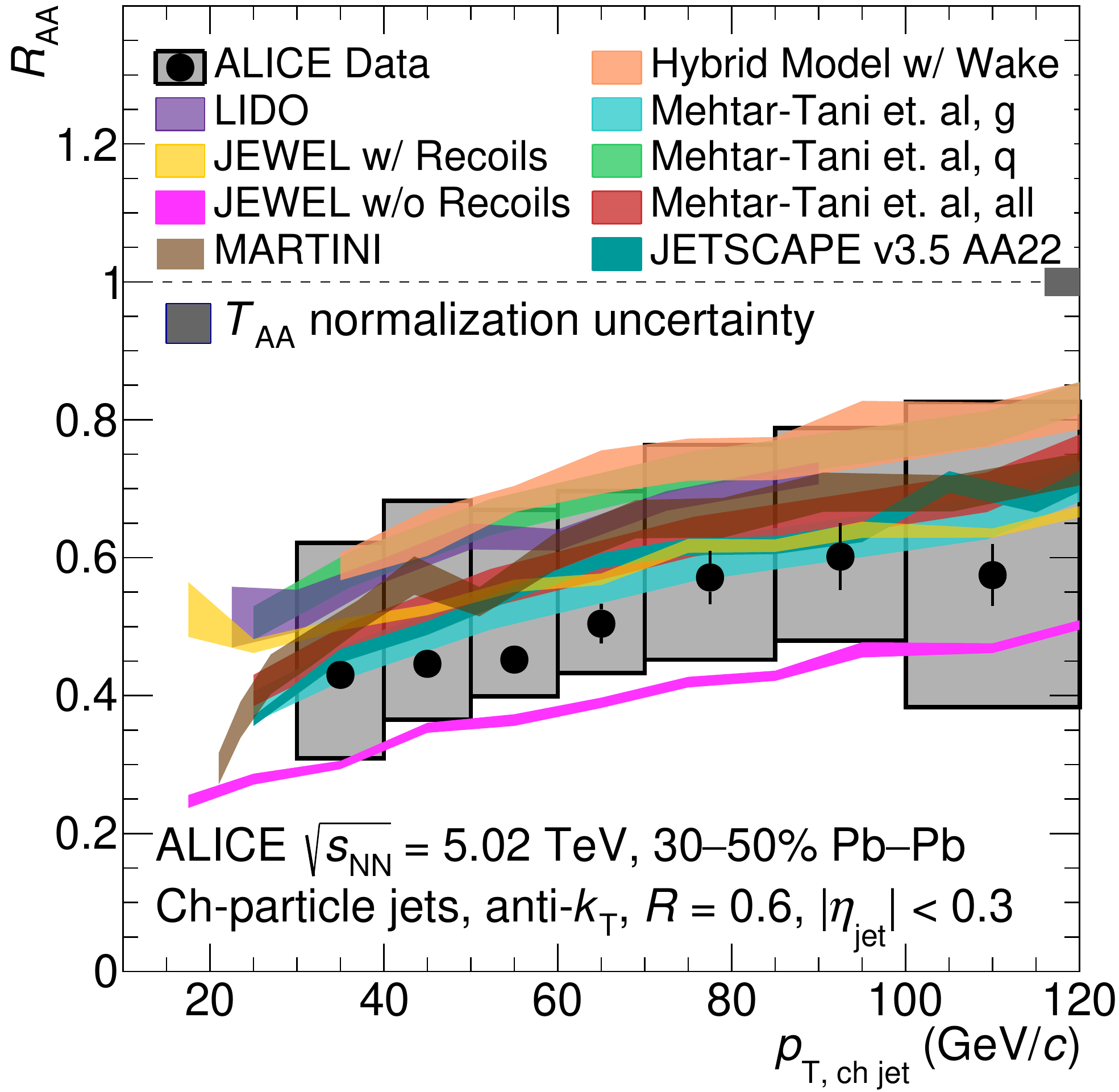}
  \caption{Nuclear modification factors for $R=0.2$, $R=0.4$, and $R=0.6$, shown for 0--10\% and 30--50\% central \PbPb\ collisions compared to theoretical calculations incorporating jet quenching (see text for details).}
  \label{fig:RAA_3}
\end{figure}

\begin{figure}[t!]
  \includegraphics[width=0.469\textwidth]{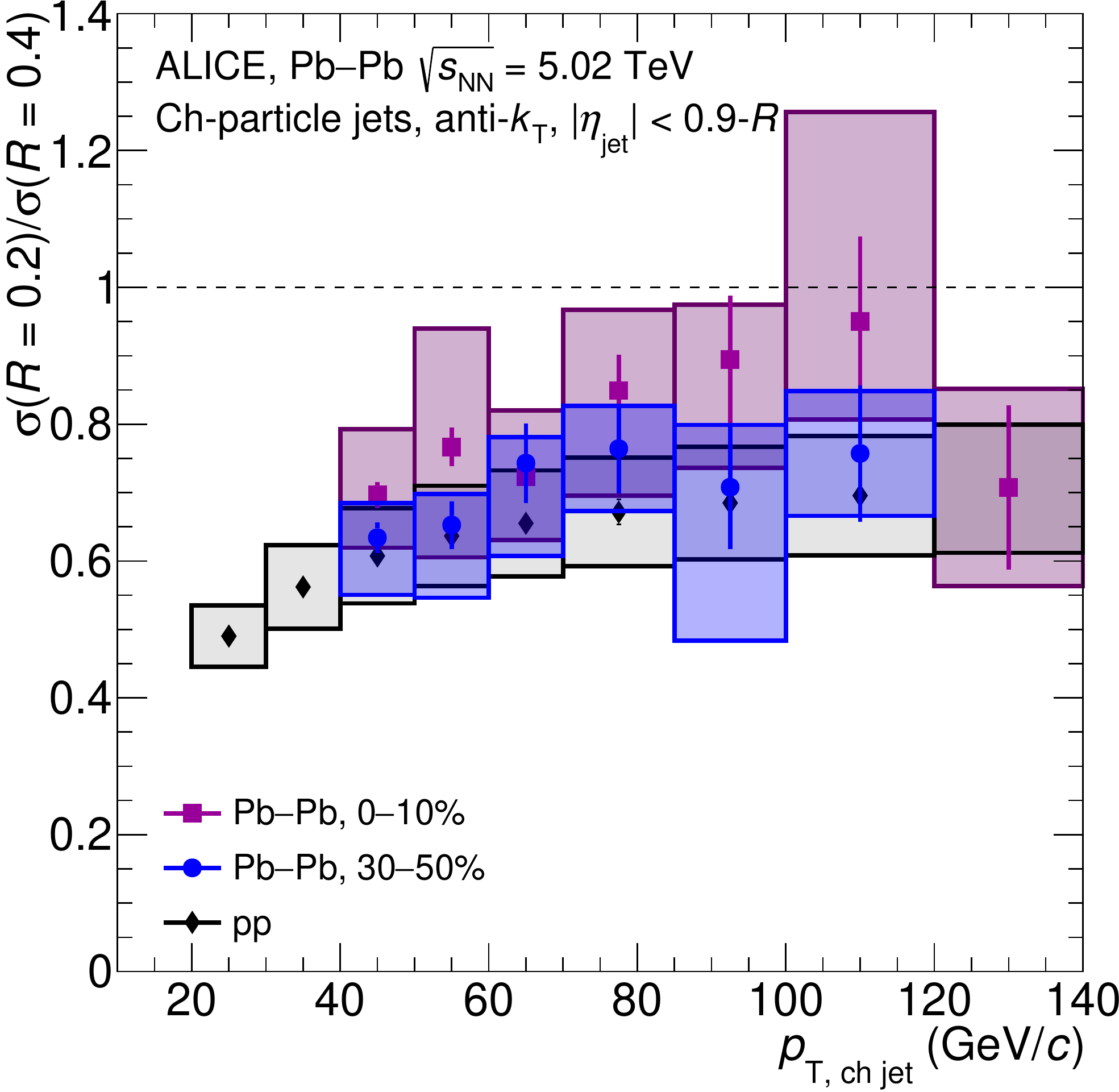}
  \hspace{0.2cm}
  \includegraphics[width=0.469\textwidth]{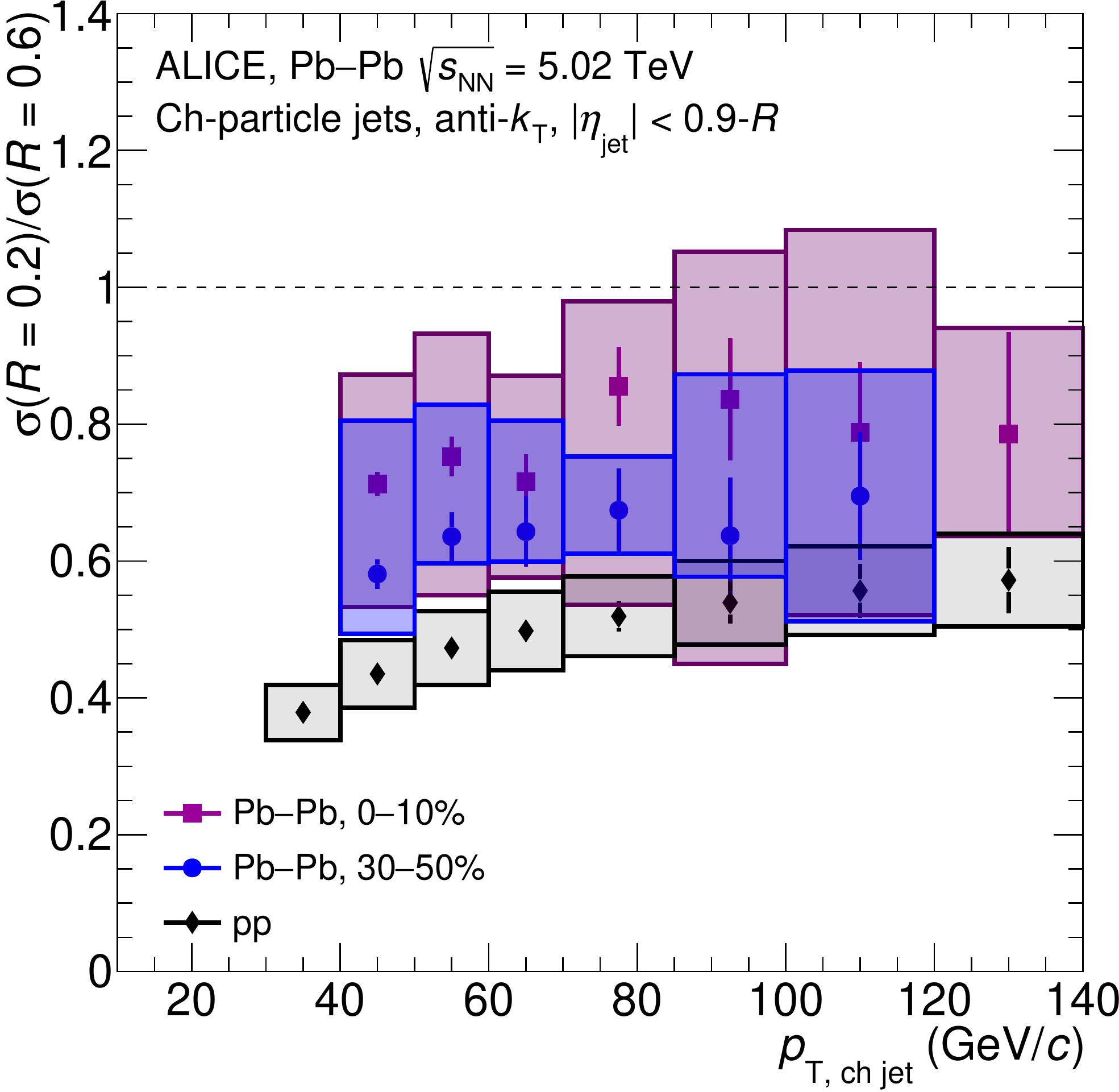}
  \caption{Jet cross section ratios for $\sigma(R = 0.2)/\sigma(R = 0.4)$ (left) and $\sigma(R = 0.2)/\sigma(R = 0.6)$ (right).}
  \label{fig:CSR}
\end{figure}

\begin{figure}[t!]
  \includegraphics[width=0.469\textwidth]{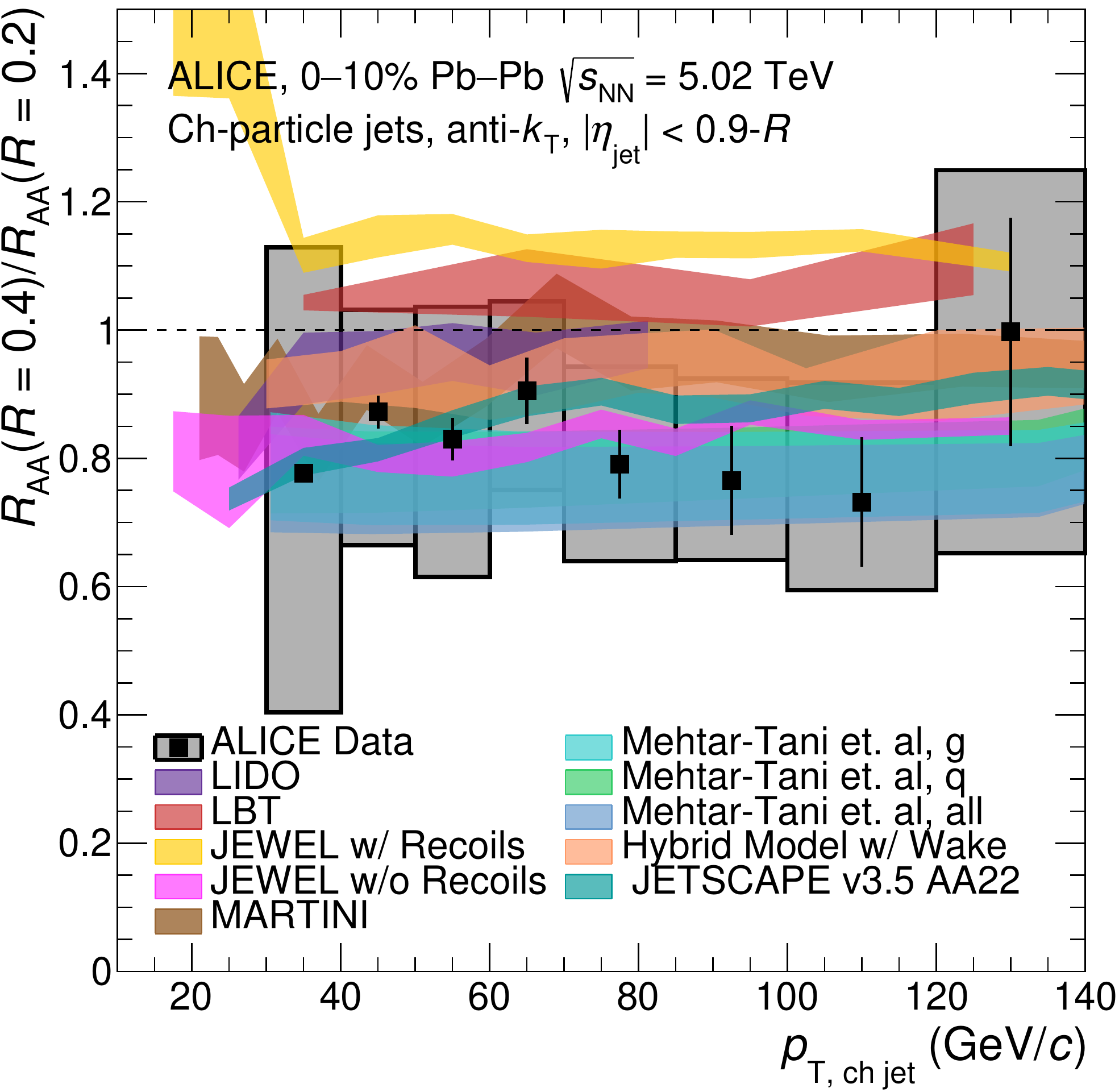}
  \hspace{0.2cm}
  \includegraphics[width=0.469\textwidth]{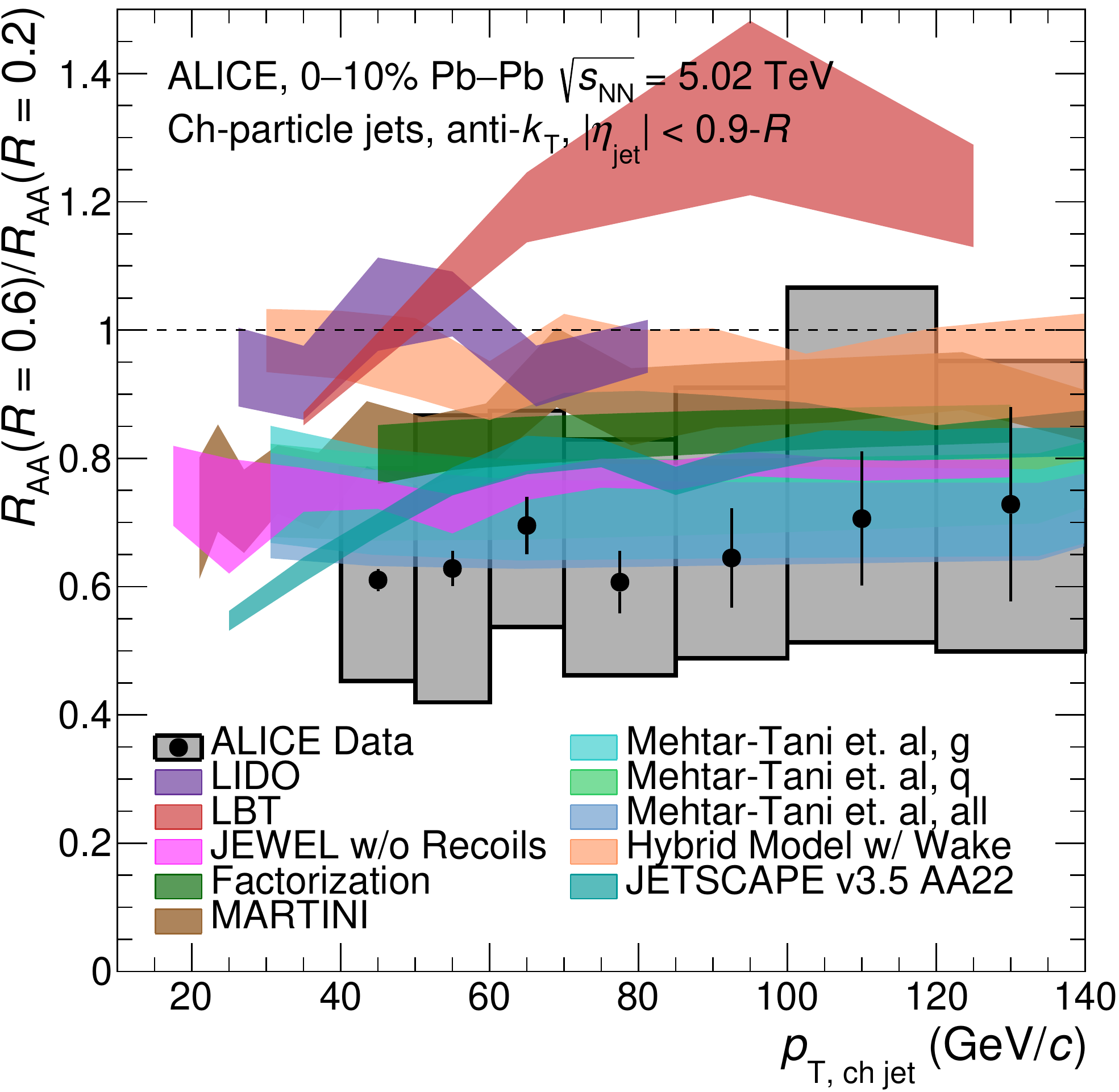}
\includegraphics[width=0.469\textwidth]{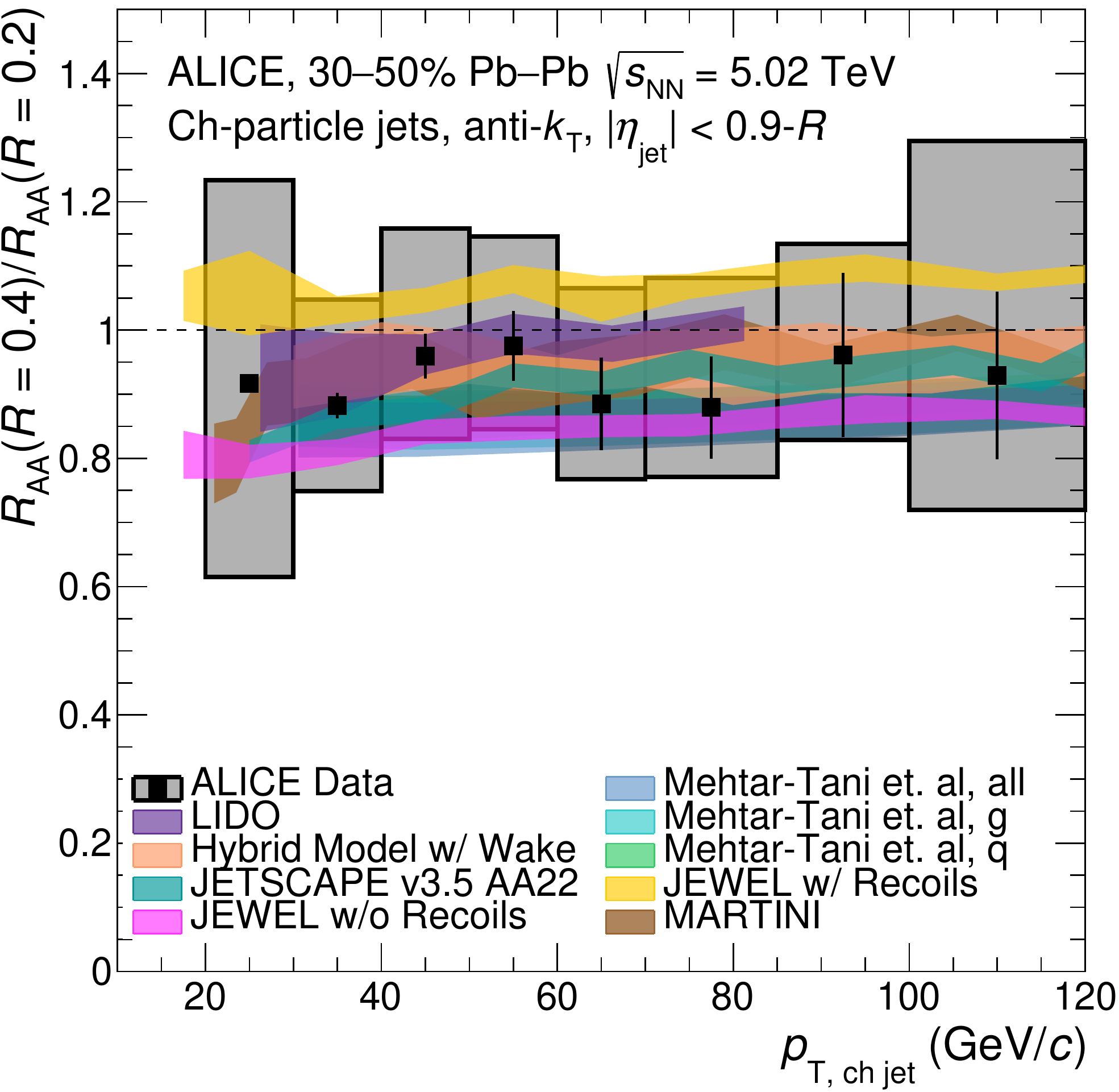}
  \hspace{0.2cm}
  \includegraphics[width=0.469\textwidth]{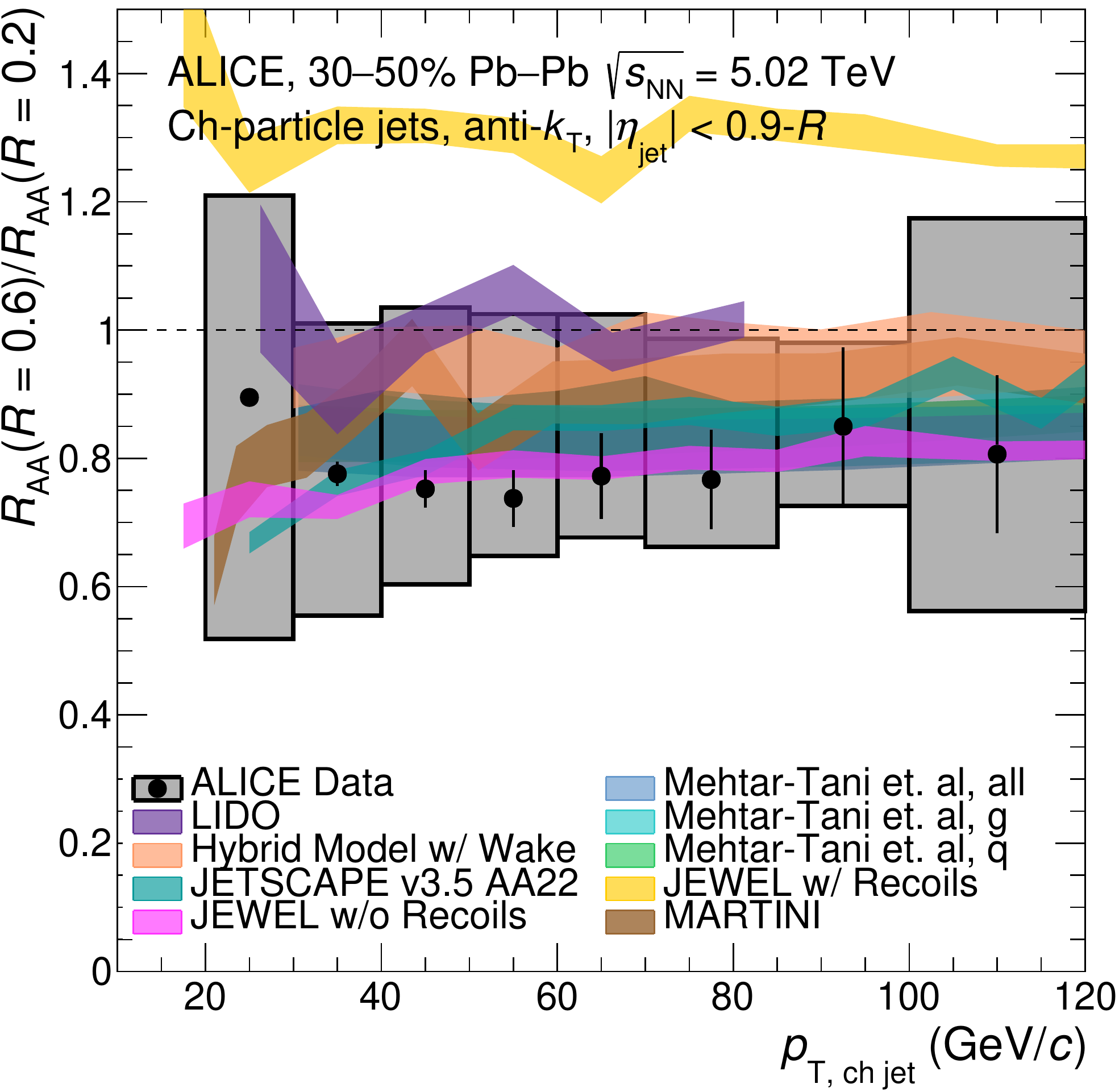}
  \centering
  \caption{Double ratio of jet nuclear modification factors using $R_\mathrm{AA}^{R=0.2}$ as the denominator and using $R = 0.4$ (left) and $R = 0.6$ (right) as the numerator compared to model predictions for central (top row) and semi-central (bottom row) collisions. Note that a comparison to JEWEL with recoils was omitted from the top right plot as its prediction is out of scale.}
  \label{fig:DoubleRatio}
\end{figure}

%%%%%%%%%%%%%%%%%%%%%%%%%%%%%%%%%%%%%%%%%%%%%%%%%%%%%%%%%%%%%%%%%%%%%%%%%%%%%%%%%%%%%%%%%%%%%%%%%%%%
In this section, the \pT\/-differential inclusive charged-particle jet production yields~(\Fig{fig:Spectrum_1}), 
the nuclear modification factors $R_\mathrm{AA}$ (Figs.~\ref{fig:RAA_2},~\ref{fig:RAA_1},~\ref{fig:RAA_3}), the ratios of the jet cross sections for different $R$ (\Fig{fig:CSR}), and the $R_\mathrm{AA}$ double ratios representing the change of nuclear modification with respect to resolution parameter of $R=0.2$ (Fig.~\ref{fig:DoubleRatio}) are reported.

The \pT\/-differential charged-particle jet cross section in \pp\ collisions is shown in the bottom right panel of Fig.~\ref{fig:Spectrum_1} for a broad range of $R$ values from $R=0.2$ to $R=0.6$. Jets with larger $R$ capture more of the jet's energy, shifting the spectra to the right and resulting in an increased yield at fixed jet $p_{\rm T}$. 
The effect is largest at low transverse momenta. The charged-particle jet spectra in Pb--Pb collisions are presented as an event-normalized yield divided by the average nuclear thickness $\left<T_\mathrm{AA}\right>$~\cite{Loizides:2017ack} of the given centrality class
\begin{equation}
\label{eq:spectrum}
\frac{1}{\left<T_{\rm AA}\right> } \frac{\mathrm{d}^{2}N_\mathrm{ch\;jet}}{\mathrm{d}{p}_\mathrm{T,\;ch\;jet} \mathrm{d}\eta_\mathrm{jet}} [\text{mb}\;(\mathrm{GeV}/c)^{-1}].
\end{equation}
These spectra are shown in the first three panels of Fig.~\ref{fig:Spectrum_1} for the three considered centrality classes and the three values of $R$. The uncertainty from the Glauber calculation to derive $\left<T_\mathrm{AA}\right>$ is included in the normalization uncertainty of the measurements. 
The spectra for central and semi-central collisions are measured using the ML-based method, while the area-based method is used for peripheral (60--80\%) collisions. 
While the performance of the ML-based correction is comparable to the AB method in peripheral collisions, the AB correction has the benefit of reduced systematic uncertainties due to the absence of fragmentation uncertainties for this method. 
The area-based method does not include these fragmentation uncertainties because the effect of the fragmentation biases is small~\cite{Adam:2015doa, Adamczyk:2017yhe, STAR:2020xiv}. 

Figure~\ref{fig:RAA_2} shows the nuclear modification factors using both the new ML-based estimator and the established area-based estimator for the 0--10\%,  30--50\%, and 60--80\% centrality classes. The inclusive yield suppression measured using these two approaches is consistent within uncertainties in their region of overlap.
The ML-based method enables measurements at lower transverse momenta and for large $R$ ($R$ = 0.6) in most central \PbPb\ collisions. 

\Figure{fig:RAA_1} shows the variation in the central value of the inclusive yield suppression for $R$ = 0.6 and the
different fragmentation models described in \Sec{sec:fragdependence}.
The systematic uncertainties with and without the fragmentation systematic uncertainty are shown to illustrate its relative contributions to the total. As discussed in Sec.~\ref{sec:systematics}, the fragmentation systematic uncertainties were treated as asymmetric. As demonstrated in Fig.~\ref{fig:RAA_1}, this typically results in a positive contribution to the yield suppression with the exception of the Quarks Only variation.
Typically the Fractional Collinear variation gives the largest contribution to the systematic uncertainty.

In Fig.~\ref{fig:RAA_3}, the nuclear modification factors for $R=0.2$, $R=0.4$, and $R=0.6$ are compared to theoretical models incorporating jet quenching. 
Results are compared with:
the Hybrid Model~\cite{Casalderrey-Solana:2014bpa}, which implements an energy loss with an AdS-CFT-inspired
dependence on the path length, as well as a response of the medium to the lost energy;
the Linear Boltzmann Transport (LBT) model~\cite{He:2015pra, He:2018xjv} and LIDO~\cite{Ke:2020clc}, which use linear Boltzmann equations to describe the transport of partons in the QGP; 
JETSCAPE~\cite{Putschke:2019yrg}, which includes a medium-modified parton shower at high parton virtuality via MATTER~\cite{Cao:2017qpx}, switching to the LBT model at low virtuality~\cite{He:2015pra} (JETSCAPEv3.5 AA22 tune); Mehtar-Tani et al.~\cite{Mehtar-Tani:2021fud}, which is a first-principles analytical calculation of the single-inclusive jet spectrum using quenching factors; MARTINI~\cite{Schenke:2009gb}, which embeds partons into a hydrodynamic medium with a modified parton shower; 
and JEWEL~\cite{Zapp:2012ak, Zapp:2013vla}, which consists of a Monte Carlo implementation of BDMPS-based medium-induced gluon radiation in a medium modeled with a Bjorken expansion, including calculations both with and without enabling recoils~\cite{KunnawalkamElayavalli:2017hxo}.
Some model calculations do not cover the full phase space of the measurements.

The calculations generally describe the data in central collisions for the smaller resolution parameters ($R = $ 0.2 and 0.4), except for JEWEL with recoils which overestimates the $R=0.4$ result at low jet $p_{\rm T}$. 
Additionally, the JEWEL with recoils predicts significantly higher values for the result at $R=0.6$ than the measurement. 
The calculations span a larger range for the semi-central collisions but still mostly describe the data within uncertainties, although JEWEL shows some slight tension. 
Overall, these comparisons demonstrate the importance of comparing with models over a wide $p_{\rm T}$ interval and for different values of the $R$ parameter, particularly at large $R$.

The jet cross section ratio is defined as the ratio of the per-event jet yields measured in the same collision system for different resolution parameters:
\begin{equation}
\label{eq:CSR}
\sigma(R = R_1)/\sigma(R = R_2) = \frac{\mathrm{d}N_\mathrm{R_1}}{\mathrm{d}p_\mathrm{T,\;ch\;jet}}\Big/\frac{\mathrm{d}N_\mathrm{R_2}}{\mathrm{d}p_\mathrm{T,\;ch\;jet}}.
\end{equation}
In the ratio, the tracking and fragmentation uncertainties are highly correlated,
while all other uncertainties are considered as uncorrelated. \Sec{sec:systematics} describes the calculation of the systematic uncertainties in the ratio.
The inclusive jet cross section ratios are a key observable for jet shapes and have been measured both at RHIC and the LHC~\cite{Abelev:2013fn, ALICE:2019qyj, CMS:2016uxf, CMS:2016jip, ALICE:2013dpt, STAR:2020xiv}.
The jet cross section ratios are shown in Fig.~\ref{fig:CSR} for \pp\ collisions and for \PbPb\ collisions in the 0--10\% and 30--50\% centrality intervals. The left panel presents the ratios for $R=0.2$ and $R=0.4$, and the right panel for $R=0.2$ and $R=0.6$. 
The ratio for $\sigma(R=0.2)$/$\sigma(R=0.6)$ in Pb--Pb collisions is slightly larger than
for pp collisions, taking into account the uncertainties. This suggests a narrowing of the intra-jet energy distribution in Pb--Pb collisions.
No significant centrality dependence is observed in the jet cross section ratios within measurement uncertainties. 
There is a small $p_{\rm T}$-dependence in the jet cross section ratios in \pp\ collisions, which becomes stronger at lower jet $p_{\rm T}$, resulting from the $p_{\rm T}$ spectrum being steeper for larger $R$ than from smaller $R$ in pp collisions. 
However, the jet cross section ratios in \PbPb\ collisions are consistent with no dependence on the jet $p_{\rm T}$. 
This indicates that there is an $R$-dependence to the evolution of the jet cross section in pp collisions with $p_{\rm T}$ which may impact the dependence of the $R_{\rm AA}$ on $R$ and $p_{\rm T}$, as discussed below.

The double ratio of the nuclear modification factor, which compares $R_\mathrm{AA}^{R}$ for different $R$ to $R_\mathrm{AA}^{R=0.2}$, is used to quantify the variation of the nuclear modification factor with respect to the jet resolution parameter. 
It is defined as
\begin{equation}
\label{eq:DoubleRatio}
R_\mathrm{AA}^{R/0.2} = \frac{R_\mathrm{AA}^{R}}{R_\mathrm{AA}^{0.2}} = \frac{\sigma_\mathrm{AA}(R)}{\sigma_\mathrm{AA}(0.2)} \Big/ \frac{\sigma_\mathrm{pp}(R)}{\sigma_\mathrm{pp}(0.2)}.
\end{equation}
Thus, the observable is not only a double ratio of nuclear modification factors but also of jet cross section ratios as defined in Eq. (~\ref{eq:CSR}).
The $R_{\rm AA}$ double ratio is a key observable to quantify the $R$-dependence of energy loss: 
when this ratio is less than unity, jets with larger $R$ are more suppressed; 
when it is consistent with unity, there is no $R$-dependence~(or a cancellation of effects); and when it is greater than unity, larger $R$ jets are less suppressed.

The measured $R_{\rm AA}$ double ratios are shown in Fig.~\ref{fig:DoubleRatio} for 0--10\% and 30--50\% central \PbPb\ collisions.
The $R_\mathrm{AA}^{0.4/0.2}$ ratio in 0--10\% and 30--50\% central collisions and the $R_\mathrm{AA}^{0.6/0.2}$ ratio in 30--50\% central collisions are consistent with unity, indicating no significant $R$-dependence or $\pT{}$-dependence.
In contrast, the $R_\mathrm{AA}^{0.6/0.2}$ ratio in 0--10\% central collisions is below unity at lower jet $\pT$ values, indicating a hint of an $R$-dependence within uncertainties.
Assuming a 50\% (100\%) correlation of the systematic uncertainties, the deviation from unity in 0--10\% central collisions is approximately 2.8 (1.6) sigma for the $R_\mathrm{AA}^{0.6/0.2}$ ratio, and approximately 1.6 (1.0) sigma for $R_\mathrm{AA}^{0.4/0.2}$ ratio.

Many competing effects can be considered when interpreting this ratio.
As observed in Ref.~\cite{CMS:2018zze}, the jet energy is transferred mostly to soft particles, and a significant fraction of these particles are found at large angles relative to the jet axis. 
Thus, as the jet $R$ increases, the energy lost outside of a smaller $R$ jet cone should be recovered in a larger $R$ jet cone. 
Additionally, the medium responds to the jet as the jet propagates through it, which is hypothesized to cause a wake that pushes particles back inside the jet cone~\cite{Casalderrey-Solana:2014bpa}.
This could cause an increase in the $R_{\rm AA}$ with increasing $R$. 
Recent jet substructure measurements, such as those in Ref.~\cite{ALargeIonColliderExperiment:2021mqf, ATLAS:2022vii}, show that the yield of jets with a more complex substructure, for example, wider jets, is suppressed at a fixed jet $p_{\rm T}$ compared to narrower jets. 
The larger $R$ jets at a given $\pT$ could be a population whose transverse distribution is broader in pp collisions but which, consequently, are more strongly quenched in the medium.
Such an effect would cause the $R_{\rm AA}$ to decrease with increasing $R$. 
Finally, in vacuum, the jet spectrum is slightly steeper for large $R$ jets as seen in Fig.~\ref{fig:CSR} and Ref.~\cite{CMS:2020caw}.
Thus, even for the same energy loss, some small decrease in the $R_{\rm AA}$ with increasing $R$ would be observed.

The double ratio is also compared in Fig.~\ref{fig:DoubleRatio} to calculations that can exhibit different dependencies with the jet $R$ depending on the relative contributions of various energy loss mechanisms, making this a potentially discriminating observable.
In addition to the models previously compared to the jet $R_{\rm AA}$, further comparisons are made with Qiu et al.~\cite{Qiu:2019sfj} calculations, which are based on a factorization approach inspired by phenomenological considerations.
JEWEL with recoils shows an increasing $R_{\rm AA}$ with increasing $R$ due to the medium response, which contrasts with the data, especially for large $R$.
LBT also shows an increasing trend with increasing $R$, as well as a jet $p_{\rm T}$ dependence at large $R$, which is not supported by the data.
The Hybrid Model and LIDO predict values of the double ratio close to unity, indicating a very mild dependence on $R$ in these models.
JETSCAPE, JEWEL without recoils, MARTINI, all variations of Mehtar-Tani et al., and the factorization (Qiu et al.) model show the $R_{\rm AA}$ decreasing with increasing $R$, describing the trend in the data.
This could indicate that wider jets with a more complex substructure experience more suppression. 
This explanation may be compatible with the CMS measurement~\cite{CMS:2021vui}, which did not show an $R$-dependence, as there may be a $p_{\rm T}$-dependence to the substructure of the jet~\cite{CMS:2020caw}. 
Additionally, quenching effects are expected to be larger at lower jet $p_{\rm T}$. The double ratio in Fig.~\ref{fig:DoubleRatio} is in contrast with the $R_{\rm cp}$ results from ATLAS~\cite{ATLAS:2012tjt}, though many differences in the jet populations may contribute to these differences. 
Assuming a purely fractional energy loss scheme where each jet loses a specified fraction of its energy, only an $\sim$ 3\% difference in energy loss between $R = 0.6$ and $R = 0.2$ is needed to create a difference in the $R_{\rm AA}$ values compatible with what is observed in Fig.~\ref{fig:DoubleRatio}. 

%%%%%%%%%%%%%%%%%%%%%%%%%%%%%%%%%%%%%%%%%%%%%%%%%%%%%%%%%%%%%%%%%%%%%%%%%%%%%%%%%%%%%%%%%%%%%%%%%%%%
\section{Summary}
\label{sec:summary}
%%%%%%%%%%%%%%%%%%%%%%%%%%%%%%%%%%%%%%%%%%%%%%%%%%%%%%%%%%%%%%%%%%%%%%%%%%%%%%%%%%%%%%%%%%%%%%%%%%%%
A new ML-based background estimator was applied to measure inclusive charged-particle jets with resolution parameters up to $R=0.6$ in \PbPb\ collisions at the LHC. 
The ML-based approach leads to significantly reduced residual background fluctuations compared to the area-based method ~(Fig.~\ref{fig:deltaPt}).
This improvement is achieved at the expense of an additional systematic uncertainty from the fragmentation dependence of the background estimator~(Figs.~\ref{fig:ToyStudies} and~\ref{fig:RAA_1}, Tabs.~\ref{tab:Systematics1},~\ref{tab:Systematics2}, and~\ref{tab:Systematics3}). Still, this method allowed for the measurement of $R=0.6$ jets down to a low jet $\pT$ of 40 GeV/$c$ in central (0--10\%) \PbPb\ collisions for the first time.
Using the new estimator, the transverse momentum spectra (Fig.~\ref{fig:Spectrum_1}),  nuclear modification factors~(Figs.~\ref{fig:RAA_2} and~\ref{fig:RAA_3}), 
cross section ratios (Fig.~\ref{fig:CSR}), and nuclear modification double ratios~(Fig.~\ref{fig:DoubleRatio}) of charged-particle jets in \PbPb\ collisions at $\sqrt{s_\mathrm{NN}} = 5.02$ TeV could be measured. 
Comparing the nuclear modification factors for different resolution parameters indicates increased jet suppression with increasing $R$, most significantly for the $R=0.6$ jets.
This is also reflected in the ratios of jet cross sections reconstructed with different $R$ values measured in \PbPb\ collisions, which also deviate from the \pp\ reference for most central collisions. 
The data are consistent with a variety of theoretical descriptions, including JETSCAPE, the analytical calculations of Mehtar-Tani and Qiu et al., and JEWEL without recoiling thermal medium particles.

%==========================================================%
%======================ACK+BIBLIO==========================%
%==========================================================%

\newenvironment{acknowledgement}{\relax}{\relax}
\begin{acknowledgement}
\section*{Acknowledgements}
% Version: 2023-02-10

The ALICE Collaboration would like to thank all its engineers and technicians for their invaluable contributions to the construction of the experiment and the CERN accelerator teams for the outstanding performance of the LHC complex.
The ALICE Collaboration gratefully acknowledges the resources and support provided by all Grid centres and the Worldwide LHC Computing Grid (WLCG) collaboration.
The ALICE Collaboration acknowledges the following funding agencies for their support in building and running the ALICE detector:
A. I. Alikhanyan National Science Laboratory (Yerevan Physics Institute) Foundation (ANSL), State Committee of Science and World Federation of Scientists (WFS), Armenia;
Austrian Academy of Sciences, Austrian Science Fund (FWF): [M 2467-N36] and Nationalstiftung f\"{u}r Forschung, Technologie und Entwicklung, Austria;
Ministry of Communications and High Technologies, National Nuclear Research Center, Azerbaijan;
Conselho Nacional de Desenvolvimento Cient\'{\i}fico e Tecnol\'{o}gico (CNPq), Financiadora de Estudos e Projetos (Finep), Funda\c{c}\~{a}o de Amparo \`{a} Pesquisa do Estado de S\~{a}o Paulo (FAPESP) and Universidade Federal do Rio Grande do Sul (UFRGS), Brazil;
Bulgarian Ministry of Education and Science, within the National Roadmap for Research Infrastructures 2020-2027 (object CERN), Bulgaria;
Ministry of Education of China (MOEC) , Ministry of Science \& Technology of China (MSTC) and National Natural Science Foundation of China (NSFC), China;
Ministry of Science and Education and Croatian Science Foundation, Croatia;
Centro de Aplicaciones Tecnol\'{o}gicas y Desarrollo Nuclear (CEADEN), Cubaenerg\'{\i}a, Cuba;
Ministry of Education, Youth and Sports of the Czech Republic, Czech Republic;
The Danish Council for Independent Research | Natural Sciences, the VILLUM FONDEN and Danish National Research Foundation (DNRF), Denmark;
Helsinki Institute of Physics (HIP), Finland;
Commissariat \`{a} l'Energie Atomique (CEA) and Institut National de Physique Nucl\'{e}aire et de Physique des Particules (IN2P3) and Centre National de la Recherche Scientifique (CNRS), France;
Bundesministerium f\"{u}r Bildung und Forschung (BMBF) and GSI Helmholtzzentrum f\"{u}r Schwerionenforschung GmbH, Germany;
General Secretariat for Research and Technology, Ministry of Education, Research and Religions, Greece;
National Research, Development and Innovation Office, Hungary;
Department of Atomic Energy Government of India (DAE), Department of Science and Technology, Government of India (DST), University Grants Commission, Government of India (UGC) and Council of Scientific and Industrial Research (CSIR), India;
National Research and Innovation Agency - BRIN, Indonesia;
Istituto Nazionale di Fisica Nucleare (INFN), Italy;
Japanese Ministry of Education, Culture, Sports, Science and Technology (MEXT) and Japan Society for the Promotion of Science (JSPS) KAKENHI, Japan;
Consejo Nacional de Ciencia (CONACYT) y Tecnolog\'{i}a, through Fondo de Cooperaci\'{o}n Internacional en Ciencia y Tecnolog\'{i}a (FONCICYT) and Direcci\'{o}n General de Asuntos del Personal Academico (DGAPA), Mexico;
Nederlandse Organisatie voor Wetenschappelijk Onderzoek (NWO), Netherlands;
The Research Council of Norway, Norway;
Commission on Science and Technology for Sustainable Development in the South (COMSATS), Pakistan;
Pontificia Universidad Cat\'{o}lica del Per\'{u}, Peru;
Ministry of Education and Science, National Science Centre and WUT ID-UB, Poland;
Korea Institute of Science and Technology Information and National Research Foundation of Korea (NRF), Republic of Korea;
Ministry of Education and Scientific Research, Institute of Atomic Physics, Ministry of Research and Innovation and Institute of Atomic Physics and Universitatea Nationala de Stiinta si Tehnologie Politehnica Bucuresti, Romania;
Ministry of Education, Science, Research and Sport of the Slovak Republic, Slovakia;
National Research Foundation of South Africa, South Africa;
Swedish Research Council (VR) and Knut \& Alice Wallenberg Foundation (KAW), Sweden;
European Organization for Nuclear Research, Switzerland;
Suranaree University of Technology (SUT), National Science and Technology Development Agency (NSTDA) and National Science, Research and Innovation Fund (NSRF via PMU-B B05F650021), Thailand;
Turkish Energy, Nuclear and Mineral Research Agency (TENMAK), Turkey;
National Academy of  Sciences of Ukraine, Ukraine;
Science and Technology Facilities Council (STFC), United Kingdom;
National Science Foundation of the United States of America (NSF) and United States Department of Energy, Office of Nuclear Physics (DOE NP), United States of America.
In addition, individual groups or members have received support from:
European Research Council, Strong 2020 - Horizon 2020, Marie Sk\l{}odowska Curie (grant nos. 950692, 824093, 896850), European Union;
Academy of Finland (Center of Excellence in Quark Matter) (grant nos. 346327, 346328), Finland;
Programa de Apoyos para la Superaci\'{o}n del Personal Acad\'{e}mico, UNAM, Mexico.
        %%%%%%% get the latest version before submitting
\end{acknowledgement}
\bibliographystyle{utphys}
\bibliography{biblio}{}

\newpage
% \appendix
%%%%%%%%%%%%%%%%%%%%%%%%%%%%%%%%%%%%%%%%%%%%%%%%%%%%%%%%%%%%%%%%%%%%%%%%%%%%%%%%%%%%%%%%%%%%%%%%%%%%
\appendix
\section{Machine learning methods} 
\label{sec:machineLearningMethods}
%%%%%%%%%%%%%%%%%%%%%%%%%%%%%%%%%%%%%%%%%%%%%%%%%%%%%%%%%%%%%%%%%%%%%%%%%%%%%%%%%%%%%%%%%%%%%%%%%%%%
A novel background estimator based on machine learning is used to correct the $\pT$-smearing effects caused by the background, utilizing the approach described in Ref.~\cite{Haake:2018hqn}.
In this treatment, correcting the background effects is framed as a regression task which aims to predict a reconstructed jet $\pT$ value for each jet candidate.
Following this approach, several ML algorithms have been evaluated and compared, such as neural networks~\cite{Haykin1998}, random forests~\cite{Breiman2001}, and linear regression.
While these algorithms differ in performance, they all lead to similar, fully corrected results. 
For this analysis, we chose a shallow neural network model, which (as in Ref.~\cite{Haake:2018hqn}) demonstrates a slightly improved performance compared to other explored algorithms.
This shallow neural network is implemented as a three-layer perceptron with 100 nodes in the first two layers and 50 in the last. The activation function chosen for the nodes is the ReLU~\cite{NairHinton2010}, while the ADAM optimizer~\cite{KingmaB14} is employed in the neural network training.

The quality of the training dataset plays a crucial role in the applicability of corrections based on a machine-learning technique. 
To simulate events with jets in a heavy-ion background for the training of the ML estimator, reconstructed PYTHIA 8 events simulated to the detector level are embedded into a thermally-distributed background.
The thermal background is created by randomly distributing charged particles according to a uniform particle multiplicity distribution ranging from $0$ to $3000$ tracks with a uniform $\eta$ distribution and a realistic (quasi-thermal) transverse momentum distribution based on a Tsallis fit to data.
The transverse momentum distribution is tuned to describe the reconstructed track momentum distribution of Pb--Pb minimum bias data at low $\pt$.
For particles with $\pT \geq$ 4~\GeVc, the transverse momentum distribution in the thermal background is steeper than in data since, by construction, the thermal background does not include jets.
As a cross-check, real minimum bias \PbPb\ events were also used as background heavy-ion events. The difference in the distribution of final jet observables was negligible for different
choices of the background training distribution. This indicates that the model is robust to changes in the background used in training. 

In order to find a suitable combination of jet and event properties as input features to the neural network, the analysis was repeated for a large variety of configurations of the input parameters/features. 
The number of features used was kept small to ensure a generalizable model. 
These features were chosen by iteratively removing unimportant or highly correlated features as long as the performance was not significantly reduced, ensuring a minimal list with good performance was reached. 
Note that the metric for considering a parameter unimportant is a relatively low Gini~\cite{https://doi.org/10.48550/arxiv.1407.7502} importance in the random forest estimator, which is used as a proxy for important features in the neural network. 
The degree of correlation between variables was calculated using the Pearson coefficient. 
For example, the uncorrected jet transverse momentum was excluded from this list due to its correlation with the area-based corrected transverse momentum.  Based on these considerations, the following input features were selected:
the jet transverse momentum corrected by the standard area-based method,
the first radial moment of constituent momenta~(jet angularity), the number of constituents within the jet, and the transverse momenta of the eight leading~(highest $\pT$) particles within the jet.
For the training, the applied supervised learning techniques need a target value assigned to each sample, i.e.\ to each jet.
The regression target that is approximated by the correction method is the  jet \pT\ at detector level,  $p^{\rm target}_{\rm T, jet}$.
This is defined as the reconstructed jet transverse momentum multiplied by the jet momentum fraction that is carried by the jet constituents originating from the PYTHIA simulated event,
\begin{equation}
  p^{\rm target}_{\rm T, jet} = \jetraw \, \sum_i p_{\mathrm{T,\;const}\;i}^\mathrm{PYTHIA}/ \sum_i p_{\mathrm{T,\;const}\;i}\,,
\end{equation}
where \jetraw\ is the reconstructed jet transverse momentum before any background correction. 
With this definition of the target jet \pT, the background is also defined implicitly; it consists of all the particles from the thermal model.
As an alternative definition, the target jet \pT\ could also be defined as the detector-level jet \pT\ described in \Sec{sec:reconstruction}.
Since the background influences the jet finding algorithm, these matched jets are conceptually closer to the perfectly corrected jets but have other disadvantages, such as potential mismatches to jets present in the minimum bias data sample, in particular at low transverse momentum.
However, models trained on the two target definitions were shown to have a similar performance within the uncertainty of the measurement.

\section{Supplementary Figures}
\label{sec:appendix}
\begin{figure}[ht!]
   \includegraphics[width=0.49\textwidth]{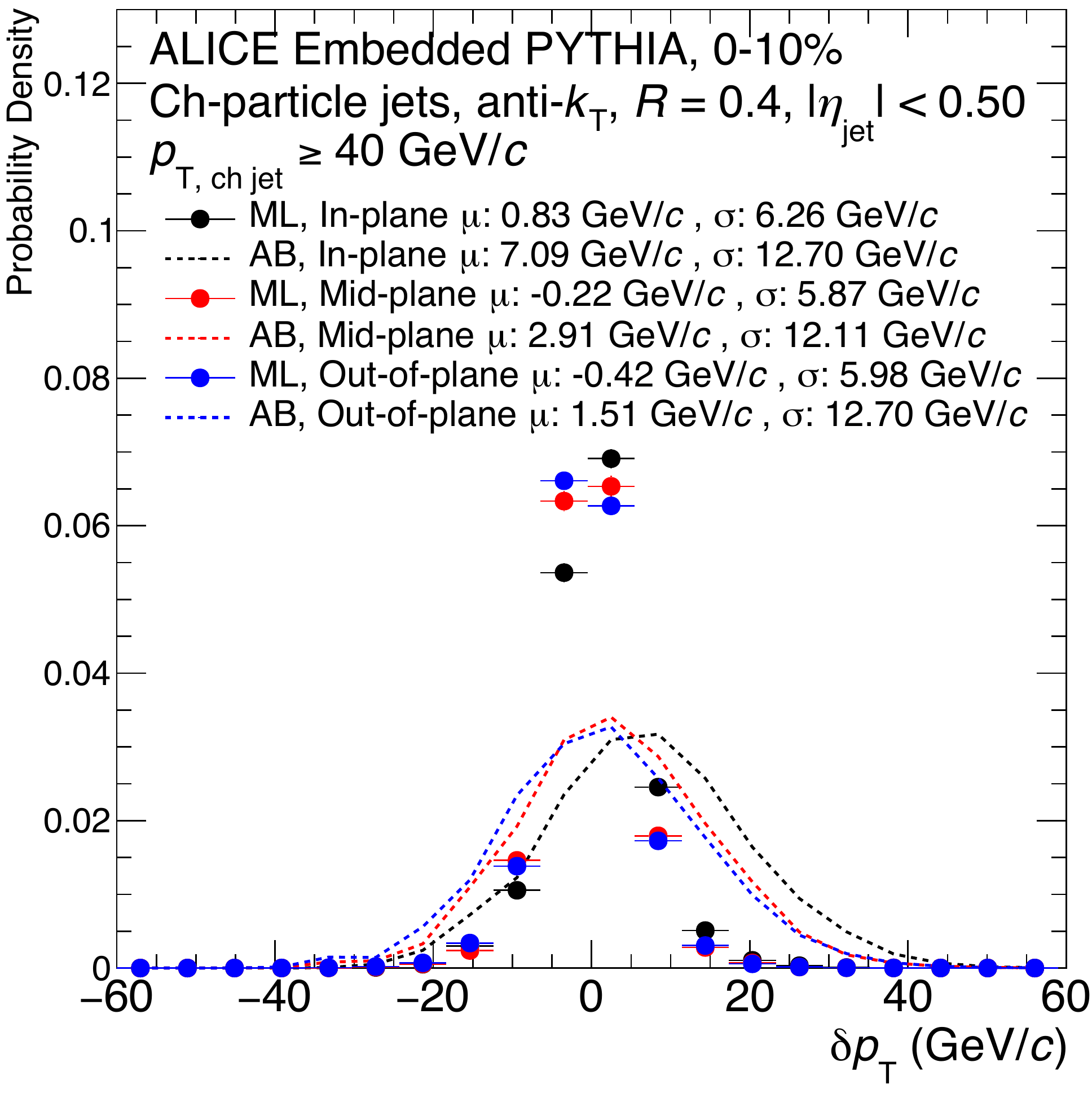}
  \includegraphics[width=0.49\textwidth]{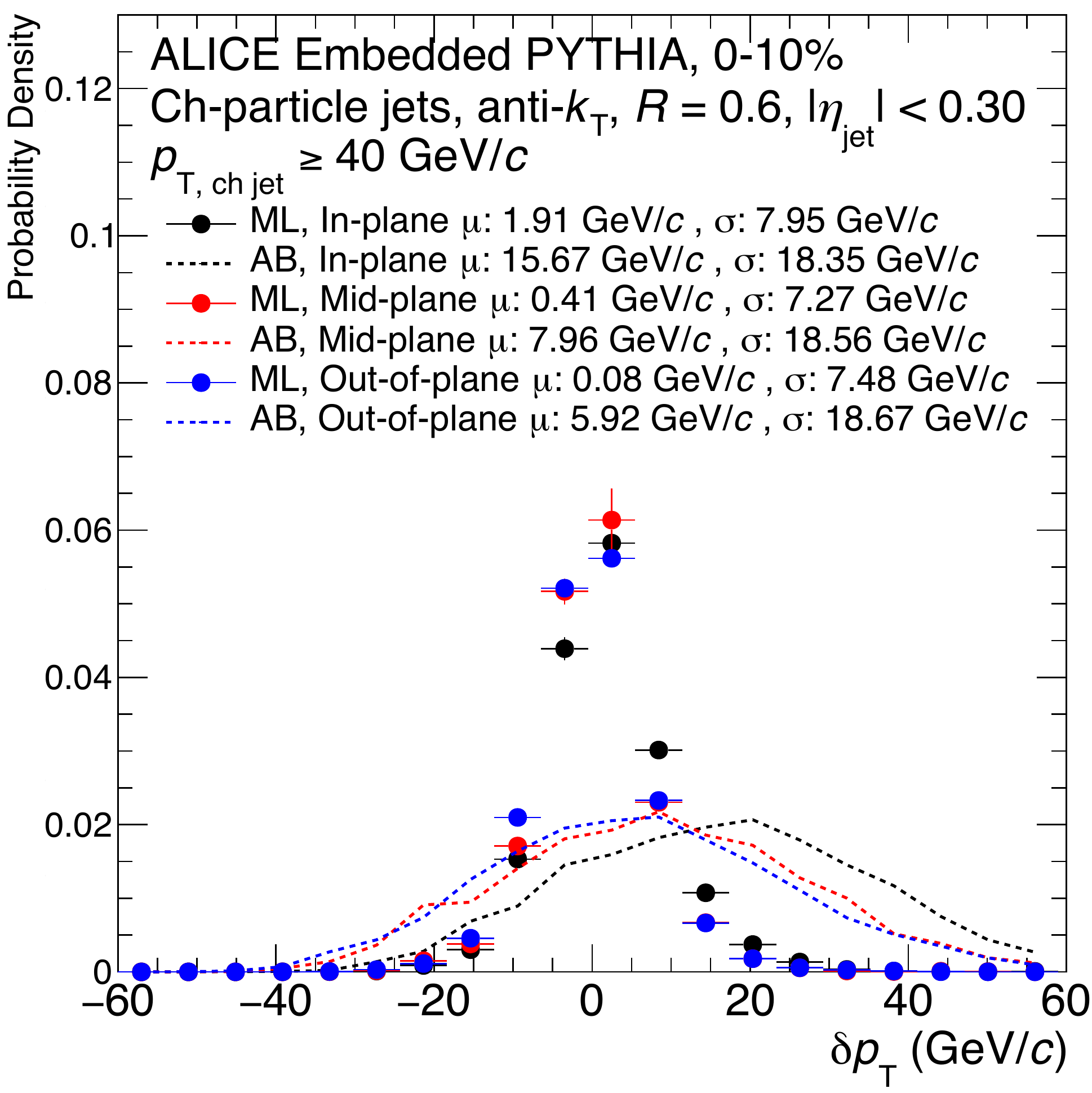}
  \centering
  \caption{The $\delta p_{\rm T}$ distributions for $R = 0.4$ (left) and $R = 0.6$ (right) jets in central (0-10\%) collisions using the area-based (dashed lines) and ML-based (solid markers) corrections.}
  \label{fig:FlowCrosscheck}
\end{figure}

%%%%%%%%%%%%%%%%%%%%%%%%%%%%%%%%%%%%%%%%%%%%%%%%%%%%%%%%%%%%%%%%%%%%%%%%%%%%%%%%%%%%%%%%%%%%%%%%%%%%
One source of fluctuations in the heavy-ion background is correlated fluctuations due to flow-like effects. These flow-like effects generate a momentum-space anisotropy of particles, which are oriented with respect to the event plane. The magnitude of this correlated background grows with the radius of the jet and depends on the angle between the jet axis and the event plane, given by  $\Delta\phi = \phi_{\rm jet} - \psi_{\rm EP}$, where $\psi_{\rm EP}$ represents the event plane orientation. Jets are classified as in-plane $\Delta \phi < 30^{o}$, mid-plane ( $  60^{o} >\Delta \phi > 30^{o}$), or out-of-plane ($\Delta \phi > 60^{o}$). The AB method does not correct explicitly for flow-like effects, which manifests itself as a dependence of the $\delta p_{\rm T}$ distributions on $\Delta\phi$. This can be seen in the dashed lines in \Fig{fig:FlowCrosscheck}, where the mean of the $\delta p_{\rm T}$ distributions is larger for the in-plane case. Such an observation is consistent with not fully accounting for correlated fluctuations. The performance of the ML-based method is shown as solid markers in \Fig{fig:FlowCrosscheck}, where the $\delta p_{\rm T}$ distribution exhibits minimal event plane dependence. This indicates that the ML is able to properly correct for these effects despite omitting the event plane from training. 

\clearpage

\section{The ALICE Collaboration}
\label{app:collab}
% ALICE Collaboration author list for 2023-02-10
\begin{flushleft} 
\small

S.~Acharya\,\orcidlink{0000-0002-9213-5329}\,$^{\rm 126}$, 
D.~Adamov\'{a}\,\orcidlink{0000-0002-0504-7428}\,$^{\rm 86}$, 
A.~Adler$^{\rm 70}$, 
G.~Aglieri Rinella\,\orcidlink{0000-0002-9611-3696}\,$^{\rm 32}$, 
M.~Agnello\,\orcidlink{0000-0002-0760-5075}\,$^{\rm 29}$, 
N.~Agrawal\,\orcidlink{0000-0003-0348-9836}\,$^{\rm 51}$, 
Z.~Ahammed\,\orcidlink{0000-0001-5241-7412}\,$^{\rm 134}$, 
S.~Ahmad\,\orcidlink{0000-0003-0497-5705}\,$^{\rm 15}$, 
S.U.~Ahn\,\orcidlink{0000-0001-8847-489X}\,$^{\rm 71}$, 
I.~Ahuja\,\orcidlink{0000-0002-4417-1392}\,$^{\rm 37}$
% ...

\section*{Affiliation Notes}

$^{\rm I}$ Deceased\\
$^{\rm II}$ Also at: Max-Planck-Institut fur Physik, Munich, Germany\\
$^{\rm III}$ Also at: Italian National Agency for New Technologies, Energy and Sustainable Economic Development (ENEA), Bologna, Italy\\
$^{\rm IV}$ Also at: Dipartimento DET del Politecnico di Torino, Turin, Italy\\
$^{\rm V}$ Also at: Department of Applied Physics, Aligarh Muslim University, Aligarh, India\\
$^{\rm VI}$ Also at: Institute of Theoretical Physics, University of Wroclaw, Poland\\
$^{\rm VII}$ Also at: An institution covered by a cooperation agreement with CERN\\

\section*{Collaboration Institutes}

$^{1}$ A.I. Alikhanyan National Science Laboratory (Yerevan Physics Institute) Foundation, Yerevan, Armenia\\
$^{2}$ AGH University of Krakow, Cracow, Poland\\
$^{3}$ Bogolyubov Institute for Theoretical Physics, National Academy of Sciences of Ukraine, Kiev, Ukraine\\
$^{4}$ Bose Institute, Department of Physics  and Centre for Astroparticle Physics and Space Science (CAPSS), Kolkata, India\\
$^{5}$ California Polytechnic State University, San Luis Obispo, California, United States\\
$^{6}$ Central China Normal University, Wuhan, China\\
$^{7}$ Centro de Aplicaciones Tecnol\'{o}gicas y Desarrollo Nuclear (CEADEN), Havana, Cuba\\
$^{8}$ Centro de Investigaci\'{o}n y de Estudios Avanzados (CINVESTAV), Mexico City and M\'{e}rida, Mexico\\
$^{9}$ Chicago State University, Chicago, Illinois, United States\\
$^{10}$ China Institute of Atomic Energy, Beijing, China\\
$^{11}$ Chungbuk National University, Cheongju, Republic of Korea\\
$^{12}$ Comenius University Bratislava, Faculty of Mathematics, Physics and Informatics, Bratislava, Slovak Republic\\
$^{13}$ COMSATS University Islamabad, Islamabad, Pakistan\\
$^{14}$ Creighton University, Omaha, Nebraska, United States\\
$^{15}$ Department of Physics, Aligarh Muslim University, Aligarh, India\\
$^{16}$ Department of Physics, Pusan National University, Pusan, Republic of Korea\\
$^{17}$ Department of Physics, Sejong University, Seoul, Republic of Korea\\
$^{18}$ Department of Physics, University of California, Berkeley, California, United States\\
$^{19}$ Department of Physics, University of Oslo, Oslo, Norway\\
$^{20}$ Department of Physics and Technology, University of Bergen, Bergen, Norway\\
$^{21}$ Dipartimento di Fisica, Universit\`{a} di Pavia, Pavia, Italy\\
$^{22}$ Dipartimento di Fisica dell'Universit\`{a} and Sezione INFN, Cagliari, Italy\\
$^{23}$ Dipartimento di Fisica dell'Universit\`{a} and Sezione INFN, Trieste, Italy\\
$^{24}$ Dipartimento di Fisica dell'Universit\`{a} and Sezione INFN, Turin, Italy\\
$^{25}$ Dipartimento di Fisica e Astronomia dell'Universit\`{a} and Sezione INFN, Bologna, Italy\\
$^{26}$ Dipartimento di Fisica e Astronomia dell'Universit\`{a} and Sezione INFN, Catania, Italy\\
$^{27}$ Dipartimento di Fisica e Astronomia dell'Universit\`{a} and Sezione INFN, Padova, Italy\\
$^{28}$ Dipartimento di Fisica `E.R.~Caianiello' dell'Universit\`{a} and Gruppo Collegato INFN, Salerno, Italy\\
$^{29}$ Dipartimento DISAT del Politecnico and Sezione INFN, Turin, Italy\\
$^{30}$ Dipartimento di Scienze MIFT, Universit\`{a} di Messina, Messina, Italy\\
$^{31}$ Dipartimento Interateneo di Fisica `M.~Merlin' and Sezione INFN, Bari, Italy\\
$^{32}$ European Organization for Nuclear Research (CERN), Geneva, Switzerland\\
$^{33}$ Faculty of Electrical Engineering, Mechanical Engineering and Naval Architecture, University of Split, Split, Croatia\\
$^{34}$ Faculty of Engineering and Science, Western Norway University of Applied Sciences, Bergen, Norway\\
$^{35}$ Faculty of Nuclear Sciences and Physical Engineering, Czech Technical University in Prague, Prague, Czech Republic\\
$^{36}$ Faculty of Physics, Sofia University, Sofia, Bulgaria\\
$^{37}$ Faculty of Science, P.J.~\v{S}af\'{a}rik University, Ko\v{s}ice, Slovak Republic\\
$^{38}$ Frankfurt Institute for Advanced Studies, Johann Wolfgang Goethe-Universit\"{a}t Frankfurt, Frankfurt, Germany\\
$^{39}$ Fudan University, Shanghai, China\\
$^{40}$ Gangneung-Wonju National University, Gangneung, Republic of Korea\\
$^{41}$ Gauhati University, Department of Physics, Guwahati, India\\
$^{42}$ Helmholtz-Institut f\"{u}r Strahlen- und Kernphysik, Rheinische Friedrich-Wilhelms-Universit\"{a}t Bonn, Bonn, Germany\\
$^{43}$ Helsinki Institute of Physics (HIP), Helsinki, Finland\\
$^{44}$ High Energy Physics Group,  Universidad Aut\'{o}noma de Puebla, Puebla, Mexico\\
$^{45}$ Horia Hulubei National Institute of Physics and Nuclear Engineering, Bucharest, Romania\\
$^{46}$ HUN-REN Wigner Research Centre for Physics, Budapest, Hungary\\
$^{47}$ Indian Institute of Technology Bombay (IIT), Mumbai, India\\
$^{48}$ Indian Institute of Technology Indore, Indore, India\\
$^{49}$ INFN, Laboratori Nazionali di Frascati, Frascati, Italy\\
$^{50}$ INFN, Sezione di Bari, Bari, Italy\\
$^{51}$ INFN, Sezione di Bologna, Bologna, Italy\\
$^{52}$ INFN, Sezione di Cagliari, Cagliari, Italy\\
$^{53}$ INFN, Sezione di Catania, Catania, Italy\\
$^{54}$ INFN, Sezione di Padova, Padova, Italy\\
$^{55}$ INFN, Sezione di Pavia, Pavia, Italy\\
$^{56}$ INFN, Sezione di Torino, Turin, Italy\\
$^{57}$ INFN, Sezione di Trieste, Trieste, Italy\\
$^{58}$ Inha University, Incheon, Republic of Korea\\
$^{59}$ Institute for Gravitational and Subatomic Physics (GRASP), Utrecht University/Nikhef, Utrecht, Netherlands\\
$^{60}$ Institute of Experimental Physics, Slovak Academy of Sciences, Ko\v{s}ice, Slovak Republic\\
$^{61}$ Institute of Physics, Homi Bhabha National Institute, Bhubaneswar, India\\
$^{62}$ Institute of Physics of the Czech Academy of Sciences, Prague, Czech Republic\\
$^{63}$ Institute of Space Science (ISS), Bucharest, Romania\\
$^{64}$ Institut f\"{u}r Kernphysik, Johann Wolfgang Goethe-Universit\"{a}t Frankfurt, Frankfurt, Germany\\
$^{65}$ Instituto de Ciencias Nucleares, Universidad Nacional Aut\'{o}noma de M\'{e}xico, Mexico City, Mexico\\
$^{66}$ Instituto de F\'{i}sica, Universidade Federal do Rio Grande do Sul (UFRGS), Porto Alegre, Brazil\\
$^{67}$ Instituto de F\'{\i}sica, Universidad Nacional Aut\'{o}noma de M\'{e}xico, Mexico City, Mexico\\
$^{68}$ iThemba LABS, National Research Foundation, Somerset West, South Africa\\
$^{69}$ Jeonbuk National University, Jeonju, Republic of Korea\\
$^{70}$ Johann-Wolfgang-Goethe Universit\"{a}t Frankfurt Institut f\"{u}r Informatik, Fachbereich Informatik und Mathematik, Frankfurt, Germany\\
$^{71}$ Korea Institute of Science and Technology Information, Daejeon, Republic of Korea\\
$^{72}$ KTO Karatay University, Konya, Turkey\\
$^{73}$ Laboratoire de Physique Subatomique et de Cosmologie, Universit\'{e} Grenoble-Alpes, CNRS-IN2P3, Grenoble, France\\
$^{74}$ Lawrence Berkeley National Laboratory, Berkeley, California, United States\\
$^{75}$ Lund University Department of Physics, Division of Particle Physics, Lund, Sweden\\
$^{76}$ Nagasaki Institute of Applied Science, Nagasaki, Japan\\
$^{77}$ Nara Women{'}s University (NWU), Nara, Japan\\
$^{78}$ National and Kapodistrian University of Athens, School of Science, Department of Physics , Athens, Greece\\
$^{79}$ National Centre for Nuclear Research, Warsaw, Poland\\
$^{80}$ National Institute of Science Education and Research, Homi Bhabha National Institute, Jatni, India\\
$^{81}$ National Nuclear Research Center, Baku, Azerbaijan\\
$^{82}$ National Research and Innovation Agency - BRIN, Jakarta, Indonesia\\
$^{83}$ Niels Bohr Institute, University of Copenhagen, Copenhagen, Denmark\\
$^{84}$ Nikhef, National institute for subatomic physics, Amsterdam, Netherlands\\
$^{85}$ Nuclear Physics Group, STFC Daresbury Laboratory, Daresbury, United Kingdom\\
$^{86}$ Nuclear Physics Institute of the Czech Academy of Sciences, Husinec-\v{R}e\v{z}, Czech Republic\\
$^{87}$ Oak Ridge National Laboratory, Oak Ridge, Tennessee, United States\\
$^{88}$ Ohio State University, Columbus, Ohio, United States\\
$^{89}$ Physics department, Faculty of science, University of Zagreb, Zagreb, Croatia\\
$^{90}$ Physics Department, Panjab University, Chandigarh, India\\
$^{91}$ Physics Department, University of Jammu, Jammu, India\\
$^{92}$ Physics Program and International Institute for Sustainability with Knotted Chiral Meta Matter (SKCM2), Hiroshima University, Hiroshima, Japan\\
$^{93}$ Physikalisches Institut, Eberhard-Karls-Universit\"{a}t T\"{u}bingen, T\"{u}bingen, Germany\\
$^{94}$ Physikalisches Institut, Ruprecht-Karls-Universit\"{a}t Heidelberg, Heidelberg, Germany\\
$^{95}$ Physik Department, Technische Universit\"{a}t M\"{u}nchen, Munich, Germany\\
$^{96}$ Politecnico di Bari and Sezione INFN, Bari, Italy\\
$^{97}$ Research Division and ExtreMe Matter Institute EMMI, GSI Helmholtzzentrum f\"ur Schwerionenforschung GmbH, Darmstadt, Germany\\
$^{98}$ Saga University, Saga, Japan\\
$^{99}$ Saha Institute of Nuclear Physics, Homi Bhabha National Institute, Kolkata, India\\
$^{100}$ School of Physics and Astronomy, University of Birmingham, Birmingham, United Kingdom\\
$^{101}$ Secci\'{o}n F\'{\i}sica, Departamento de Ciencias, Pontificia Universidad Cat\'{o}lica del Per\'{u}, Lima, Peru\\
$^{102}$ Stefan Meyer Institut f\"{u}r Subatomare Physik (SMI), Vienna, Austria\\
$^{103}$ SUBATECH, IMT Atlantique, Nantes Universit\'{e}, CNRS-IN2P3, Nantes, France\\
$^{104}$ Sungkyunkwan University, Suwon City, Republic of Korea\\
$^{105}$ Suranaree University of Technology, Nakhon Ratchasima, Thailand\\
$^{106}$ Technical University of Ko\v{s}ice, Ko\v{s}ice, Slovak Republic\\
$^{107}$ The Henryk Niewodniczanski Institute of Nuclear Physics, Polish Academy of Sciences, Cracow, Poland\\
$^{108}$ The University of Texas at Austin, Austin, Texas, United States\\
$^{109}$ Universidad Aut\'{o}noma de Sinaloa, Culiac\'{a}n, Mexico\\
$^{110}$ Universidade de S\~{a}o Paulo (USP), S\~{a}o Paulo, Brazil\\
$^{111}$ Universidade Estadual de Campinas (UNICAMP), Campinas, Brazil\\
$^{112}$ Universidade Federal do ABC, Santo Andre, Brazil\\
$^{113}$ Universitatea Nationala de Stiinta si Tehnologie Politehnica Bucuresti, Bucharest, Romania\\
$^{114}$ University of Cape Town, Cape Town, South Africa\\
$^{115}$ University of Houston, Houston, Texas, United States\\
$^{116}$ University of Jyv\"{a}skyl\"{a}, Jyv\"{a}skyl\"{a}, Finland\\
$^{117}$ University of Kansas, Lawrence, Kansas, United States\\
$^{118}$ University of Liverpool, Liverpool, United Kingdom\\
$^{119}$ University of Science and Technology of China, Hefei, China\\
$^{120}$ University of South-Eastern Norway, Kongsberg, Norway\\
$^{121}$ University of Tennessee, Knoxville, Tennessee, United States\\
$^{122}$ University of the Witwatersrand, Johannesburg, South Africa\\
$^{123}$ University of Tokyo, Tokyo, Japan\\
$^{124}$ University of Tsukuba, Tsukuba, Japan\\
$^{125}$ Universit\"{a}t M\"{u}nster, Institut f\"{u}r Kernphysik, M\"{u}nster, Germany\\
$^{126}$ Universit\'{e} Clermont Auvergne, CNRS/IN2P3, LPC, Clermont-Ferrand, France\\
$^{127}$ Universit\'{e} de Lyon, CNRS/IN2P3, Institut de Physique des 2 Infinis de Lyon, Lyon, France\\
$^{128}$ Universit\'{e} de Strasbourg, CNRS, IPHC UMR 7178, F-67000 Strasbourg, France, Strasbourg, France\\
$^{129}$ Universit\'{e} Paris-Saclay, Centre d'Etudes de Saclay (CEA), IRFU, D\'{e}partment de Physique Nucl\'{e}aire (DPhN), Saclay, France\\
$^{130}$ Universit\'{e}  Paris-Saclay, CNRS/IN2P3, IJCLab, Orsay, France\\
$^{131}$ Universit\`{a} degli Studi di Foggia, Foggia, Italy\\
$^{132}$ Universit\`{a} del Piemonte Orientale, Vercelli, Italy\\
$^{133}$ Universit\`{a} di Brescia, Brescia, Italy\\
$^{134}$ Variable Energy Cyclotron Centre, Homi Bhabha National Institute, Kolkata, India\\
$^{135}$ Warsaw University of Technology, Warsaw, Poland\\
$^{136}$ Wayne State University, Detroit, Michigan, United States\\
$^{137}$ Yale University, New Haven, Connecticut, United States\\
$^{138}$ Yonsei University, Seoul, Republic of Korea\\
$^{139}$  Zentrum  f\"{u}r Technologie und Transfer (ZTT), Worms, Germany\\
$^{140}$ Affiliated with an institute covered by a cooperation agreement with CERN\\
$^{141}$ Affiliated with an international laboratory covered by a cooperation agreement with CERN.\\

\end{flushleft} 
      %%%%%%% get the latest version before submitting

%==========================================================%
\end{document}